\documentclass[12pt]{article}
\usepackage[pdftex]{graphicx}
\usepackage{setspace}
\usepackage{booktabs}
\usepackage{mathpazo}

\usepackage[utf8]{inputenc}
\DeclareUnicodeCharacter{FB01}{fi} % added based on error, according to https://tex.stackexchange.com/questions/417749/package-inputenc-error-unicode-char-%EF%AC%81-ufb01

\usepackage[T1]{fontenc}
\usepackage{times}
\usepackage{geometry}
\usepackage{color}
\usepackage{amsthm}
\usepackage{float}
\usepackage{amsmath}
\usepackage{graphicx}
\usepackage{scrextend}

\usepackage{array}
\newcolumntype{H}{>{\setbox0=\hbox\bgroup}c<{\egroup}@{}}

\usepackage{lscape}
\usepackage{tikz}
\usetikzlibrary{matrix}
\makeatletter
\usepackage[linewidth=1pt]{mdframed}
\usepackage{dcolumn}
\usepackage[many]{tcolorbox}
\usepackage{multirow}
\usepackage{tikz}
\usetikzlibrary{patterns,decorations.pathreplacing}
\usepackage[font=small,labelfont={bf,sc}, skip=2pt]{caption}
\usetikzlibrary{arrows}
\usepackage{rotating}
\usepackage{pgfplots}
\usepackage{adjustbox}
\usepackage{subcaption}
\usepackage{enumitem}
\usepackage{pdflscape}
\usepackage{makecell}
\usepackage{pdflscape}
\usepackage{booktabs}
\usepackage{tabularx}
\usepackage{geometry}
\usepackage{chngcntr}
\usepackage{cancel}
\usepackage{csvsimple}
\usepackage{siunitx}

	\newcommand{\pval}[1]{%
		\ifnum\fpeval{(#1) < 0.001}=1 $<$0.001\else
		\num[round-mode=places,round-precision=3]{#1}\fi}
\usepackage[style=authoryear-comp,firstinits=true,citestyle=authoryear,natbib=true,backend=bibtex,doi=false,isbn=false,url=false,maxcitenames=2,maxbibnames=99,date=year]{biblatex}
\AtBeginBibliography{\normalsize }

\DeclareNameAlias{sortname}{last-first}

\renewbibmacro{in:}{}
\renewbibmacro*{volume+number+eid}{%
	\printfield{volume}%
	\setunit*{\addnbspace}% NEW (optional); there's also \addnbthinspace
	\printfield{number}%
	\setunit{\addcomma\space}%
	\printfield{eid}}
\DeclareFieldFormat[article]{number}{\mkbibparens{#1}}

\bibliography{biblio}

\makeatletter
\def\thmhead@plain#1#2#3{%
	\thm@notefont{}% same as heading font
	\thmname{#1}\thmnumber{\@ifnotempty{#1}{ }\@upn{#2}}%
	\thmnote{ {\the\thm@notefont#3}}}
\let\thmhead\thmhead@plain
\itshape % body font
\makeatother

\DeclareCiteCommand{\citeyear}
{}
{\bibhyperref{\printdate}}
{\multicitedelim}
{}

\usepackage{bbm}
\newtheorem*{definition*}{Definition}
\newtheorem*{assumption*}{Assumption}

\newtheorem*{lemma*}{Lemma}
\newtheorem{lemma}{Lemma}
\newtheorem*{proposition*}{Proposition}
\newtheorem{proposition}{Proposition}
\newtheorem*{conjecture*}{Conjecture}
\newtheorem*{theorem*}{Theorem}
\newtheorem{theorem}{Theorem}
\newtheorem*{corollary*}{Corollary}
\newtheorem{corollary}{Corollary}
\newtheorem*{remark*}{Remark}
\newtheorem{remark}{Remark}
\newtheorem*{example*}{Example}
\newtheorem{example}{Example}

\usepackage{hyperref}

\newcommand{\indep}{\perp\!\!\!\!\perp}

\usepackage{nameref, zref-xr,zref-hyperref,zref-user}
\zxrsetup{toltxlabel}
\DeclareSymbolFont{operators}   {OT1}{lmr} {m}{n}
\DeclareSymbolFont{letters}     {OML}{cmm} {m}{it}
\DeclareSymbolFont{symbols}     {OMS}{cmsy}{m}{n}
   \SetSymbolFont{operators}{bold}{OT1}{cmr}{b}{n}
   \SetSymbolFont{letters}{bold}{OML}{cmm}{b}{it}
   \SetSymbolFont{symbols}{bold}{OMS}{cmsy}{b}{n}
\usetikzlibrary{arrows, decorations.markings}

\tikzstyle{vecArrow} = [thick, decoration={markings,mark=at position
   1 with {\arrow[semithick]{open triangle 60}}},
   double distance=1.4pt, shorten >= 5.5pt,
   preaction = {decorate},
   postaction = {draw,line width=1.4pt, white,shorten >= 4.5pt}]
\tikzstyle{innerWhite} = [semithick, white,line width=1.4pt, shorten >= 4.5pt]
\newcolumntype{Y}{>{\centering\arraybackslash}X}
\begin{document}

\title{\fontsize{24}{28}{Partial Identification from Bunching at Kinks and Notches}}
\author{\Large{Leonard Goff}\thanks{Department of Economics, University of Calgary, \href{mailto:leonard.goff@ucalgary.ca}{leonard.goff@ucalgary.ca}. This paper represents a new version of the econometric theory portion of \citet{goff2024treatmenteffectsbunchingdesigns}: ``Treatment Effects in Bunching Designs: The Impact of Mandatory Overtime Pay on Hours''. I thank Daniel Feenberg for facilitating data access via the NBER, and Tao Fu for excellent research assistance. This paper draws on research supported by the Social Sciences and Humanities Research Council of Canada. Some ancillary materials are available in an Online Supplement at \url{www.leonardgoff.com/resources/bunching_onlinesupplement.pdf}.}}
\date{}%

\maketitle
\thispagestyle{empty}

\begin{abstract}
This paper proposes a partial identification approach to using choices around a kink or notch as a means to overcome endogeneity in a general nonparametric choice model. I show that observed choices are informative about the joint distribution of two counterfactual choices, which leads naturally to analyzing bunching using tools from causal inference. I define as the parameter of interest an average treatment effect among bunchers, which nests the traditional elasticity parameter while allowing for heterogeneity in responsiveness. Identification of the buncher ATE (and hence the elasticity) requires the researcher to extrapolate the distribution of each counterfactual choice beyond where it is observed. I introduce a flexible family of nonparametric shape restrictions to obtain partial identification, and propose a method for empirical validation of this approach using changes in the location of the kink/notch or using the distribution from another comparison group. I find that a log-concavity version of the distributional assumption is widely---though not universally---supported across the empirical literature, and leads to informative bounds in a canonical tax kink application. The approach accommodates diffuse bunching, at the expense of wider bounds.
\end{abstract}

\newpage
\pagenumbering{arabic}

\section{Introduction}

Beginning with the work of \citet{saez2010}, \citet{chetty2011} and \citet{kleven2013}, a large empirical literature over the last fifteen years has used bunching around choice-set kinks or notches to identify behavioral elasticities. The classic examples estimate the responsiveness of income to the marginal tax rate, using an observed spike (or ``bunching'') in the distribution of income near the threshold between two tax brackets. Given assumptions about the distribution of heterogeneity, observed choices around the threshold identify the elasticity parameter from a simple isoelastic model of labor supply, despite the very strong endogeneity that confounds the relationship between individuals' incomes and their tax rates.

The ``bunching design'', as it is sometimes called, has intuitive appeal because bunching provides compelling evidence of \textit{some} behavioral response to the change in incentives that occurs at a kink or notch, provided that the distribution under a smooth tax schedule would have also been smooth. \citet{blomquist_bunching_2019} show however that to point identify the magnitude of the elasticity parameter in the traditional model, the bunching design requires much more of the econometrician. One must be willing to use strong (typically parametric) assumptions to extrapolate the distribution of choices observed on one side of the threshold to the other, where it is not observed.

In this paper I suggest a means of implementing the bunching design for partial identification that makes only nonparametric assumptions, both in the choice model and in this extrapolation step. In a general choice model, responsiveness to the incentives that change at a kink or notch can be heterogeneous. I therefore focus on the average response among bunchers, who comprise the subpopulation for whom the distribution of choices around the threshold is most informative. Drawing on ideas from the causal inference literature, I refer to this parameter as the ``buncher ATE''. I show that the buncher ATE is bounded from above and from below if the researcher makes a concavity-type assumption about the distribution of counterfactual choices, and imposes a local restriction on deviations from a rank invariance property. I propose a means to assess this extrapolation assumption empirically by appealing to a reference distribution, for example the distribution of choices before a reform that introduces the kink/notch or moves the threshold.

I proceed by first laying out in Section \ref{sec:policyenviro} a general policy environment with a kink, notch, or both at a threshold $k$, and relaxing the parametric utility model (the isoelastic model) assumed in most existing formal results for the bunching design. The key assumption of the general choice model is convexity of preferences, though I also derive a one-sided bound assuming only the weak axiom of revealed preference. I obtain a mapping that holds even in this general setup between observed choices $Y_i$ and two counterfactual choices: $Y_{0i}$ and $Y_{1i}$. At a linear tax kink where the tax rate jumps from $\tau_0$ to $\tau_1$, $Y_{0i}$ reflects the income that taxpayer $i$ would choose subject to a constant marginal tax at rate $\tau_0$, and $Y_{1i}$ the income $i$ would choose subject to a constant marginal tax at rate $\tau_1$ (less a lump sum that makes the two tax schedules cross at $k$). The buncher ATE upon a log transformation to the outcome, $\mathbbm{E}[\ln Y_{0i} -\ln Y_{1i}|Y_i=k]$, then captures an average reduced-form elasticity of income with respect to the net-of-tax rate among bunchers. The buncher ATE in levels $\mathbbm{E}[Y_{0i}-Y_{1i}|Y_i=k]$ instead gives e.g. a dollar-denominated effect.

In Section \ref{sec:concavity} I introduce a family of local concavity assumptions on the marginal distribution of each counterfactual choice to partially identify the buncher ATE (whether in logs or levels). In one extreme, this family yields point identification, recovering a small-kink approximation from the literature that assumes the distribution of counterfactual choices is locally uniform \citep{kleven_bunching_2016}. In the other extreme, the family contains all distributions and the bounds become uninformative. In between, I highlight two special cases from the family that I argue are particularly empirically relevant: (i) assuming a locally declining counterfactual density  (concavity of the CDF); or (ii) that the potential outcomes $Y_0$ and $Y_1$ are each locally ``bi log concave'' (BLC). A declining density (i) is quite natural in many applications, as bunching is often observed at tax kinks and notches located in the right tail of an income or wealth distribution. This assumption yields only a lower bound on the buncher ATE.\footnote{I show that at a kink, convex preferences are not needed for this lower bound: revealed preference alone suffices.} Local BLC (ii) assumes that the CDF and survival functions of $Y_0$ and $Y_1$ are each log-concave over their \textit{bunching interval}, and yields bounds in both directions given a local restriction on deviations from rank invariance.

While there are theoretical reasons to expect BLC to hold in many applications, I propose a workflow for researchers to test whether BLC (or another of the related assumptions from the family) is empirically well-supported in a given setting. When the researcher has only a single cross section of data, one can only test the implications of BLC on one side of the threshold for each potential outcome, and then assume that the restriction continues to hold on the other side. However, researchers often also have access to a reference distribution to compare against, such as the distribution of income before or after a tax threshold changes, or the distribution of income in another time period or jurisdiction where the kink or notch is absent. Such reference distributions are already used informally in the empirical literature, interpreted as providing a reasonable counterfactual in the absence of the kink/notch.

I show that confirming BLC in the reference distribution is sufficient to conclude that BLC holds for $Y_0$ and $Y_1$ provided that non-linearities in the Q-Q (quantile vs. quantile) relationship between the observed and reference distributions are compensated by slack in the corresponding inequalities that define BLC. While the relevant Q-Q relationship can only be partially observed (on one side of the kink/notch for a given potential outcome), BLC can often be tested globally in the reference distribution. As such, this method provides evidence of BLC across the bunching interval where $Y_0,Y_1$ are unobserved. The remaining assumption required of the researcher makes precise the sense in which the reference distribution must continue to provide a ``good counterfactual'': continued approximate Q-Q linearity across the bunching interval.

I apply this approach to a set of 27 published papers from leading economics journals that employ the bunching design, and find the BLC is typically, though not universally, well supported. Using image data extraction tools, I scrape numerical distributions from the plots displayed in each paper. In two thirds of the papers, I find that the authors present a reference distribution as a counterfactual, and verify approximate linearity of the Q-Q relationship across the interval where the distributions can naturally be compared. In nearly all cases, BLC in the reference distribution appears to the eye to hold when visualized graphically. I then implement a statistical test of BLC in the reference distribution using tools for testing moment inequalities. While I find evidence of violations of global BLC in a substantial number of the papers considered (though still a minority), I find that the scale of the deviations is often small, and that they mainly occur for non-monetary choice variables. In these cases more general assumptions from the family of concavity restrictions I consider can be used, which provide wider bounds. In many of these cases, the lower bound on the buncher ATE under monotonicity is also still quite plausibly valid.

To render the identification results of this paper useful to practitioners, I also adapt them to settings in which bunching presents as a diffuse excess mass (rather than an idealized point mass), a phenomenon typically explained by ``optimization errors''. I treat the case without optimization errors in the main text for brevity and to present the main ideas clearly. However, my results suggest a new approach to tackling the problem of diffuse bunching arising from optimization errors, frictions, or measurement error, which I develop in Appendix \ref{sec:optimizationerror}. The cost to maintaining tractability without parametric assumptions is twofold: (i) bounds become wider due to a larger extrapolation region; and (ii) there is no longer a sharply identified ``buncher'' sub-population. I generalize the parameter of interest to the average treatment effect over a larger population that includes all individuals who ``intend'' to bunch, recovering the buncher population exactly when there are no errors. I then adapt the bounds to this more general setting.

I apply my identification results to the canonical bunching design application of a kink in the US Earned Income Tax Credit (EITC). I use the introduction of the EITC in 1975 to provide evidence on the local BLC assumption, using the pre-EITC distribution of income as a reference distribution for the distribution in the early years of the program. I then estimate bounds using a more recent sample analyzed by \citet{saez2010}. I present bounds on the buncher average elasticity, yielding an interval estimate under BLC of the distribution of log taxable income. I also present bounds for the buncher ATE in levels, which captures a dollar-denominated reduction in earnings among bunchers due to the policy variation induced by the kink, under BLC of taxable income. The bounds relax \citet{saez2010}'s strong extrapolation assumption of a linear density, and also reflect imprecise knowledge due to diffuse bunching. This empirical application also illustrates the identifying power of successive relaxations of assumptions regarding the degree of heterogeneity in responsiveness, moving beyond the strong rank invariance assumption from the standard isoelastic model. 

This paper contributes to debates regarding the credibility of the bunching design. Empirical researchers often acknowledge warnings of \citet{blomquist_bunching_2019} and \citet{bertanha_better_2018} that inferences about the elasticity parameter hinge on an extrapolation assumption, but then proceed with the strongest version of this assumption to obtain point identification: following a polynomial interpolation method proposed by \citet{chetty2011}. As often applied, this method is not internally consistent, though it can be rescued (\citealt{song2025identificationinferencegeneralbunching}, see also \citealt{pollinger_kinks_2020}). My approach to identification instead acknowledges the difficulty of extrapolation explicitly through partial identification. Relative to \cite{blomquist_bunching_2019} and \citet{bertanha_better_2018} my setup generalizes away from the isoelastic model, and my partial identification approach differs from ones that they each propose. \citet{blomquist_bunching_2019} propose a version of density monotonicity, but their bounds rely on additional structure specific to the isoelastic model. \citet{bertanha_better_2018} propose a Lipschitz condition, which could also be adapted to my general choice model as an identifying assumption. However while BLC is inherited from a reference distribution under approximate Q-Q linearity, the same condition does not resolve the identifying parameter value in a Lipschitz approach. To my knowledge, mine is the first paper to give a unified treatment of both kinks and notches in a nonparametric choice model, and to give formal identification results for notches with optimization errors and without a small notch approximation. 

\section{Policy environment and choice model} \label{sec:policyenviro}
\subsection{Policy environment}
I begin by introducing a general policy environment that abstracts from the conventional piece-wise linear setting familiar from tax kinks.  A population of observational units is indexed by $i$. For each $i$, a decision-maker chooses a pair $(z,\mathbf{x})$ subject to a constraint of the form:
\begin{equation} \label{eq:bc}
	z \ge \textrm{max} \{T_{0i}(\mathbf{x}), T_{1i}(\mathbf{x})\}
\end{equation}
where the scalar $z$ is a ``bad'' to the decision-maker (e.g. a cost such as tax liability). The functions $T_{0i}(\mathbf{x})$ and $T_{1i}(\mathbf{x})$ are continuous and weakly convex functions of $\mathbf{x}$, which can be a vector. There exist continuous scalar functions $y_i(\mathbf{x})$ and a scalar $k$ such that:
\begin{equation} \label{eq:kinkcharacterization}
	T_{0i}(\mathbf{x}) > T_{1i}(\mathbf{x}) \textrm{ whenever } y_i(\mathbf{x}) < k \quad \textrm{ and } \quad T_{0i}(\mathbf{x}) < T_{1i}(\mathbf{x}) \textrm{ whenever } y_i(\mathbf{x}) > k
\end{equation}
\noindent The functions $T_{0i}$, $T_{1i}$ feature two regimes of costs that are separated by the condition $y_i(\mathbf{x})=k$, and the above conditions imply that $T_{0i}(\mathbf{x})=T_{1i}(\mathbf{x})$ iff $y_i(\mathbf{x})=k$. 

\begin{figure} 
	\centering
	\begin{subfigure}[b]{0.48\textwidth}
		\centering
		\includegraphics[height=2.5in]{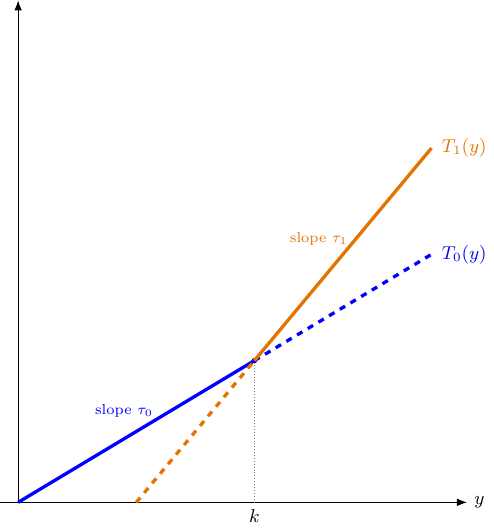}
		\caption{A standard tax kink}
	\end{subfigure}
	\hfill
	\begin{subfigure}[b]{0.48\textwidth}
		\centering
		\includegraphics[height=2.5in]{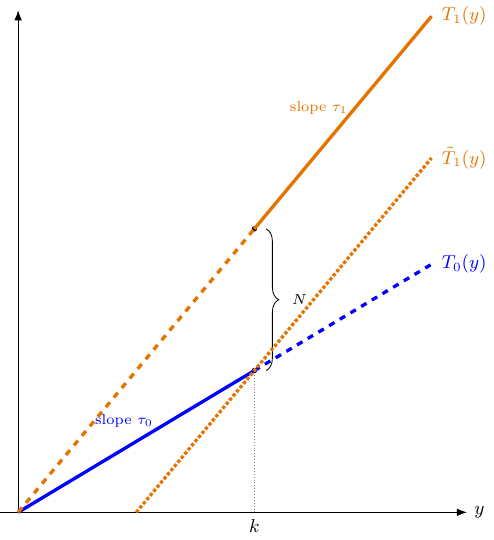}
		\caption{A proportional tax notch}
	\end{subfigure}
	\caption{Examples of a standard univariate kink (left) and a proportional notch (right), which combines a kink with a notch with $N = k(\tau_1-\tau_0)$. In both cases the solid lines yield the actual cost function $T(y)$, while the dashed and dotted lines depict counterfactual cost functions. \label{fig:kinknotch}}
\end{figure}

Define $T_{i}(\mathbf{x}) := \textrm{max} \{T_{0i}(\mathbf{x}), T_{1i}(\mathbf{x})\}$, denoting the actual constraint function that applies to $z$. The function $T_{i}(\mathbf{x})$ is typically ``kinked'' because while the function $T_{i}(\mathbf{x})$ is continuous, it will generally be non-differentiable at the $\mathbf{x}$ for which $y_i(\mathbf{x})=k$. As a maximum of two convex functions, it is itself convex. While the functions $T_0$, $T_1$ and $y$ can all depend on $i$, I will typically suppress this dependency for ease of exposition. The value $k$ is assumed to be known by the researcher.\footnote{I also take $k$ to be common to all units, but this comes at no cost to generality if heterogeneous $k_i$ were observed by the researcher, since one could then redefine $y_i(\mathbf{x})\rightarrow y_i(\mathbf{x})-k_i$ and then set $k=0$.} The cost functions need not be linear or monotone.

\begin{example*}[(standard tax kink)]
	$y_i(\mathbf{x})$ denotes pre-tax income which may depend on several choices $\mathbf{x}$ made by the worker (hours of work, employer, holding multiple jobs, effort). Meanwhile $T_{0i}(\mathbf{x})=\tau_0 \cdot y_i(\mathbf{x})+t_0$ is tax liability at a lower tax rate $\tau_0$, where $t_0$ is tax liability at zero income, and $T_{1i}(\mathbf{x})=\tau_1 \cdot y_i(\mathbf{x})+t_0-(\tau_1-\tau_0) \cdot k$ is tax at a higher tax rate $\tau_1 >\tau_0$ but with the same tax liability when $y_i(\mathbf{x})=k$. In this example $T_0$ and $T_1$ only depend on $\mathbf{x}$ through $y_i(\mathbf{x})$. We can thus write \eqref{eq:bc} as $z \ge T(y)$ where $T(y)=\tau_0 \cdot y + (\tau_1-\tau_0) \cdot  (y-k)\cdot \mathbbm{1}(y \ge k)+t_0$ where $y = y_i(\mathbf{x})$. In this example $N=0$. The settings of \citet{saez2010} and \citet{chetty2011} are of this form.
\end{example*}
%\noindent Note in the standard tax example above that the cost schedule $T_{1}$ involves a lump-sum transfer of $(\tau_1-\tau_0)\cdot k$ to the taxpayer. This makes the schedules cross when income is equal to $k$.

To accommodate a ``notch'' occurring alongside a kink, let $T_{0i}$ and $\tilde{T}_{1i}$ be continuous, weakly convex cost functions satisfying Eq. \eqref{eq:kinkcharacterization} (i.e. crossing when $y_i(\mathbf{x})=k$), but let the budget constraint for each individual $i$ now be
\begin{equation} \label{eq:bcnotch}
	z \ge \textrm{max} \{T_{0i}(\mathbf{x}), \tilde{T}_{1i}(\mathbf{x})\} + N_i\cdot \mathbbm{1}(y_i(\mathbf{x}) > k)
\end{equation}
for some $N_i \ge 0$.\footnote{I focus on cases in which a notch occurs alongside a \textit{convex} kink (if there is any kink at all). Most notches fall into this category. However, while a concave kink on its own does not lead to bunching, a concave kink coupled with a notch can, for example in the case of the Danish foreigner tax scheme studied by \citet{kleven2014}.} in In Eq. \eqref{eq:bcnotch}, $\mathbbm{1}(y_i(\mathbf{x}) > k)$  invokes a strict inequality in order to yield a prediction of bunching exactly at $k$, rather than ``just below'' $k$. This is for ease of notation and makes no practical difference relative to $\mathbbm{1}(y_i(\mathbf{x}) \ge k)$. %To generalize this type of setting, one would replace the max operator in \eqref{eq:bcnotch} with a min, and allow $N_i <0$ (or redefine $y \rightarrow -y$).

Define $T_i(\mathbf{x})$ to be the RHS of Eq. \eqref{eq:bcnotch}, again giving the actual constraint function facing the decision-maker, and let $T_{1i}(\mathbf{x}) = \tilde{T}_{1i}(\mathbf{x})+N$. Then $T_{i}(\mathbf{x}) = T_{0i}(\mathbf{x})$ when $y_i(\mathbf{x}) < k$ and $T_{i}(\mathbf{x}) = T_{1i}(\mathbf{x})$ when $y_i(\mathbf{x}) > k$, by \eqref{eq:kinkcharacterization}. In the case of $N_i=0$ for all $i$, $\tilde{T}_1=T_1$ and we recover the setting of a kink without a notch, which I call a ``pure kink''. On the other hand, this framework accommodates a ``pure notch'' (i.e. without an accompanying kink) when $\tilde{T}_1 = T_0$. The framework also accommodates $N_i$ varying by individual, though most applications $N$ is common.\footnote{As an example with heterogeneous $N_i$, \citet{defusco2017} studies the notch introduced by preferential mortgage interest rates for loans below a regulatory cap. The notch $N_i$ in interest rates at $k$ varies by individual $i$.}
\begin{example*}[(standard tax notch)]
	As before, $y_i(\mathbf{x})$ denotes pre-tax income which may depend on several choices $\mathbf{x}$ made by the worker (hours of work, employer, holding multiple jobs, effort), and again $T_0$ and $T_1$ only depend on $\mathbf{x}$ through $y_i(\mathbf{x})$. In particular, $T_{0i}(\mathbf{x})=\tau_0 \cdot y_i(\mathbf{x})+t_0$, $\tilde{T}_{1i}(\mathbf{x})=\tau_1 \cdot y_i(\mathbf{x})+t_0-(\tau_1-\tau_0) \cdot k$ with $\tau_1 >\tau_0$, and now $N>0$ represents an additional lump-sum tax levied when $y_i(\mathbf{x}) \ge k$. A common special case (e.g. \citealt{kleven2013}) is a ``proportional notch'', in which $N = (\tau_1-\tau_0) \cdot k$, and we can write the budget constraint \eqref{eq:bcnotch} as $z \ge T(y)$ where $T(y)=\tau_0 \cdot y + (\tau_1-\tau_0) \cdot  y\cdot \mathbbm{1}(y > k)+t_0$ with $y = y_i(\mathbf{x})$. For the proportional tax notch, $T_{1i}(\mathbf{x}) = \tau_1 \cdot y_i(\mathbf{x})+t_0$.
\end{example*}

Let $X_i$ be $i$'s realized outcome of $\mathbf{x}$, and $Y_i = y_i(X_i)$. $Y_i$ is observed by the econometrician, but $X_i$ need not be. In income tax settings this means that the econometrician observes pre-tax income, but not necessarily all choices made by a worker (or their employer) that pin down that income. In most instances from the literature, no distinction is made between the variable $y$ and any underlying choice variables $\mathbf{x}$. In standard tax settings such as in the examples given above, the functions $T_0$ and $T_1$ are linear and only depend on a worker's choices $\mathbf{x}$ though their pre-tax income $y_i(\mathbf{x})$. When making assumptions about decision-makers' preferences however, the distinction between taxable income $y$ and underlying margins of choice $\mathbf{x}$ can be important, as described in the next section.  

Distinguishing between $y$ and $\mathbf{x}$ also allows for more general settings in which tax liability $T$ depends directly on multiple components of $\mathbf{x}$, such as in the following example.

\begin{example*}[(non-scalar tax kink)]
	\citet{best2015} study a corporate tax setting in which firms pay the maximum of a tax on output and a tax on reported profits: $T_i(\mathbf{x})=\max \{\tau_r r,\tau_\pi(r-w)\}$
	where $\mathbf{x}=(r,w)$ with $r$ denoting firm revenue and $w$ reported costs, and $\tau_r < \tau_\pi$. Define $y_i(\mathbf{x}) = \frac{r-w}{r}$. Some algebra shows that firms pay the output tax $T_{0i}(\mathbf{x})=\tau_r r$ when $y_i(\mathbf{x}) < \tau_r/\tau_\pi$, and pay the profit tax $T_{1i}(\mathbf{x})=\tau_\pi (r-w)$ when $y_i(\mathbf{x}) > \tau_r/\tau_\pi$, fitting the form of Eq. \eqref{eq:bc}
\end{example*}

\subsection{Choice model and potential outcomes}
Define a pair of potential outcomes as what would occur if the decision-maker faced either of the functions $T_0$ or $T_1$ globally, without the kink or notch.
\begin{definition*}[(potential outcomes)]
	Let $Y_{0i}$ be the value of $y_i(\mathbf{x})$ that would occur for unit $i$ if instead of $z \ge T(\mathbf{x})$, the decision-maker faced the constraint $z \ge T_0(\mathbf{x})$, and similarly let $Y_{1i}$ be the value that would occur under the constraint $z \ge T_1(\mathbf{x})$.
\end{definition*}
\noindent In cases with notches $N>0$, one could further define $\tilde{Y}_{1i}$ to be the value of $y_i(\mathbf{x})$ that would occur subject to the constraint $z \ge \tilde{T}_1$. However, we will see that the observable distribution of $Y$ is only directly informative about the counterfactual choices $Y_0$ and $Y_1$, and not about $\tilde{Y}_1$.

My first main assumption is that preferences of each decision-maker are convex in their choice variables $\mathbf{x}$ and $z$, and that they dislike costs $z$:
\begin{assumption*}[CONVEX] Potential outcomes for a given unit $i$ and cost function $T$ maximize $u_{i}(z, \mathbf{x})$ subject to $z \ge T(\mathbf{x})$, where $u_{i}$ is strictly quasiconcave in $(z,\mathbf{x})$ and decreasing in $z$.
\end{assumption*}		
\noindent Implicit in CONVEX is an assumption that potential outcomes reflect choices made by the decision-maker. For now, this rules out optimization error limiting the decision-maker's ability to exactly manipulate values of $\mathbf{x}$ and hence $y$. I relax this assumption in Appendix \ref{sec:optimizationerror}.

An important special case of CONVEX is a parametric specification of utility as isoelastic in $y$ and quasilinear, as supposed by \citet{saez2010} and most of the subsequent empirical literature. In what I call ``the isoelastic model'', preferences only depend $\mathbf{x}$ through $y_i(\mathbf{x})$ and 
\begin{equation} \label{eq:isoelastic}
	u_i(z,y) = y-z - \frac{\eta_i}{1+\frac{1}{e}} \left( \frac{y}{\eta_i}\right)^{1+\frac{1}{e}}
\end{equation}
In this model, heterogeneity in preferences enters only through the scalar $\eta_i$. In the context of labor supply, $\eta_i$ is often interpreted as ``ability''. Convexity in the labor-leisure tradeoff is captured through an elasticity parameter $e > 0$, assumed to be common across individuals. 

\begin{remark*}
	Relative to existing literature, Assumption CONVEX relaxes the parametric specification of utility in \eqref{eq:isoelastic}. This is most closely related to \citet{blomquist_individual_2015}, who consider a nonparametric choice model in which workers facing an income tax kink determine their earnings by choosing two quantities (hours and effort). However, the way that I accommodate multiple margins of choice differs from that of \citet{blomquist_individual_2015}. Those authors define an effective utility function in terms of consumption and earnings alone by concentrating out all but the observed choice variable, and then assuming quasi-concavity of this concentrated utility function. CONVEX instead assumes convexity of preferences defined directly over the primitive margins of choice. This assumption can be evaluated on choice-theoretic grounds alone, requiring no assumptions on how $y$ depends on $\mathbf{x}$ beyond continuity. \citet{bklnnber} also discusses a nonparametric choice model for the bunching design, but takes the choice variable to be scalar.
\end{remark*}

A weaker assumption than CONVEX that still has identifying power is simply that decision-makers' choices do not violate the weak axiom of revealed preference:
\begin{assumption*}[WARP (rationalizable choices)]
	Consider two cost functions $T$ and $T'$. If a decision-maker's choice under $T'$ is feasible under $T$, then they would choose it under $T$.
\end{assumption*}
\noindent Assumption WARP again presumes that the decision-maker has precise control over $Y$. CONVEX and WARP also make implicit an assumption of no extensive margin effects. I make the stronger assumption CONVEX for most of the identification results, but Assumption WARP still allows a one-sided version of some results, indicating a degree of robustness with respect to departures from convex preferences. 

\subsection{Parameter of interest: the buncher ATE} \label{sec:buncherate}
I focus my identification analysis on the average treatment effect among units who locate at the kink, a parameter I call the ``buncher ATE''. I consider two versions of the buncher ATE. First:
$$\Delta^*_k = \mathbbm{E}[Y_{0i}-Y_{1i}|Y_{i}=k],$$
Define $\Delta_i = Y_{0i}-Y_{1i}$ to be unit $i$'s ``treatment effect''. $\Delta^*_k$ represents the average value of $\Delta_{i}$ among bunchers who bunch in response to the kink or notch. Unit $i$'s treatment effect $\Delta_i$ represents the causal effect of a counterfactual change from the choice set under $T_1$ to the choice set under $T_0$. Note that the treatment effects $\Delta_i$ are ``reduced form'' in the sense that when the decision-maker has multiple margins of response $\mathbf{x}$ to the incentives introduced by the kink/notch, they will be bundled together in the treatment effect $\Delta_i$. This is a strength when the goal is robustness over underlying structural models, and a limitation inherent in the bunching design itself when the goal is an elasticity with a specific structural interpretation \citep{einav2015}.

Alternatively, one can consider the buncher ATE after applying a log transformation to $Y$: 
$$\delta^*_k = \mathbbm{E}[\ln Y_{0i}-\ln Y_{1i}|Y_{i}=k] =  \mathbbm{E}[\ln(Y_{0i}/Y_{1i})|Y_{i}=k]$$
When it is important distinguish the two, I call $\delta^*_k$ the \textit{buncher ATE in logs} and $\Delta^*_k$ the \textit{buncher ATE in levels}. When the distinction is unimportant, I use $\Delta_k^*$ and call it simply the \textit{buncher ATE}.

To reduce notational clutter, I phrase the main identification results of this paper in terms of $\Delta_k^*$ rather than repeating them for $\delta_k^*$. With a strictly positive outcome variable $Y_i$, all identification results for $\Delta_k^*$ carry over for $\delta_k^*$ with $Y_i$ replaced by $\ln Y_i$, and with distributional assumptions about $Y_{0i}$ and $Y_{1i}$ replaced by the analogous assumptions made for $\ln Y_{0i}$ and $\ln Y_{1i}$. In Section \ref{sec:concavity} I discuss the relationship between the versions of my extrapolation assumptions for $Y$ and for $\ln Y$.

\begin{remark} \label{rem:elasticity}
Suppose that $T_0$ and $T_1$ are parameterized by values $\rho_0$ and $\rho_1 < \rho_0$ respectively (at a standard tax kink, the natural definition is the net-of-tax rate $\rho_d = 1-\tau_d$, with $T_{i,\rho}(y) = (1-\rho) \cdot y + t_0-(1-\rho-\tau_0) \cdot k$).  In the isoelastic model, the buncher ATE is logs is equal to $\delta_k^* = \ln(\rho_0/\rho_1) \cdot e$. More generally, we have that $\delta^*_k=\ln(\rho_0/\rho_1)\cdot \mathbbm{E}[\bar{e}_{i}|Y_{i}=k]$, where $e$ has been replaced by a weighted ``arc'' elasticity with respect to $\rho$ averaged among the bunchers, and integrated over hypothetical intermediate rates $\rho$ between $\rho_0$ and $\rho_1$. To see this, let $Y_{i}(\rho)$ be the outcome that unit $i$ would choose if they faced the cost schedule at parameter value $\rho$, with $Y_{1i} = Y_{i}(\rho_1)$ and $Y_{0i} = Y_{i}(\rho_0)$. Let $\bar{e}_{i}: = \left(\int_{\rho_1}^{\rho_0} \frac{\frac{d}{d\rho}\ln Y_{i}(\rho)}{\frac{d}{d\rho}\ln \rho} \cdot \frac{1}{\rho} \cdot d\rho\right)/\left(\int_{\rho_1}^{\rho_0} \frac{1}{\rho} \cdot d\rho\right)$, which is a weighted average elasticity of $Y$ with respect to $\rho$, integrated over $\rho$ in $[\rho_0,\rho_1]$.\footnote{What $\bar{e}_i$ averages over is a compensated elasticity $\frac{\rho}{Y}\frac{\partial Y}{\partial \rho} - \rho \frac{\partial Y}{\partial R}$ (with $R$ virtual income), up to a factor $k/Y(\rho, R)$ on the income effect term. This factor reflects that a buncher need not choose $k$ for all $\rho$ between $\rho_1$ and $\rho_0$.} Then:
\begin{equation} \label{eq:blcelasticity}
	\delta^*_k=\mathbbm{E}\left[\left.\int_{\rho_1}^{\rho_0} \frac{d}{d\rho}\ln Y_{i}(\rho) \cdot d\rho \right|Y_{i}=k\right]= \ln(\rho_0/\rho_1)\cdot \mathbbm{E}[\bar{e}_{i}|Y_{i}=k]
\end{equation}
\end{remark}

Which version of the buncher ATE is more interesting depends on whether the researcher is interested in a treatment effect expressed in units of the outcome variable, or rather an elasticity of response. Remark \ref{rem:elasticity} above shows that given a scalar parameterization of cost functions (e.g. by the implied net-of-tax rate in standard tax settings) the buncher ATE in logs $\delta^*_k$ is exactly proportional to an average ``arc'' elasticity among the bunchers, averaged over intermediate parameter values between those on either side. If the elasticity of response with respect to the chosen parameterization is homogenous over $i$ and over the parameter (as it is for the net-of-tax rate in the isoelastic model), then $\delta^*_k$ identifies this elasticity. The buncher ATE in logs therefore generalizes the elasticity parameter $e$ from the isoelastic model, which has been the traditional parameter of interest in bunching analyses. It's important to note that if preferences are not quasilinear, this elasticity reflects income as well as substitution effects (as the buncher ATE generally does as well).

Whether in logs or in levels, I suggest the buncher ATE as the natural parameter in the bunching design because it respects key aspects of the natural experiment afforded by bunching. First, the buncher ATE uses the actual policy variation at a kink or notch: $T_0$ versus $T_1$, and does not require the researcher to parameterize what individuals' responses would be to policies that are nowhere observed (e.g. tax rates between $\tau_0$ and $\tau_1$). If a researcher wishes to extrapolate from this observed policy variation to anticipate the effects of further policies, additional assumptions (e.g. the isoelastic model) can be layered on later. The basic identification result remains ``reduced-form''. 

Secondly, the buncher ATE respects the \textit{local} nature of the budget set discontinuity. Intuitively, bunching behavior near a kink or notch should not generally be informative about the responsiveness of individuals who locate far from $k$. A global parameter such as the overall ATE (or the average elasticity) would require learning about these individuals, which is only possible by imposing strong homogeneity assumptions and/or a parametric utility function. The buncher ATE instead focuses on a local population that is directly affected by the kink/notch itself, and whose size is exactly identified by the bunching probability (in common with the ``compliers'' in IV analysis).

I maintain the buncher ATE as the parameter of interest whether or not the researcher is faced with a pure kink, or a kink accompanied by a notch. Note that in the case of a pure notch, where $N_i > 0$ but $T_0 = \tilde{T}_1$, treatment effects will be zero if individuals' utility is quasilinear. This is in spirit with how the bunching design has been applied, where researchers are interested in an elasticity with respect to marginal incentives (and typically assume away by specifying a quasilinear utility function). In the Online Supplement, I consider the alternative parameter $\mathbbm{E}[Y_{0i}-Y_i|Y_i=k]$ which may be of interest for a pure notch, even when the buncher ATE is equal to zero.

\begin{remark*} The notation of Remark \ref{rem:elasticity} also allows us to compare to a decomposition of the bunching probability with nonparametric utility by \citet{blomquist_individual_2015}. Their Theorem 4 shows that $\mathcal{B} = \int_{\rho_1}^{\rho_0} e(\rho) \cdot \frac{k}{\rho} \cdot f_{\rho}(k) \cdot d\rho $, where $f_{\rho}(y)$ reflects the density of the counterfactual $Y_{i}(\rho)$ and $e(\rho)$ is an average compensated elasticity among those with  $Y_{i}(\rho)=k$. Since the quantity $\int_{\rho_1}^{\rho_0} \frac{k}{\rho} \cdot f_\rho(k) \cdot d\rho$ is not identified, this expression does not pin down a convex average of elasticities like Eq. \eqref{eq:blcelasticity} does. And since $f_\rho(k)$ is \textit{only} identified for the isolated values of $\rho=\rho_0$ and $\rho = \rho_1$, there is no principled way of extrapolating from the data to capture the magnitude of $f_\rho(k)$ for intermediate $\rho \in (\rho_0,\rho_1)$. To identify the buncher ATE, one instead must constrain the functions $f_0(y)$ and $f_1(y)$ over $y$. Here one can turn to the distribution for $Y \ne k$ or a reference distribution to motivate the extrapolation assumptions used. This type of evidence is unavailable for $f_\rho(k)$ over $\rho$.
\end{remark*} 

\section{Identification as a pair of extrapolation problems} \label{sec:idsetup}

\subsection{Observables around a kink} \label{sec:observables}

Lemma \ref{propobserve} outlines the core consequence of Assumption CONVEX for the relationship between observed $Y_i$ and the potential outcomes introduced in the last section, at a pure kink (N=0):
\begin{lemma}[(realized choices as truncated potential outcomes: kinks)] \label{propobserve}
	Under Assumption CONVEX, the outcome observed given the constraint Eq. \eqref{eq:bc} is:
	\begin{equation} \label{hcases}
		Y_i=
		\begin{cases}
			Y_{0i} & \hspace{.2cm} \textrm{if } \hspace{.2cm} Y_{0i}<k\\
			k & \hspace{.2cm} \textrm{if } \hspace{.2cm} Y_{1i} \le k \le Y_{0i} \hspace{.2cm}\\
			Y_{1i} & \hspace{.2cm} \textrm{if } \hspace{.2cm} Y_{1i}>k
		\end{cases}
	\end{equation}
	With WARP rather than CONVEX, $Y_{0i}\le k \implies Y_{i}=Y_{0i}$ and $Y_{1i} \ge k \implies Y_{i}=Y_{1i}$ still hold.
\end{lemma}

\noindent Lemma \ref{propobserve} implies that with convex preferences, the pair of counterfactual outcomes $(Y_{0i}, Y_{1i})$ is sufficient to pin down actual choice $Y_i$. $Y_i$ yields an observation of one or the other potential outcome, or $k$, depending on how the potential outcomes relate to the kink point $k$.

\begin{figure}[!h]
	\begin{center}
		\includegraphics[height=2.75in]{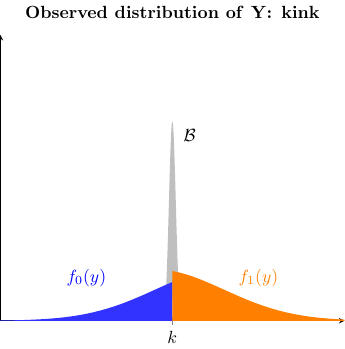}	\quad \quad 
		\includegraphics[height=2.75in]{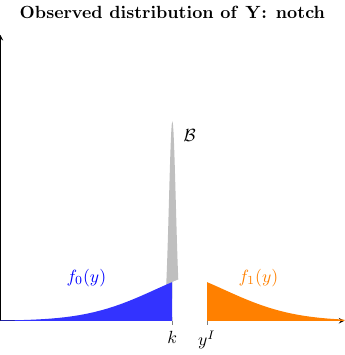}	
	\end{center}
	\caption{Left panel: observables around a kink, given Equation (\ref{hcases}). To the left of the kink at $k$, the researcher observes the density $f_0(y)$ of the counterfactual $Y_{0i}$, up to values $y=k$. To the right of the kink, the researcher observes the density $f_1(y)$ of $Y_{1i}$ for values $y>k$. At the kink, one observes a point-mass of size $\mathcal{B} :=P(Y_{i}=k)= P(Y_{1i} \le k \le Y_{0i})$. Right panel: observables around a notch, given the additional Assumption NOTCH (see Section \ref{sec:notchgeneralization}).}
	\label{observablesgraphhcolor}
\end{figure}

In particular, the individual will choose $Y_{0i}$ when the counterfactual choice $Y_{0i}$ is less than $k$, and $Y_{1i}$ when $Y_{1i}$ is greater than $k$. They will be found at the corner solution of $k$ if and only if the two counterfactual outcomes fall on either side, ``straddling'' the kink.\footnote{``Straddling'' can only occur in one direction, i.e. $Y_{1i} \le k \le Y_{0i}$. The other direction: $Y_{0i} \le k \le Y_{1i}$ with at least one inequality strict, violates WARP.} Figure \ref{observablesgraphhcolor} depicts the implications of Eq. (\ref{hcases}) for what is observable by the researcher in the bunching design: censored distributions of $Y_0$ and of $Y_1$, and a point-mass of $\mathcal{B} = P(Y_{1i} \le k \le Y_{0i})$ at the kink.

Equation (\ref{hcases}) represents a central departure from the main previous approach to the bunching design, which characterizes bunching in terms of the counterfactual $Y_0$ only. This is a simplification afforded by the benchmark isoelastic utility model, but in a generic choice model, both $Y_0$ and $Y_1$ are necessary to pin down actual choices $Y_{i}$. Lemma \ref{propobserve} relies only on convexity of preferences: the key to its generality is to express the observable data in terms of counterfactual \textit{choices}, rather than in terms of primitives of the utility function. 

Let $\mathcal{B}=P(Y_i = k)$ be the observable probability that the decision-maker chooses to locate exactly at $k$. The expression below for the bunching probability follows directly from Lemma \ref{propobserve}:
\begin{proposition}[(relation between bunching and treatment effects at a kink)] \label{thmstraddle} Under WARP,
	\begin{equation} \label{bunchprob}
		\mathcal{B} \le P(Y_{1i}\le k \le Y_{0i}) = P(Y_{0i} \in [k, k+\Delta_{i}]) = P(Y_{1i} \in [k-\Delta_{i}, k]),
	\end{equation}
	 with equality under CONVEX.
\end{proposition}
\noindent In words: bunchers are those units whose $Y_0$ potential outcome lies to the right of the kink, but within that unit's individual treatment effect of kink. Assuming continuity of $F_0$, we can then also write bunching in terms of the marginal distributions of $Y_{0}$ and $Y_{1}$:\footnote{To obtain this expression, write $1 = P(Y \le k) + P(Y >k) = \{P(Y_0<k) + \mathcal{B}\} + P(Y_1 > k) = P(Y_0 \le k) + \mathcal{B} + 1 - P(Y_1 \le k)$ where the first equality uses Eq. (\ref{hcases}) and the second uses CONVEX and continuity.}
\begin{equation} \label{eq:bmarginals}
	\mathcal{B}=F_1(k)-F_0(k).
\end{equation}

\begin{example}[(isoelastic model at a tax kink)] \label{ex:isoelastickink}
	$Y_{0i} = \eta_i \cdot (1-\tau_0)^e$ and $Y_{1i} = \eta_i \cdot (1-\tau_1)^e$, so $e$ can be interpreted as the elasticity of income with respect to the net-of-tax rate $1-\tau$, since $\ln Y_{di} =  \ln \eta_i + e \cdot \ln (1-\tau_d)$ for either $d \in \{0,1\}$. Note that preferences are quasilinear in consumption $y-z$, thus choices do not depend on any level-shift in tax liability arising from $t_0$.
	 
	 \begin{figure}[H]
	 	\begin{center}
	 		\begin{minipage}{.35\textwidth}
	 			\includegraphics[height=2.5in]{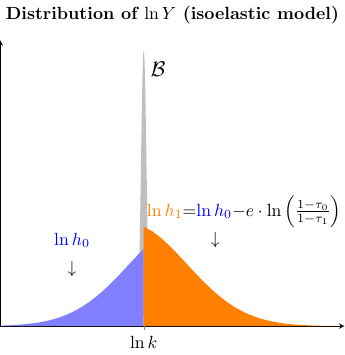}	
	 		\end{minipage}
	 		\begin{minipage}{.1\textwidth}$\iff$\end{minipage}
	 		\begin{minipage}{.4\textwidth}
	 			\includegraphics[height=2.5in]{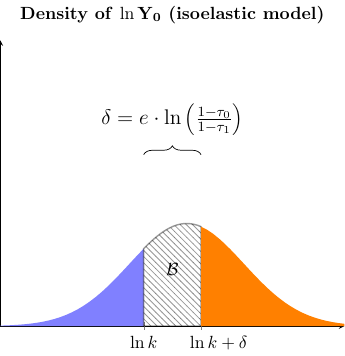}
	 		\end{minipage}
	 	\end{center}
	 	\caption{In the isoelastic model: the left panel depicts the distribution of the log of $Y$, and the right panel depicts the underlying full density of $\ln Y_{0}$. Treatment effects in logs are homogeneous across individuals and equal to $\delta:= e \cdot \ln \frac{1-\tau_0}{1-\tau_1}$. The density of $\ln Y_0$ is related to the observed distribution by ``sliding'' the observed distribution for $\ln Y >\ln k$ to the right by the unknown distance $\delta$, leaving a missing region in which $f_0$ is unobserved. The total area in the missing region from $\ln k$ to $\ln k+\delta$ must equal the observed bunching mass $\mathcal{B}$.}
	 	\label{observablesgraphh}
	 \end{figure} 	
	 
	  \noindent Standard approaches in the bunching design make parametric assumptions that interpolate $f_0$ through the missing region of Figure \ref{observablesgraphh} to point-identify $e$. If in the other extreme the researcher is unwilling to assume anything about the density of $h_0$ in the missing region of Figure \ref{observablesgraphh}, then the data are compatible with any finite $e > 0$ as emphasized by \citet{blomquist_bunching_2019} and \citet{bertanha_better_2018}. An arbitrarily small $e$ could be rationalized by a density that spikes sufficiently high just to the right of $k$, while an arbitrarily large $e$ can be reconciled with the data by supposing that the density of $Y_0$ drops quickly to some very small level throughout the missing region. 
\end{example}

\subsection{Observables around a notch} \label{sec:notchgeneralization}
To analyze the buncher ATE given a possible notch in addition to a possible kink, our first step will be to derive a generalization of Lemma \ref{propobserve}, relating observed choice $Y_i$ to features of $i$'s preferences. 

\begin{lemma}[(realized choices under a notch)] \label{propobservenotch}
	Under Assumption CONVEX, the outcome observed given the constraint Eq. \eqref{eq:bcnotch} is:
	\begin{equation} \label{hcasesnotch}
		Y_i=
		\begin{cases}
			Y_{0i} & \hspace{.2cm} \textrm{if } \hspace{.2cm} Y_{0i}<k\\
			k & \hspace{.2cm} \textrm{if } \hspace{.2cm} Y_{0i} \ge k \textrm{ and } \{Y_{1i} \le k \textrm{ or } A_i\}  \hspace{.2cm}\\
			Y_{1i} & \hspace{.2cm} \textrm{if } \hspace{.2cm} Y_{1i}>k \textrm{ and not } A_i
		\end{cases}
	\end{equation}
	where $A_i$ is the event $\{Y_{1i} > k \textrm{ and } \max_{\mathbf{x}: y_i(\mathbf{x})=k} u_i(\tilde{T}_{1i}(\mathbf{x}),\mathbf{x}) \ge u_i(\tilde{T}_{1i}(\mathbf{x}_{1i})+N_i,\mathbf{x}_{1i})\}$, with $\mathbf{x}_{1i}$ $i's$ choice of $\mathbf{x}$ under $T_{1i}$.
\end{lemma}
\noindent Lemma \ref{propobservenotch} differs from Lemma \ref{propobserve} in two respects. Firstly, rather than involving the counterfactual choice $\tilde{Y}_{1i}$ according to the cost function $\tilde{T}_{1}$ that intersects $T_0$ at $k$ (cf. Figure \ref{fig:kinknotch}), it involves $Y_{1i}$, which captures any income effects from the lump-sum cost $N_i$ in addition to any response to a change in marginal incentives. Note however that $\tilde{Y}_{1i}=Y_{1i}$ if preferences are quasilinear in $y$.

The second difference between Lemmas \ref{propobservenotch} and \ref{propobserve} is the event $A_i$ appearing in Eq. \eqref{hcasesnotch}. The event $A_i$ occurs when $i$ prefers an $\mathbf{x}$ that puts them at the notch to their most preferred choice under $T_{1i}$, which happens to lie ``beyond'' the notch. Individuals counted in event $A_i$ are those who would not bunch due to a counterfactual shift from cost function $T_0$ to $\tilde{T}_1$ alone (i.e. the change in marginal tax rate in standard settings), but would bunch given the additional cost $N$ of crossing the notch.

Unfortunately, $A_i$ cannot in full generality be phrased in terms of the potential outcomes $Y_{1i}$ and $Y_{0i}$ alone. The reason is that whether an individual chooses to locate at a notch depends not only on their preferences ``local'' to the $\mathbf{x}$'s near the budget set discontinuity, but also how the utility they would achieve at at $Y_{1i}$ under $T_1$ compares with their utility at the notch. With pure kinks these local features can always be inverted into simple inequalities involving only $Y_{1i}$ and $Y_{0i}$.

The following additional assumption makes settings with a notch particularly tractable:
\begin{assumption*}[NOTCH (origin of notch bunchers)]
	For some $y^I \ge k$: $A_i = \{Y_{1i} \in (k,y^I]\}$ with probability one.
\end{assumption*}
\noindent Assumption NOTCH says that those who benefit from bunching at the notch despite having a $Y_{1i} > k$ are exactly those whose $Y_{1i}$ are not ``too much'' above $k$. Note that NOTCH is trivially satisfied with $y^I = k$ given CONVEX in the case of a pure kink $(N_i=0)$.

For a notch, Assumption NOTCH is restrictive, but is satisfied by the isoelastic model. In Section \ref{sec:ranknotchsuff} I also give a nonparametric generalization of the isoelastic model for which it remains satisfied. To move forward with minimum possible structure, one can instead of NOTCH assume:
\begin{assumption*}[NOTCHBOUND]
	For some $y^I \ge k$: $P(Y_1 \le y^I|A) = 1$
\end{assumption*}
\noindent NOTCHBOUND assumes only that the support of $Y_1|A$ is bounded from above, i.e. that those individuals with $Y_1>k$ who nevertheless prefer the notch location come from bounded values of $Y_1$. With NOTCHBOUND, one can still bound the average treatment effect over a superset of the bunchers, provided that the researcher stipulates or can identify a value $y_U \ge y^I$. Because the structure of this result is the same of that which allows for optimization errors, I defer this result to Appendix \ref{sec:optimizationerror}. I maintain the stronger assumption NOTCH in the main presentation of ideas, as it better elucidates the relationship between notches and kinks. 

Proceeding under NOTCH, we can rewrite Eq. \eqref{hcasesnotch} as:
\begin{equation} \label{hcasesranknotch}
	Y_i=
	\begin{cases}
		Y_{0i} & \hspace{.2cm} \textrm{if } \hspace{.2cm} Y_{0i}<k\\
		k & \hspace{.2cm} \textrm{if } \hspace{.2cm} Y_{0i} \ge k \textrm{ and } Y_{1i} \le y^I\\
		Y_{1i} & \hspace{.2cm} \textrm{if } \hspace{.2cm} Y_{1i}> y^I
	\end{cases}
\end{equation}
allowing $Y_i$ to be pinned down entirely by $Y_{1i}$ and $Y_{0i}$ (as with a pure kink), given $y^I$. Note that we recover Eq. \eqref{hcases} when $y^I=k$. Under CONVEX and NOTCH, there will be a ``hole'' in the distribution of $Y_i$ between $k$ and $y^I$: i.e. $P(Y_i \in (k,y^I])=0$ (see Fig. \ref{observablesgraphhcolor}). This allows $y^I$ to in principle be identified, and provides a testable implication of NOTCH. The notation $y^I$ comes from \citet{kleven2013}: in their isoelastic model (and more generally under univariate heterogeneity---see Sec. \ref{sec:ranknotchsuff}), individuals with $Y_i=y^I$ are indifferent between $y^I$ and the notch.

In practice, the prediction of a hole in the distribution of $Y$ between $k$ and $y^I$ for some $y^I \ge k$ may be falsified by the data. The absence of a hole can be indicative of a failure of NOTCH, but can also arise due to optimization errors/frictions (a failure of CONVEX). A useful diagnostic is that if bunching is exact but there is no hole above $k$, then NOTCH is the culprit. If bunching is diffuse and there is no hole, as is typically the case, this can be explained by optimization error/frictions. As discussed in Appendix \ref{sec:optimizationerror}, partial identification with optimization errors requires an upper bound on $y^I$, which can be obtained by inspecting where $Y_1$ ``breaks trend'' due to excess mass.

By Eq. \eqref{hcasesranknotch}, Eq. \eqref{eq:bmarginals} generalizes under NOTCH to:
\begin{equation} \label{eq:bmarginalsnotch}
	F_1(y^I)=F(y^I)=F(k)=F_0(k)+\mathcal{B}
\end{equation}
with $\mathcal{B}=P(Y_i=k)$, where we now let $\mathcal{B} := P(Y_{0i} \ge k, Y_{1i} \le y^I)$.

\subsection{Identification of basic features of $F_0$ and $F_1$} \label{app:basicid}
\noindent The following combines the above results for kinks and notches to establish identification of some basic features of the distributions of $Y_0$ and $Y_1$, which will be useful in the identification of treatment effects to follow. Consider a random sample of observations of $Y_i$. Under i.i.d. sampling of $Y_i$, the distribution $F(y)$ of $Y_i$ is identified. Let $F_0(y) = P(Y_{0i} \le y)$ be the distribution function of the random variable $Y_0$, and $F_1(y)$ the distribution function of $Y_1$. 

From Eq. \eqref{hcasesnotch} it follows that $F_{0}(y) = F(y)$ for all $y<k$, and $F_1(y) = F(y)$ for $Y>y^I$, given CONVEX and NOTCH. Thus, observations of $Y_i$ are informative about the marginal distributions of $Y_{0i}$ and $Y_{1i}$ on either side of the kink/notch, as well as the feature $\mathcal{B} = P(Y_{0i} \ge k, Y_{1i} \le y^I)$ of their joint distribution. To discuss identification in terms of densities, I introduce a mild restriction on the distributions. Let $Q_d(u) := \inf\{y: F_d(y) \ge u\}$ be the quantile function of $Y_{di}$.
\begin{assumption*}[CONT (continuity)]
	(i) $F$ admits a density within a neighborhood below $k$ and above $y^I$; (ii) $F_0$ admits a density $f_0=F_0'$ within a neighborhood below $k$ and $F_1$ admits a density $f_1=F_1'$ within a neighborhood above $y^I$. Furthermore (iii) $F_0$ admits a density in a neighborhood of $Q_0(F_1(y^I))$ and $F_1$ admits a density in a neighborhood of $Q_1(F_0(k))$.
\end{assumption*}
\noindent Note that under CONVEX and NOTCH, part (i) of CONT implies part (ii) and vice-versa by Eq. \eqref{hcasesnotch}. Part (iii) is not used in Proposition \ref{corr:dens} but I introduce it now for later use (in Proposition \ref{prop:buncherATE}). 

For any function $\phi(y)$, let $\phi(y^+):=\lim_{t \downarrow y}$ and $\phi(y^-):=\lim_{t \uparrow y}$.
\begin{proposition}[(identification of truncated densities)] \label{corr:dens}
	Suppose WARP and CONT hold. Then for a pure kink: $f_0(y) \le f(y)$ for $y<k$ and $f_0(k) \le f(k^-)$, and $f_1(y) \le f(y)$ for $y>k$ and $f_1(k)\le  f(k^+)$, for $y$ in a neighborhood of $k$. Adding CONVEX and NOTCHBOUND: $f_0(k) = f(k^-)$ and $f_1(y^I)= f(y^{I^+})$, while $f_0(y) = f(y)$ for $y$ in a neighborhood below $k$ and $f_1(y) = f(y)$ for $y$ in a neighborhood above $y^I$. 
\end{proposition}
\noindent By Proposition \ref{corr:dens}, we have under CONT, CONVEX, and NOTCHBOUND that:
\begin{equation} \label{eq:iddensnotch}
	F_{0}(k) = F(k^-), \quad \quad F_{1}(y^I) = F(y^I), \quad \quad  f_{0}(k) = f(k^-), \quad \textrm{ and } \quad f_{1}(y^I) = f(y^{I+})
\end{equation}
where in the special case of a pure kink recall that $y^I=k$.

\begin{remark} \label{rem:yI} One can obtain the implication that $\mathbbm{E}[Y_0-Y_1|Y=k] = \mathbbm{E}[Y_0-Y_1|Y_{0} \ge k, Y_{1} \le y^I]$ for some value $y^I \ge k$ even without NOTCHBOUND by an intermediate value theorem argument, under monotonicity assumptions that strictly generalize the isoelastic model. This result is not immediately useful however, because it lacks the implication that  $F_1(y^I)$ and $f_1(y^I)$ are identified.
\end{remark}

\subsection{Treatment effects among the bunchers} \label{app:tesinbunching}
Proposition \ref{thmstraddle} already suggests that bunching at a kink can be informative about features of the distribution of treatment effects, for example in the case of a pure kink $\mathcal{B}>0$ implies that $P(\Delta_i > 0) > 0$.  This forms a basic link between bunching and treatment effects. Our basic goal is to invert Eq. (\ref{bunchprob}) in some way to learn about the buncher ATE from the observable bunching $\mathcal{B}$.

To do so it is useful to introduce an auxiliary parameter I call the ``buncher QTE'':\footnote{If $\mathcal{B}=0$, I take $Q_k^*$ to be defined by its limit as $\mathcal{B} \downarrow 0$, which yields $Q_k^*=0$.}
\begin{align}
	Q_k^*&:= \frac{1}{\mathcal{B}}\int_{F_0(k)}^{F_0(k)+\mathcal{B}} Q_0(u)\cdot du - \frac{1}{\mathcal{B}}\int_{F_1(y^I)-\mathcal{B}}^{F_1(y^I)} Q_1(u)\cdot du = \frac{1}{\mathcal{B}}\int_{F(k)-\mathcal{B}}^{F(k)} \{Q_0(u)-Q_1(u)\}\cdot du \label{eqdeltak}
\end{align}
The rightmost equality in \eqref{eqdeltak} holds by \eqref{eq:bmarginalsnotch} under CONVEX (and NOTCH in the case of a notch). Recall that in the case of a pure kink, $y^I=k$ and \eqref{eq:bmarginalsnotch} simplifies to $F_1(k)=F_0(k)+\mathcal{B}$.

\begin{figure}[h!]
	\begin{center}
		\begin{minipage}{.4\textwidth}
			\includegraphics[height=2.5in]{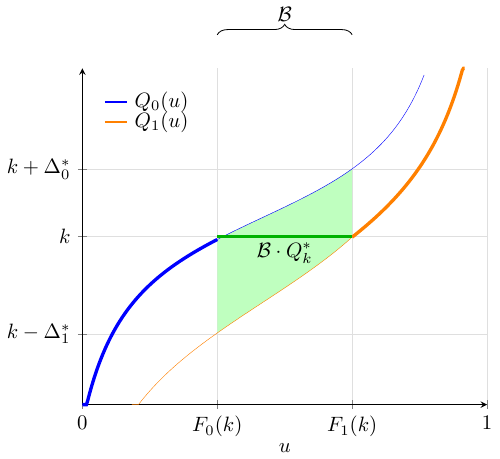}	
		\end{minipage}
		\begin{minipage}{.1\textwidth}\quad \quad \end{minipage}
		\begin{minipage}{.4\textwidth}
			\includegraphics[height=2.5in]{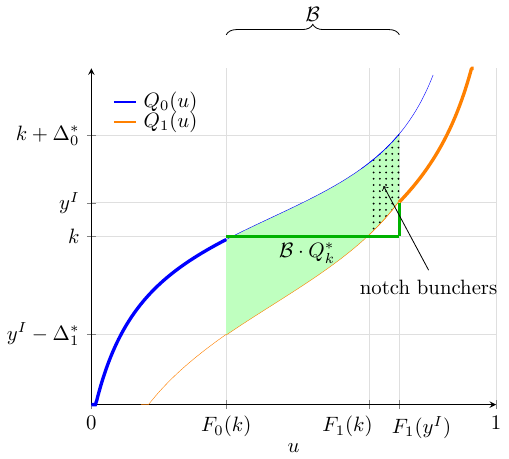}
		\end{minipage}
	\end{center}
	\caption[]{Left panel (kink): the observed portions of each quantile function are depicted by thick curves, and the unobserved portions by thin curves. The buncher QTE is equal to the area shaded in green, divided by the bunching probability $\mathcal{B}$. The right panel depicts a setting with a notch in addition to a kink, under NOTCH. Individuals with $Y_0 \ge k$ and $Y_1 \in (k,y^I)$ represent additional bunchers arising from the notch $N>0$, while those with $Y_0 \ge k, Y_1 \le k$ would bunch anyways due to the kink. In both panels, the observed distribution of $Y$ follows the thick curves from blue to green (through the bunching region) to orange. \label{figquantiles}} 
\end{figure}

Define the values
\begin{align} 
	\Delta^*_0 &:= Q_0(F_1(y^I)) - k \quad \quad \quad \quad \textrm{ and } &\Delta_1^* := y^I-Q_1(F_0(k)) \label{eq:deltasdef}
\end{align}
Substituting \eqref{eq:bmarginalsnotch}, we can rewrite $\Delta_1^* = y^I-Q_1(F_1(y^I)-\mathcal{B})$ and $\Delta^*_0 = Q_0(F_0(k)+\mathcal{B})-k$. Under CONT, $\Delta_0^*$ and $\Delta_1^*$ translate the limits of integration for $Q_k^*$ from quantile space to $Y$ space; the two terms in \eqref{eqdeltak} are equal to $\mathbbm{E}[Y_{0i}|Y_{0i} \in [k, Q_0(F_0(k)+\mathcal{B})]]$ and $\mathbbm{E}[Y_{1i}|Y_{1i} \in Q_1(F_1(y^I)-\mathcal{B}), y^I]]$, respectively.

The quantity $Q_k^*$ represents an average quantile treatment effect $Q_0(u)-Q_1(u)$ among individuals at the ranks that are associated with bunching, i.e. $u \in [F_0(k),F_1(k)]$ in the case of a pure kink, and $u \in [F_0(k),F_1(y^I)]$ in the case with a notch. However, while quantile treatment effects are causal quantities, they do not represent directly an average of individual-level treatment effects. An additional assumption allows us to interpret $Q_k^*$ as the average quantile effect among \textit{bunchers}, with $Y_i=k$. As we'll see, this assumption can be relaxed at the expense of wider bounds.
\begin{assumption*}[RANK] With probability one: $F_0(Y_{0i}) \in [F_0(k), F_1(y^I)] \textrm{ iff } F_1(Y_{1i}) \in [F_0(k), F_1(y^I)]$.
\end{assumption*}
\noindent Assumption RANK amounts to a weakened version of \textit{rank invariance} between $Y_{0}$ and $Y_{1}$. Rank invariance (\citealt{chernozhukov_iv_2005}) says that $F_0(Y_{0i})=F_1(Y_{1i})$ for all units. That is, switching from cost function $T_1$ to $T_0$ would not change any unit's rank in the distribution of $Y_d$. Rank invariance is satisfied by the isoelastic model and more generally any model in which there is perfect positive co-dependence between $Y_{0i}$ and $Y_{1i}$.

\begin{figure}[h!]
	\begin{subfigure}{.45\textwidth}
		\includegraphics[width=2.8in]{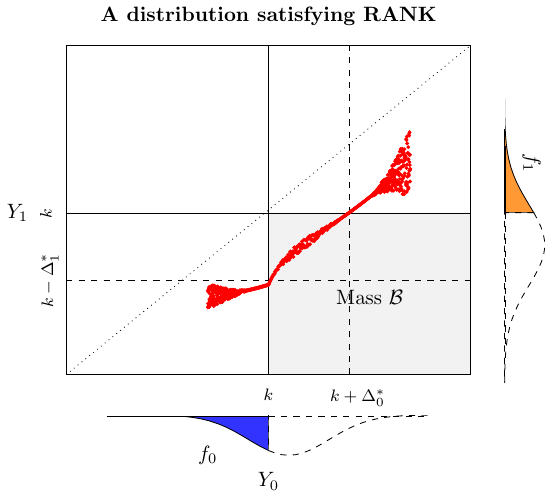}\vspace{.4cm}
	\end{subfigure} \hspace{.5in}
	\begin{subfigure}{.45\textwidth}
		\includegraphics[width=2.9in]{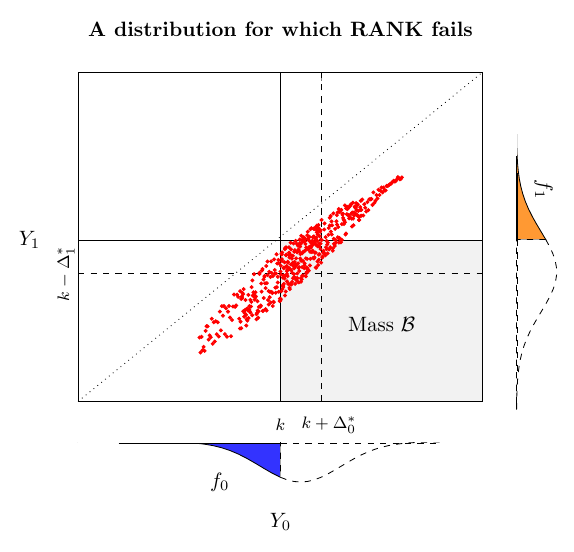}
	\end{subfigure}
	\caption{The joint distribution of $(Y_{0i}, Y_{1i})$ (in red) around a pure kink, comparing an example satisfying Assumption RANK (left) and one that does not (right). RANK allows the support of the joint distribution to ``fan-out'' from perfect co-dependence of $Y_0$ and $Y_1$, except when either outcome is exactly equal to $k$. Under CONVEX the observable data identifies the shaded portions of each outcome's marginal distribution (depicted along the bottom and right edges), and the total mass $\mathcal{B}$ in the (shaded) south-east quadrant. In the general case, the horizontal line at $Y_1=k$ defining this quadrant simply moves to $Y_1=y^I$ under NOTCH.} \label{observables}
\end{figure}

In Appendix \ref{sec:ranknotchsuff}, I give a set of sufficient conditions for Assumptions RANK and NOTCH to both hold, which relaxes the parametric structure of the isoelastic model. The key feature of this model in standard tax settings is that heterogeneity in $Y_0,Y_1$ is generated by a scalar, implying rank invariance. I show how this structure provides a tractable model for settings with notches in a wealth tax, tailored to applications such as that of \citet{londonovelez2025}.

Full rank invariance is a strong assumption---one that can be rejected in some contexts, e.g. where the treatment variable is exposure to a welfare reform \citep{klinetartari_bounding_2016}.  It is worth noting that full rank invariance is not necessary for Assumption RANK to hold. Rather, it need only hold \textit{locally} for $(Y_0,Y_1)$ near $(k,y^I-\Delta_1^*)$ and near $(k+\Delta_0^*,y^I)$. In particular, the joint distribution of $(Y_0,Y_1)$ can fan out in the tails (see e.g. the left panel of Figure \ref{observables}), as well as in the interior of the bunching quadrant $Y_0 \ge k, Y_1 \le y^I$. RANK can be maintained under NOTCHBOUND defining the value $y^I$, or under the stronger assumption of NOTCH.

Assumption RANK is thus useful because it establishes equivalence between the parameter $\Delta_k^*$, which depends on the \textit{joint} distribution of potential outcomes ($\Delta_k^* = \mathbbm{E}[Y_{i}-Y_{1i}|Y_{1i} \le k \le Y_{0i}]$), and $Q_k^*$, which depends only on the \textit{marginal} distributions of $Y_{0}$ and $Y_{1i}$. Under CONT and CONVEX, we can rewrite Assumption RANK as $Y_{0i} \in [k, k+\Delta_0^*] \textrm{ iff } Y_{1i} \in [y^I-\Delta_1^*, y^I]$. The bunching event is then equivalent under RANK to $Y_{0i} \in [k, k+\Delta_0^*]$ and to $Y_{1i} \in [y^I-\Delta_1^*, y^I]$. Note that with RANK and CONT we have that $y^I \le k + \Delta_0^*$, if $Y_0$ first-order stochastically dominates $Y_1$ (as in the isoelastic model).\footnote{To see this, note that by RANK (with CONVEX and NOTCH): $F_0(k+\Delta_0^*) - F_0(k) = F_1(y^I)-F_1(y^I-\Delta_1^*)$. Substitute Eq. \eqref{eq:deltasdef} to obtain $y^I = Q_1(F_0(k+\Delta_0^*))$ under CONT. Then $Q_1(F_0(k+\Delta_0^*)) \le Q_0(F_0(k+\Delta_0^*))$ by stochastic dominance. That $y^I \le [k, k + \Delta_0^*]$ is implicit in foundational bunching work (see e.g. in Figure Ib of \citealt{kleven2013}), though this ranking may not hold in general settings if RANK fails or sufficient bunchers have negative treatment effects. Whether or not $y^I \le k + \Delta_0^*$ does not affect my analysis of identification.} 

However, the buncher ATE is always weakly larger than $Q_k^*$, even if RANK fails to hold:
\begin{proposition} \label{prop:buncherATE} \label{prop:buncherATEnotch}
	Assume that WARP and CONT hold. Then:
	\begin{enumerate}[noitemsep]
		\item In the case of a pure kink, $\Delta^*_k \ge Q_k^*$ with equality if RANK and CONVEX hold.
		\item Under CONVEX and NOTCH, $\Delta^*_k \ge Q_k^*$ with equality if RANK holds.
	\end{enumerate} 
\end{proposition}
\noindent The proof of Proposition \ref{prop:buncherATE} also extends the conclusion $\Delta^*_k \ge Q_k^*$ to the case where NOTCH is replaced by NOTCHBOUND, given a redefinition of $\Delta^*_k$ to include a superset of the bunchers.

To understand the inequality in Proposition \ref{prop:buncherATE}, consider first the case of a pure kink, and assume CONVEX. All bunchers have $Y_{0i} \ge k$ and $Y_{1i} \le k$. $Q_k^*$ coincides with $\Delta_k^*$ when the mass $\mathcal{B}$ of bunchers occurs among those $i$ with the \textit{smallest} possible such values of $Y_{0i}$---that is those between $k$ and $Q_0(F_0(k)+\mathcal{B})$---and among those $i$ with the \textit{largest} possible values of $Y_{1i}$---those between $Q_1(F_1(k)-\mathcal{B})$ and $k$. Any deviation from this scenario can only make $\Delta_k^*$ larger. The slackness of the inequality in Proposition \ref{prop:buncherATE} is thus related to how ``thin'' the distribution of $(Y_0,Y_1)$ is as it crosses either $Y_0=k$ or $Y_1=y^I$ (compare the left and right panels of Figure \ref{observables}), an observation that I make use of in Proposition \ref{prop:bATEUB1} below. The Online Supplement provides a visualization.

Now consider item 2., treating the case of a notch. Again, the buncher QTE provides a lower bound on the buncher ATE without Assumption RANK. However unlike the case of a pure kink, we cannot make this conclusion under WARP alone without CONVEX. The reason is that with $N_i>0$ bunchers can have arbitrarily large values of $Y_1$ without violating WARP.

Given that an upper bound on $Q_k^*$ only provides an automatic upper bound on $\Delta_k^*$ under the somewhat restrictive assumption RANK, it is desirable to provide alternative upper bounds that do not invoke it. I provide two such bounds, both of which rely on an assumption of stochastic increasingness between $Y_0$ and $Y_1$ instead of RANK:
\begin{assumption*}[SI (stochastic increasingness)]
	$P(Y_{0i} \le t|Y_{1i} = y)$ and $P(Y_{1i} \le t|Y_{0i} = y)$ are both non-increasing in $t$, for every $y$.
\end{assumption*}
\noindent Stochastic increasingness offers an alternative generalization of rank invariance to RANK. Consider two ranks $u'>u$. Under rank invariance, an individual with rank $u'$ in $Y_0$ has exactly rank $u'$ in $Y_1$ too. Under SI, we only assume that among those with rank $u'$ in $Y_0$, the distribution of $Y_1$ ranks is higher (in the stochastic dominance sense) than among those with $Y_0$ rank $u$, and vice versa.

The first upper bound on  $\Delta_k^*$ supplements Assumption SI with a known bound on the width of the distribution of treatment effects, conditional on the value of one potential outcome. The second bound is given in Section \ref{sec:concavity}, as it also relies on the distributional shape restrictions introduced there.
\begin{proposition} \label{prop:bATEUB1}
	Let $w(Z|X):=\operatorname{esssup} \operatorname{supp}\{Z|X\} - \operatorname{essinf} \operatorname{supp}\{Z|X\}$ denote the width of the (convex hull of) conditional support of $Z$ given $X$. Suppose $w(\Delta|Y_0=k)$ and $w(\Delta|Y_1=y^I)$ are bounded from above by $M$ with probability one, and SI, CONVEX, CONT and NOTCH hold. Then
	$$\Delta_k^* \le Q_k^* + M \cdot \left( \min \left\{\frac{F(y^I) \cdot (1-F(y^I))}{\mathcal{B}},1\right\} + \min \left\{\frac{F(k^-) \cdot (1-F(k^-))}{\mathcal{B}},1\right\}\right)$$
	This implies the simpler (but possibly larger bound) $\Delta_k^* \le Q_k^* + M \cdot \min \left\{\frac{1}{2 \mathcal{B}},2\right\}$.
\end{proposition} 
\noindent The proof of Proposition \ref{prop:bATEUB1} shows that NOTCH can again be weakened to NOTCHBOUND if the parameter of interest is $\mathbbm{E}[Y_0-Y_1|Y_0 \ge 0, Y_1 \le y^I]$, which may then differ from $\mathbbm{E}[Y_0-Y_1|Y=k]$. I also give in the proof a tighter upper bound on $\Delta_k^*$ that holds if SI is strengthened to assume that $Y_0$ and $Y_1$ are \textit{affiliated} \citep{MilgromWeber}.

Proposition \ref{prop:bATEUB1} shows that the gap between $\Delta_k^*$ and $Q_k^*$ is related to the dispersion in treatment effects for a given level of each potential outcome (equivalently, the width of the conditional support of each potential outcome given the value of the other). Note that if an isoelastic model with heterogeneous elasticities $e_i$ holds approximately, then a log transformation to $Y$ will make this hold with a smaller value of $M$. In a simple tax setting, $\{\Delta_i|Y_{1i}=y\} = \left(\left(\frac{1-\tau_0}{1-\tau_1}\right)^{e_i} - 1\right) \cdot y$, varying strongly with $y$, whereas $\{\delta_i|Y_{1i}=y\} = e_i  \cdot \ln \left(\frac{1-\tau_0}{1-\tau_1}\right)$. In the latter case a small value of $M$ will be sufficient provided that the dispersion in $e_i$ is small.

Combining with Proposition \ref{prop:buncherATE}, we obtain a two-sided bound $\Delta_k^* \in [Q_k^*,Q_k^* + 2M]$, given SI rather than RANK. When $M=0$, we recover RANK and hence equality between $\Delta_k*$ and $Q_k*$  

\section{Partial identification via concavity restrictions} \label{sec:concavity}
Given the results of Section \ref{sec:idsetup}, the buncher ATE can be partially identified if the researcher is willing to make assumptions about the distributions of $Y_0$ and $Y_1$ in the bunching intervals $[k,k+\Delta_0^*]$ and $[y^I-\Delta_1^*, y^I]$ respectively. These correspond to assumptions about the quantile functions $Q_0(u)$ for $u \in [F_0(k),F_0(k)+\mathcal{B}]$ and $Q_1(u)$ for $u \in [F_1(y^I)-\mathcal{B},F_1(y^I)]$ (cf. Fig. \ref{figquantiles}), both of which are equal (under CONVEX and NOTCH) to $[F_0(k),F_1(y^I)]$ by Eq. \eqref{eq:bmarginalsnotch}.

There are multiple ways that one might approach this extrapolation problem. The standard approach in the empirical literature is to make parametric assumptions that lead to point identification of the buncher ATE in logs ($e$ in the context of the isoelastic model). In practice, the density of $Y_0$ is typically taken to be a polynomial, allowing it to be perfectly extrapolated across the bunching region. A popular methodology introduced by \cite{chetty2011} additionally relies on an assumption that $f_1$ and $f_0$ are proportional beyond the kink. This turns the problem from an extrapolation problem to an easier \textit{interpolation} problem (as in Figure \ref{observablesgraphh}), and uses the strong polynomial assumption to also overcome diffuse rather than perfect bunching. The proportional density assumption is incoherent as typically applied (see Remark \ref{rem:incoherent} below). Though it can be corrected \citep{song2025identificationinferencegeneralbunching}, the polynomial approach still inherits the drawbacks of extrapolating polynomial functions: the extrapolated value of $f_0(y)$ for each value $y$ in the bunching window may be driven almost entirely by points far from the bunching interval \citep{GelmanImbens}.

\citet{bertanha_better_2018} suggest instead assuming a bound on the derivative the density of $Y_0$. I describe an analogous approach for the buncher ATE using my quantile representation in the Online Supplement.\footnote{The basic idea is to use the derivatives of $Q_d(u)$ on either side of the kink/notch to construct a Taylor series. An advantage of the quantile representation is that the Taylor expansion is in the known dimensionless quantity $\mathcal{B} < 1$.} In practice, this approach ultimately relies heavily on smoothness of $F_0$ and $F_1$. In the extreme case that $Y_0$ and $Y_1$ are analytic and the Taylor approximation extends to infinite order, it in fact yields point-identification of the buncher ATE (see also \citet{pollinger_kinks_2020, song2025identificationinferencegeneralbunching} for related results). Even obtaining informative bounds by this method typically requires the distributions of $Y_d$ to admit derivatives to a high order, and for them to be precisely estimable.

I propose to instead approach the extrapolation problem for $Q_k^*$ by placing restrictions on the \textit{concavity} of the CDFs $F_0$ and $F_1$ and their associated survival functions $1-F_0$ and $1-F_1$. By contrast, this family of assumptions only requires the densities of $Y_0$ and $Y_1$ to be once differentiable, and estimation only requires estimating their CDFs and densities at a point. I introduce a nested family of such assumptions that include two important special cases: (i) densities that are monotonic on the bunching interval; and (ii) distributions that satisfy the shape constraint of ``bi-log-concavity'' on the bunching interval. In the leading cases of (i) and (ii), no further tuning parameters need to be specified by the researcher. However, these leading cases are nested within a family of restrictions that can be made arbitrarily general, offering a transparent map between the strength of identifying assumptions and the size of the identified set \citep{KKKL18}.

An important advantage of the concavity-based approach relative to existing approaches is that it has a natural \textit{portability} across distributions. I show in Section \ref{sec:validate} that the concavity approach offers a principled way to justify the identifying assumptions by appeal to a comparison distribution. This offers a way to formalize, for example, why it is useful to look at the distribution of choices before or after a kink or notch moves, without assuming that the counterfactual distribution is the same in both periods. 

\begin{remark}[(incoherence of a proportional-density approach)] \label{rem:incoherent}
	A method introduced by \citet{chetty2011} for kinks assumes that for $y>k$, $f_1(y) = \alpha f_0(y)$ for some number $\alpha$. This allows us to impute the counterfactual distribution $f_0(y)$ up to the factor $\alpha$, and a global normalization of $f_0$ in turn pins down the value of $\alpha =  1- \frac{\mathcal{B}}{1-F_0(k)}$, since Eq. \eqref{eq:bmarginals} implies that $F_0(k)+P(Y_1 > k)+\mathcal{B} = 1$. However there is typically nothing inherently special about the point $k$ in the distributions $F_0$ and $F_1$. Thus, if $f_1(y) = \alpha f_0(y)$ for $y>k$ then it is equally plausible to assume $f_1(y) = \alpha f_0(y)$ for $y <k$ as well. This is incompatible with positive bunching though, because $F_1(k) = \alpha \cdot F_0(k)$ implies again by Eq. \eqref{eq:bmarginals} that $\alpha = 1+\frac{\mathcal{B}}{F_0(k)}$. Both equations for $\alpha$ can hold only simultaneously if $\mathcal{B}=0$ and $\alpha = 1$. To proceed with the proportional-density assumption, one would need to believe that $f_1(y) = \alpha f_0(y)$ for all $y>k$, but that this relationship breaks down below $k$.
\end{remark}

\subsection{Identifying assumption: s-concavity of CDFs and survival functions}

A univariate function $\phi(y)$ is called \textit{s-concave} for a value $s\ne 0$ if for any $\theta \in [0,1]$ and $y,y'$: $\phi(\theta y + (1-\theta)y') \ge (\theta \phi(y)^s + (1-\theta)\phi(y')^s])^{1/s}$, and $\phi(\theta y + (1-\theta)y') \ge \phi(y)^\theta \phi(y')^{1-\theta}$ in the case of $s=0$.\footnote{Equivalently, $\phi$ is s-concave for an $s>0$ if the function $\phi(y)^s$ is concave, for $s=0$ if $\ln \phi(y)$ is concave, and for an $s < 0$ if $\phi(y)^s$ is convex.} s-concavity generalizes the notion of concavity and is stronger condition the higher the value of $s$, corresponding to conventional concavity when $s=1$, log-concavity when $s=0$, and quasi-concavity as $s \rightarrow -\infty$. If $\phi(y)$ is s-concave with $s=t$ then $\phi(y)$ is also s-concave with any $s < t$.

A CDF $F$ is called bi-s-concave (BsC) when $F$ and $1-F$ are both s-concave for the same value $s \le 1$ \citep{lahamiaowellner}. Note that for $s=1$, this requires $F$ to be linear, corresponding to distribution with a uniform density on its support. The case of $s=0$ is called bi-log-concavity (BLC), covering the broad class of distributions for which both $\ln F$ and $\ln(1-F)$ are concave. In the limit of $s \rightarrow -\infty$ the family of BsC distributions contains all CDFs, since all CDFs $F$ and survival functions $1-F$ are weakly monotonic and hence quasi-concave. Call a CDF $F$ \textit{locally} BsC on an interval $I$ if $F$ and $1-F$ are s-concave on the interval $I$, with no restriction assumed on its behavior outside of $I$.

\begin{assumption*}[BC(s,t) (local ``bi-concavity'' $(s,t)$ among bunchers)]
	$F_0$ is s-concave on the interval $[k,k+\Delta_0^*]$ and $F_1$ is s-concave on the interval $[y^I-\Delta_1^*, y^I]$. Meanwhile $1-F_0$ is t-concave on the interval $[k,k+\Delta_0^*]$ and $1-F_1$ is t-concave on the interval $[y^I-\Delta_1^*,y^I]$.
\end{assumption*}
\noindent Assumption BC(s,t) allows the researcher to assume different levels of concavity for the CDFs and the survival functions corresponding to $Y_0$ and $Y_1$. This encodes assumptions like monotonicity: if $Y_0$ and $Y_1$ each have densities that decline over the bunching interval, then $F_0$ and $F_1$ are concave over this interval and we say $\textrm{BC}(1,-\infty)$ holds (with the notation $-\infty$ denoting that no restriction is placed on the survival functions $1-F_0$ and $1-F_1$). $\textrm{BC}(-\infty,1)$ instead imposes that the densities are increasing. BC(s,t) recovers the assumption that $F_0$ and $F_1$ are locally BsC over the bunching interval when $t=s$; for example local BLC corresponds to BC(0,0).

Local BLC in particular holds when, over the bunching interval, the hazard rate $f(y)/(1-F(y))$ is increasing while the reverse hazard rate $f(y)/F(y)$ is decreasing \citep{dumbgen_bi-log-concave_2017}. What BLC rules out is thus a density that falls too fast relative to its remaining upper tail, or builds too fast relative to its accumulated lower tail. BLC includes all (globally) log-concave densities, with the normal, exponential, logistic, uniform, Gumbel and Laplace among them.\footnote{A globally log-concave density implies global BLC, and this implication generalizes beyond $s=0$: if $f_d$ is globally s-concave for some $t > -1$, then $F_d$ is globally BsC for all $s \le \frac{t}{1+t}$ \citep{lahamiaowellner}.} A log-normal $Y$ is only locally BLC for low values of $Y$; however $\ln Y$ is normal and thus globally BLC. The Pareto distribution with tail parameter $\alpha>0$ has a hazard rate $\alpha/y$ and is thus nowhere BLC, though it is BsC for all $s \le -1/\alpha$. But again, if $Y$ is Pareto, then $\ln Y$ is exponential and hence BLC everywhere. Thus BLC is satisfied for $\ln Y$---the relevant random variable when the parameter of interest is an elasticity---in the two distributions commonly used to model income and wealth.

Not every polynomial density satisfies BLC, and thus BLC does not strictly nest the polynomial approach of \cite{chetty2011}.\footnote{Section \ref{app:blc} shows that any polynomial or Lipschitz density nonzero on an interval is locally BsC for \textit{some} finite $s$. } This is a strength, as a shape constraint containing all polynomials would be undesirable: a close polynomial fit to one side of $k$ sometimes provides a wild extrapolation on the other side (see Online Supplement for an example). However, BLC contains polynomial densities of any order, including e.g. a linear density as assumed by \cite{saez2010}.

\begin{remark*}[(s-concavity of CDFs vs. the density)]
	An alternative to Assumption BC(s,t) would be to assume that the densities of $f_0$ and $f_1$ are locally s-concave for some $s$. I focus on s-concavity of CDFs rather than densities for two principal reasons. Firstly, s-concavity of CDFs and survival functions nests monotonicity of the density, and s-concavity of a density does not. Monotonicity is a very natural assumption for many applications, and BC(s,t) provides a single unified family of shape restrictions that includes it. Second, the bounds that arise from s-concavity of the density become uninformative in one direction even as the assumption remains substantive. In particular, the upper bound on $Q_k^*$ diverges as $s \downarrow -1/2$. See Online Supplement for details.	
\end{remark*}

Write the buncher QTE from Eq. \eqref{eqdeltak} as
\begin{equation} \label{eq:twoterms}
	Q_k^*= \left\{\frac{1}{\mathcal{B}}\int_{F_0(k)}^{F_0(k)+\mathcal{B}} Q_0(u)\cdot du - k\right\} + \left\{y^I- \frac{1}{\mathcal{B}}\int_{F_1(y^I)-\mathcal{B}}^{F_1(y^I)} Q_1(u)\cdot du\right\}-(y^I-k)
\end{equation}
Each term in brackets in Eq. \eqref{eq:twoterms} admits of an assumption-free lower bound that it is non-negative. Combining, we have that $Q_k^* \ge 0$. One can then increase this trivial lower bound through BC(s,t). For instance s-concavity of $F_0$ yields a lower bound for the first term in \eqref{eq:twoterms}, and t-concavity of $1-F_1$ yields a lower bound on the second. Meanwhile t-concavity of $1-F_0$ yields an upper bound for the first term in \eqref{eq:twoterms}, and s-concavity of $F_1$ yields an upper bound for the second. Neither term in \eqref{eq:twoterms} has an assumption-free upper bound, so an upper bound on $Q_k^*$ requires finite $s$ and $t$.

The following Theorem yields analytic bounds on $Q_k^*$ under BC(s,t), via Eq. \eqref{eq:twoterms}:
\begin{theorem}[(bounds on the buncher QTE from distributional concavity bounds)] \label{thmblc} Assume Assumption BC(s,t) holds. Then the buncher QTE $Q_k^*$ lies in the interval $\left[\Delta^{*L}_k , \Delta^{*U}_k\right]$, where:
	$$ \Delta^{*L}_k:= g_s(F_0(k),f_0(k), \mathcal{B}) + g_t\left(1-F_1(y^I),f_1(y^I),\mathcal{B}\right) - (y^I-k)$$
	$$ \Delta^{*U}_k:= -g_t(1-F_0(k),f_0(k), -\mathcal{B} ) -g_s\left(F_1(y^I),f_1(y^I), -\mathcal{B}\right) - (y^I-k)$$
	with $g_0(a,b,x)=\frac{a}{b}\left[\left(1+ \frac{a}{x}\right)\ln\left(1+\frac{x}{a}\right)-1\right]$ and $g_s(a,b,x)= \frac{a}{b(s+1)}\left[\left(1+ \frac{a}{x}\right) \frac{\left(1+\frac{x}{a}\right)^s-1}{s}-1\right]$ for general $s \notin \{0,-1\}$.\footnote{For $s=-1$, $g_s(a,b,x)$ evaluates to $0/0$ but with a definite limit: $\frac{a}{bx}\left[x- a\ln\left(1+\frac{x}{a}\right)\right]$.} The bounds are sharp. Note that $g_1(a,b,x) = x/2b$.
\end{theorem}
\noindent I note the limits $\lim_{s \downarrow -\infty} g_s(a,b,x) = 0$ and $\lim_{s \downarrow -\infty} -g_s(a,b,-x) = \infty$ for $x \ge 0$. Thus we have $\Delta^{*L}_k = 0$ in the case of $\textrm{BC}(-\infty,-\infty)$ and $\Delta^{*U}_k = \infty$ if either $s$ or $t$ are $-\infty$.

When $t=s$, Assumption $BC(s,t)$ reflects bi-s-concavity. The bounds in Theorem \ref{thmblc} widen as $s$ decreases, from point identification of $Q_k^*$ when $s=1$, through bi-log-concavity and onwards to $[0,\infty)$ as $s \downarrow -\infty$. One can combine BsC with an additional assumption such as monotonicity. For example, if the researcher assumes that $Y_0$ and $Y_1$ are both BLC \textit{and} have declining densities, then $BC(1,0)$ holds, tightening the bounds relative to BLC alone. Log concavity of $F_d$ (from BLC) adds nothing beyond concavity of $F_d$ (from monotonicity); given the nesting property of s-concavity, the researcher can simply use the larger value of $s$ or $t$ in Assumption BC(s,t).

If $F$ is BsC then it admits a strictly positive density $f$ that is itself differentiable almost everywhere with the local bound: $ \frac{-(1-s) \cdot f(h)^2}{1-F(h)} \le f'(h) \le (1-s) \cdot \frac{f(h)^2}{F(h)}$ (see Appendix \ref{app:blc}). Intuitively, this rules out cases in which the density of $f_0$ or $f_1$ ever spikes or falls \textit{too} quickly on the interior of its support, leading to non-identification of the type discussed in \citet{blomquist_bunching_2019}. For a given value $f(y)$, BsC tends to constrain $f'(y)$ more the closer $y$ is to the median of distribution.\footnote{Meanwhile the identifying power of BsC becomes one-sided as $F$ approaches $0$ or $1$. Importantly, if the researcher conditions the analysis on a range $[r_\ell, r_u]$ of the outcome variable, whether the resulting conditional distribution $F$ satisfies BsC may depend on the values of $r_\ell$ and $r_u$. I show in Appendix \ref{app:blc} that conditioning always preserves BsC.}

\begin{remark} \label{remark:logs}
	If the random variables $\ln Y_d$ (rather than $Y_d$) for each $d \in \{0,1\}$ are assumed to satisfy BC(s,t), then Theorem \ref{thmblc} yields bounds on the buncher QTE in logs
	$$\frac{1}{\mathcal{B}}\int_{F_0(k)}^{F_1(k)} [\ln Q_0(u)-\ln  Q_1(u)]\cdot du$$
	Applying the change-of-variables formula for densities, the Theorem \ref{thmblc} bounds are simple functions of the bounds for the buncher QTE in levels. In particular, we obtain by Theorem \ref{thmblc} the bounds $[g_s/k + g_t/y^I, -g_t/k-g_s/y^I] -\ln\left(\frac{y^I}{k}\right)$ where in a slight abuse of notation $g_s$,$g_t$,$-g_t$,$-g_s$ are the terms appearing in Theorem \ref{thmblc}. The assumptions of $BC(s,t)$ for $Y$ and for $\ln Y$ are partially nested. With $F$ the CDF of $Y$ and $G$ the CDF of $\ln Y$, s-concavity of $G$ implies s-concavity of $F$ for any $s \le 1$, while for the survival function the implication runs in the opposite direction: t-concavity of $1-F$ implies t-concavity of $1-G$ for any $t \le 1$. In the case of BLC, assuming BLC for $\ln Y$ and for $Y$ are thus genuinely separate assumptions. But for a one-sided bound such as a declining density $BC(1,\infty)$, the assumption for $\ln Y_d$ is stronger. 
\end{remark}

\subsection{Partial identification results for the buncher ATE} \label{sec:summarizedbounds}
The following Corollary follows from combining Theorem \ref{thmblc} for $Q^*_k$ with the results relating $Q^*_k$ to the buncher ATE $\Delta_k^*$ from Proposition \ref{prop:buncherATEnotch} (with Proposition \ref{prop:buncherATE} as a special case). I state these results in general settings with a kink and/or notch $N>0$, unless stated otherwise.\footnote{Note that $BC(s,t)$ implies CONT with finite $s,t$ (see e.g. Theorem 3 of \citet{lahamiaowellner}).}

For simplicity, I state results for the leading assumptions of BLC and monotone densities. The generalizations to $BC(s,t)$ hold in a straightforward way with $g_s$ and $g_t$ replacing each $g_0$. Define:
$$ \Delta^{L}_k:= g_0(F(k^-), f(k^{-}), \mathcal{B}) + g_0\left(1-F(y^I),f(y^{I+}),\mathcal{B}\right) - (y^I-k)$$
$$ \Delta^{U}_k:= -g_0(1-F(k^{-}),f(k^{-}), -\mathcal{B} ) -g_0\left(F(y^I),f(y^{I+}), -\mathcal{B}\right) - (y^I-k)$$
By Eq. \eqref{eq:iddensnotch}, the bounds $\Delta^{*L}_k$ and $\Delta^{*U}_k$ for the case of BC(0,0) are identified by $\Delta^{L}_k$ and $\Delta^{U}_k$ under CONVEX, CONT (and NOTCHBOUND if $N>0$, given $y^I$).\footnote{Since the bounds depend only on the density around $k$ and the total amount mass to its left/right, point masses elsewhere in the distributions of $Y_0$ and $Y_1$ do not effect on the bounds provided that they are well-separated from $k$.}

When the shape constraint $BC(s,t)$ allows for an upper bound on $Q_k^*$, I state the corresponding upper bound on $\Delta_k^*$ below under the weak assumption SI, noting that $M=0$ yields bounds under RANK. Recall that by Proposition \ref{prop:bATEUB1} we can take $M$ to be $\max \{w(\Delta|Y_1=y^I),w(\Delta|Y_0=k)\}$. For brevity I use the simpler bound $\Delta_k^* \le Q_k^* + 2M$, since $\mathcal{B} \le 1/4$ in most applications. I take items 1.-2. and 4. below as the leading cases for empirical practice. Analogs of these results hold with NOTCH replaced by NOTCHBOUND (see Section \ref{sec:optimizationerror}).
\begin{corollary}[(leading cases of bounds on the buncher ATE)] \label{corr:blc} Combining Theorem \ref{thmblc} with Proposition \ref{prop:buncherATE} and Remark \ref{remark:logs}, we have the following partial identification results:
	\begin{enumerate}[itemsep=1pt]
		\item Assume CONVEX, NOTCH, SI and BLC (i.e. BC(0,0)). Then $\Delta_k^* \in \left[\Delta^{L}_k , \Delta^{U}_k + 2M\right]$.
		\item Assume CONVEX, NOTCH, SI and that BLC (i.e. BC(0,0)) holds for $\ln Y_0$ and $\ln Y_1$. Then $\delta_k^*\in \left[\Delta^{L}_k/k , \Delta^{U}_k/k+2M\right]+\{\frac{y^I}{k}-\ln\left(\frac{y^I}{k}\right)-1\}$, bounding $\mathbbm{E}[\bar{e}_{i}|Y_{i}=k]$.\footnote{See Remark \ref{remark:logs}. Here I take $M$ to be an upper bound on $w(\ln Y_0-\ln Y_1|Y_1=y^I)$ and $w(\ln Y_0-\ln Y_1|Y_0=k)$.} Note that the term in brackets is zero for a pure kink.
		\item Assume CONVEX, NOTCH and BLC (i.e. BC(0,0)). Then $\Delta_k^* \ge \Delta^{L}_k$.
		\item Assume CONVEX, NOTCH, and that $f_{0}$ is weakly decreasing on the interval $[k, k+\Delta_0^*]$ (implied by BC(1,$-\infty$)). Then $\Delta_k^* \ge \frac{\mathcal{B}}{2 \cdot f(k^-)}-(y^I-k)$. The analagous result holds for $\delta_k^*$.
		\item Assume CONVEX, NOTCH, and that $f_{1}$ is weakly increasing on the interval $[y^I-\Delta_1^*, y^I]$ (implied by BC($-\infty$,1)). Then $\Delta_k^* \ge \frac{\mathcal{B}}{2 \cdot f(y^{I+})}-(y^I-k)$. The analagous result holds for $\delta_k^*$.
		\item Assume CONVEX, NOTCH, SI, and BLC holds with $f_0$ also weakly decreasing (i.e. $BC(1,0)$). Then $\Delta_k^* \in \left[\frac{\mathcal{B}}{2 \cdot f(k^-)}-(y^I-k), \Delta^{U}_k + 2M\right]$.
		\item In the case of a pure kink ($N_i=0$), lower bounds under monotonicity (e.g. $\Delta_k^* \ge \frac{\mathcal{B}}{2 \cdot f(k^-)}$) remain valid assuming WARP only, without CONVEX. This uses that $f(k^-) \ge f_0(k^-)$ and $f(k^+) \ge f_1(k^+)$ by Proposition \ref{propobservenotch}.
		\item Point identification holds with CONVEX, RANK and NOTCH under BC(1,1): i.e. if $f_{0}$ and $f_1$ are locally uniform. Then $\Delta_k^* =\frac{\mathcal{B}}{2} \cdot \left(\frac{1}{f(k^-)}+\frac{1}{f_1(y^{I+})}\right) - (y^I-k)$.
	\end{enumerate}
\end{corollary}
\noindent \textit{Comparison to existing results:} The key new features of Corollary \ref{corr:blc} relative to the literature are the explicit allowance for heterogeneous effects, and the form and flexibility of the shape constraints employed. The literature does contain a few results that are suggestive that bunching is informative about a local average response, when responsiveness to incentives varies by unit. For instance, \citet{kleven_bunching_2016} considers a ``small-kink'' approximation that $\mathbbm{E}[\Delta_{i}|Y_{0i}=k] \approx \mathcal{B}/f(k^-)$. This result takes $f_{0}$ to be constant throughout the region $[k, k+\Delta_{i}]$ conditional on each value of $\Delta_{i}$, an assumption that is hard to justify except in the limit that the distribution of $\Delta_{i}$ concentrates around zero. Nevertheless, even in a ``small-kink'' setting, Theorem \ref{thmblc} offers a refinement to this approximation: for example a second-order expansion of $\ln(1+\frac{x}{a})$ shows that when $\mathcal{B}$ is small, the bounds $\Delta^{L*}_k$ and $\Delta^{U*}_k$ under BLC converge around $\frac{\mathcal{B}}{2f(k^-)}+\frac{\mathcal{B}}{2f(y^{I+})}$, i.e. replacing $f_0(k)$ in the formula $\mathcal{B}/f(k^-)=\mathcal{B}/f_0(k)$ with the harmonic mean $\tilde{f} = \left(\frac{1}{2f(k^-)}+\frac{1}{2f(y^{I+})}\right)^{-1}$ of $f_0(k)=f(k^-)$ and $f_1(y^I)=f_1(y^{I+})$. This small-kink limit also coincides with the point-identified value of $Q_k^*$ under BsC as $s \rightarrow 1$ (i.e. $Y_1$ and $Y_0$ have uniform distributions over the bunching interval). Indeed, under BC(1,1): $Q_k^* =\mathcal{B}/\tilde{f} - (y^I-k)$ holds exactly regardless of the value of $\mathcal{B}$.\footnote{The isoelastic model for a pure kink further implies that $f_0(k)=f_1(k)$ under a uniform density, if $\ln Y$ is used as the outcome variable. This is a testable restriction, and recovers the formula $\mathcal{B}/f_0(k)$ of \cite{kleven_bunching_2016} for $\Delta_k^*$.} Some algebra also shows that in the case of a pure kink, $\Delta^{*L}_k>0$ when bunching $\mathcal{B}>0$, so that the buncher ATE can always be bounded away from zero given a point mass at $k$ and $s$ or $t$ > $-\infty$.\footnote{\label{fn:bklndens}The lower bound from items 4. and 5. of Corollary \ref{corr:blc} are of a different nature than bounds under a monotone density given by \citet{blomquist_bunching_2019} as their bounds leverage the full structure of the isoelastic model, which allow $f_0$ and $f_1$ to be perfectly related through the elasticity $\epsilon$. Their upper bound is not valid in the general choice model.} For a notch, the condition for $\Delta^U_k \ge 0$ under the small-$\mathcal{B}$ approximation is that $\mathcal{B} > (y^I-k) \tilde{f}$. This has a natural interpretation: $\mathcal{B}$ must be large enough to fill the hole range $[k,y^I]$ with a density greater than a uniform counterfactual at $\tilde{f}$, i.e. there must be ``excess mass''.\\

\noindent \textit{Testability and Sharpness:} Since $\Delta^{U}_k \ge \Delta^{L}_k$ always, the identified sets above are never empty and cannot be used to falsify the general choice model along with BC(s,t). For a pure kink, the buncher ATE must be positive under WARP (see Proposition \ref{propobserve}), but $\Delta^{U}_k$ will also always be positive. For a notch, $\Delta^{U}_k$ can be negative, but the buncher ATE can also be negative absent further restrictions (WARP no longer rules out $Y_{1i}>Y_{0i}$ for all bunchers). The bounds in e.g. item 1. above can be used to reject the model (including a value of $M$) at a notch if $\Delta^{U}_k+2M <0$ and one adds an explicit assumption that $\Delta_k^* \ge 0$. Given that the bounds in Corollary \ref{corr:blc} consider various overlapping sets of assumptions, I do not establish their sharpness for all cases. The proof of Theorem \ref{thmblc} shows however that e.g the BLC bounds $[\Delta_k^L, \Delta_k^U]$ are sharp for $\Delta_k^*$ under CONVEX, NOTCH and RANK.\\ 

\noindent \textit{Optimization error:} Appendix \ref{sec:optimizationerror} extends the conclusions of Theorem \ref{thmblc} and Corollary \ref{corr:blc} to settings that feature ``optimization error''. There, I assume that the researcher observes $Y_i = Y_i^* + \epsilon_i$ where $Y_i^*$ can be interpreted as any of (i) unit $i$'s \textit{intended} choice; (ii) what their choice would have been without an optimization ``friction''; or (ii) actual choice absent measurement error. When $\epsilon_i$ is continuously distributed, the distribution of observed choices will generally no longer feature a point mass at $k$, i.e. bunching will be diffuse. I proceed by assuming that $\epsilon_i$ is bounded, and by focusing on the parameter of interest $\mathbbm{E}[Y_0-Y_1|Y_0 \ge y_L,Y_1 \le y_U]$ rather than $\mathbbm{E}[Y_0-Y_1|Y_0 \ge k,Y_1 \le y^I]$, where $y_L$ and $y_U$ are values that bracket $k$ and $y^I$, and are sufficiently far from the kink/notch that observations of $Y$ there reflect $Y_1$ or $Y_0$ unambiguously. This parameter of interest recovers the buncher ATE considered in this section in the limit that the $\epsilon_i$ are small. A key feature of the approach in Appendix \ref{sec:optimizationerror} is that BLC and other concavity assumptions are made on $Y_0$ and $Y_1$, rather than on $Y_0^*$ and $Y_1^*$. Bounds involve extrapolation from observed distributions, and the considerations of Section \ref{sec:validate} below around validating shape restrictions empirically apply unchanged.\\

\noindent \textit{An alternative upper bound on $\Delta_k^*$}: The following upper bound on $\Delta_k^*$ (or $\delta_k^*$) complements Proposition \ref{prop:bATEUB1} from the last section, and can be applied analogously to the results in Corollary \ref{corr:blc}.
\begin{proposition} \label{prop:bATEUB2}
	Suppose that SI holds and that $1-F_0(y)$ is t-concave for all $y \ge k$ and $F_1(y)$ is s-concave for all $y < y^I$. Then under CONVEX and NOTCH:
	$$\Delta_k^* \le \frac{1}{1+t} \cdot \frac{1-F(k)}{f(k^-)} + \frac{1}{1+s} \cdot \frac{F(y^I)}{f(y^{I+})} + (y^I-k)$$
\end{proposition}
\noindent Proposition \ref{prop:bATEUB2} shows that assuming that the components of BC(s,t) for $1-F_0$ and $F_1$ hold everywhere in the tails (rather than only through the bunching region) is sufficient to deliver an upper bound on $\Delta_k^*$, even without RANK. The proof allows a relaxation of NOTCH to NOTCHBOUND. 

\subsection{Validating bi-log-concavity} \label{sec:validate}

In this section I describe two approaches to obtaining empirical evidence regarding the concavity assumptions in the family $BC(s,t)$. I focus primarily on the simple case of bi-log-concavity, which I find to have broad empirical support and offers two-sided bounds on $Q_k^*$. The arguments made here generalize, including to the other key empirically relevant assumption of a declining density.

First, I discuss how one can marshal evidence for BLC from the observed distribution facing a kink or notch. This offers direct evidence for BLC of $Y_0$ to the left of the kink/notch, and for BLC of $Y_1$ to the right of $y^I$. Second, I discuss using a reference distribution to serve as an informative counterfactual, such as the distribution before the kink or notch is introduced or when it moves. This is typically informative about BLC of one of $Y_1$ or $Y_0$, and can speak directly to its behavior within the bunching region. Section \ref{app:blc} discusses reasons for BLC to be preserved over time given panel data or repeated cross-sections, and to hold for one of the two potential outcomes if it holds for the other. In the Online Supplement, I discuss theoretical reasons to expect BLC to hold.\footnote{Theoretical reasons to expect BLC for income or wealth draw on models of random exchange \citep{boltzmanngibbs}, while results on transferability between $Y_0$ and $Y_1$ or over time use its closure under convolutions.}

\subsubsection{Using the distribution of choices away from the kink/notch} \label{sec:usecrosssection}

First, let us consider what evidence can be brought to bear given only the cross-sectional data $Y_i$. To simplify the discussion, I throughout this section and the next assume that CONVEX, CONT and NOTCH hold. With only $F$ at one's disposal, we can by Eq. \eqref{hcases} observe that $F_0(y)$ is identified for $y \le k$ and $F_1$ is identified for $y \ge y^I$. We can therefore directly verify whether $F_0$ is BLC on the interval $(-\infty,k)$ and whether $F_1$ is BLC on the interval $(y^I,\infty)$. Identification of $\Delta_k^*$ by BLC simply requires us to then believe that we can extend the range upon which $F_0$ and $F_1$ are BLC: the interval $[k, k+\Delta_0^*]$ for $F_0$ and the interval $[y^I-\Delta^*_1, y^I]$ for $Y_1$ (see figure in Online Supplement).

\subsubsection{Using a reference distribution}
The following result proves useful in motivating BC(s,t) using a ``reference distribution''. 
\begin{proposition} \label{prop:sandwich}
	Let $Y$ be a random variable with distribution function $F$. Let $G$ be a reference distribution, and suppose that $G$ and $F$ are both continuous and strictly increasing. Let $g=G'(y)$ be the density associated with $G$, $h_G(y) = g(y)/\{1-G(y)\}$ its hazard function and $r_G(y) = g(y)/G(y)$ its reverse hazard function. Let $T(z):=F^{-1}(G(z))$.
	
	For any interval $[a,b]$, $F$ is s-concave and $1-F$ is t-concave on $[a,b]$ if and only if
	\begin{equation} \label{eq:blcsandwichgenerals}
		\frac{d}{dz} \ln r_G(z) + s\cdot r_G(z)^2 \le \frac{d}{dz} \ln T'(z) \le \frac{d}{dz} \ln h_G(z) - t\cdot h_G(z)^2
	\end{equation}	
	for all $z \in [T^{-1}(a),T^{-1}(b)]$.
\end{proposition}
\noindent To illustrate the utility of Eq. \eqref{eq:blcsandwichgenerals}, consider the special case of BLC ($s=t=0$). $F$ is BLC iff:
\begin{equation} \label{eq:blcsandwich}
	\frac{d}{dz} \ln r_G(z) \le \frac{d}{dz} \ln T'(z) \le \frac{d}{dz} \ln h_G(z)
\end{equation}
If $G$ is itself BLC, then $\frac{d}{dz} \ln h_G(z) \ge 0$ and $\frac{d}{dz} \ln r_G(z) \le 0$, meaning that $F$ will be as well provided that the non-linearity of $T$, as measured by $\frac{d}{dy} \ln T'(y)=\frac{T''(y)}{T'(y)}$, is suitably close to zero. There is always ``slack'' between the two inequalities in \eqref{eq:blcsandwich}, whether or not $G$ is itself BLC.\footnote{The condition $\frac{d}{dz} \ln h_G(z) = \frac{d}{dz} \ln r_G(z)$ implies a contradiction and is therefore satisfied by no distribution.}

Proposition \ref{prop:sandwich} thus generalizes the fact that any affine transformation (which leads to a perfectly linear $T$) of a BLC random variable is also BLC. Similarly, if $G$ is BsC, this implies $\frac{d}{dz} \ln h_G(z) + s \cdot h_G(z)^2 \ge 0$ and $\frac{d}{dz} \ln r_G(z) + s \cdot r_G(z)^2\le 0$ and the upper and lower bounds in \eqref{eq:blcsandwichgenerals} again straddle zero. Thus $Y$ (or $\ln Y$ in the case of the buncher ATE in logs) will also inherit BsC for the same value of $s$ if $T$ is close to linear.

The shape of $T(z)$ given two distributions $F$ and $G$ can be inspected by constructing a Q-Q plot between the two random variables. With $F$ the CDF of the $Y_0$ or $Y_1$ potential outcome, this Q-Q plot can only be constructed over the region in which $Y_d$ is observed, e.g. below $k$ for $Y_0$ and above $y^I$ for $Y_1$. Although the full Q-Q relationship is censored, the distribution $G$ can generally be observed over a region that plausibly contains the entire bunching interval for $Y_d$. 

Settings where the kink or notch moves provide the most credible reference distribution for $\frac{d}{dz} \ln T'(z)$. In this case, linearity of $T(z)$ is all but ensured for $Y_0$ or $Y_1$ if the distribution is approximately stationary between the time periods, shifts by a constant factor, or experiences multiplicative growth. Further closure properties under changes are discussed in Section \ref{app:blc}.
\begin{figure}[h!]
	\begin{subfigure}{.45\textwidth}
		\includegraphics[width=3.2in]{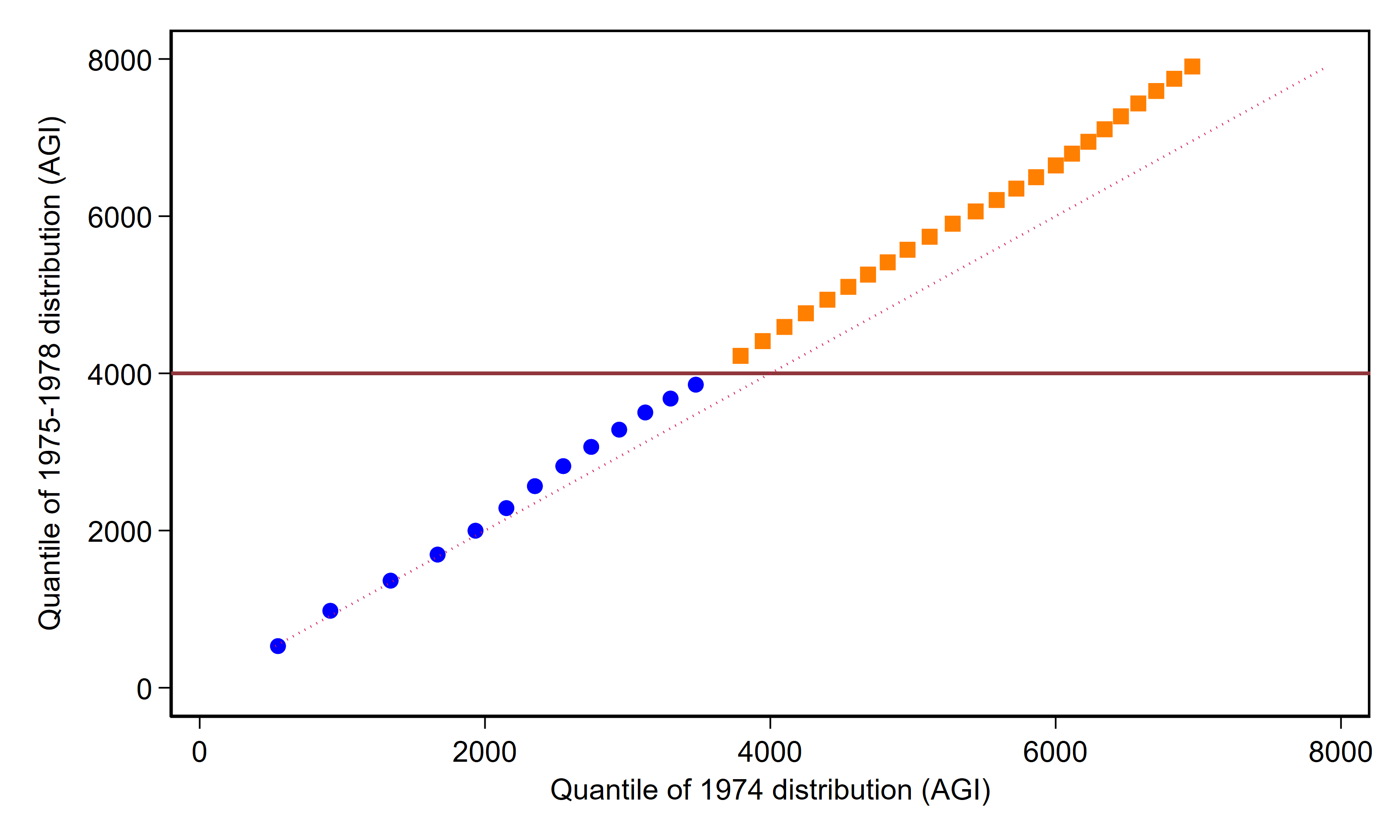}
	\end{subfigure} \hspace{.5in}
	\begin{subfigure}{.45\textwidth}
		\includegraphics[width=3.2in]{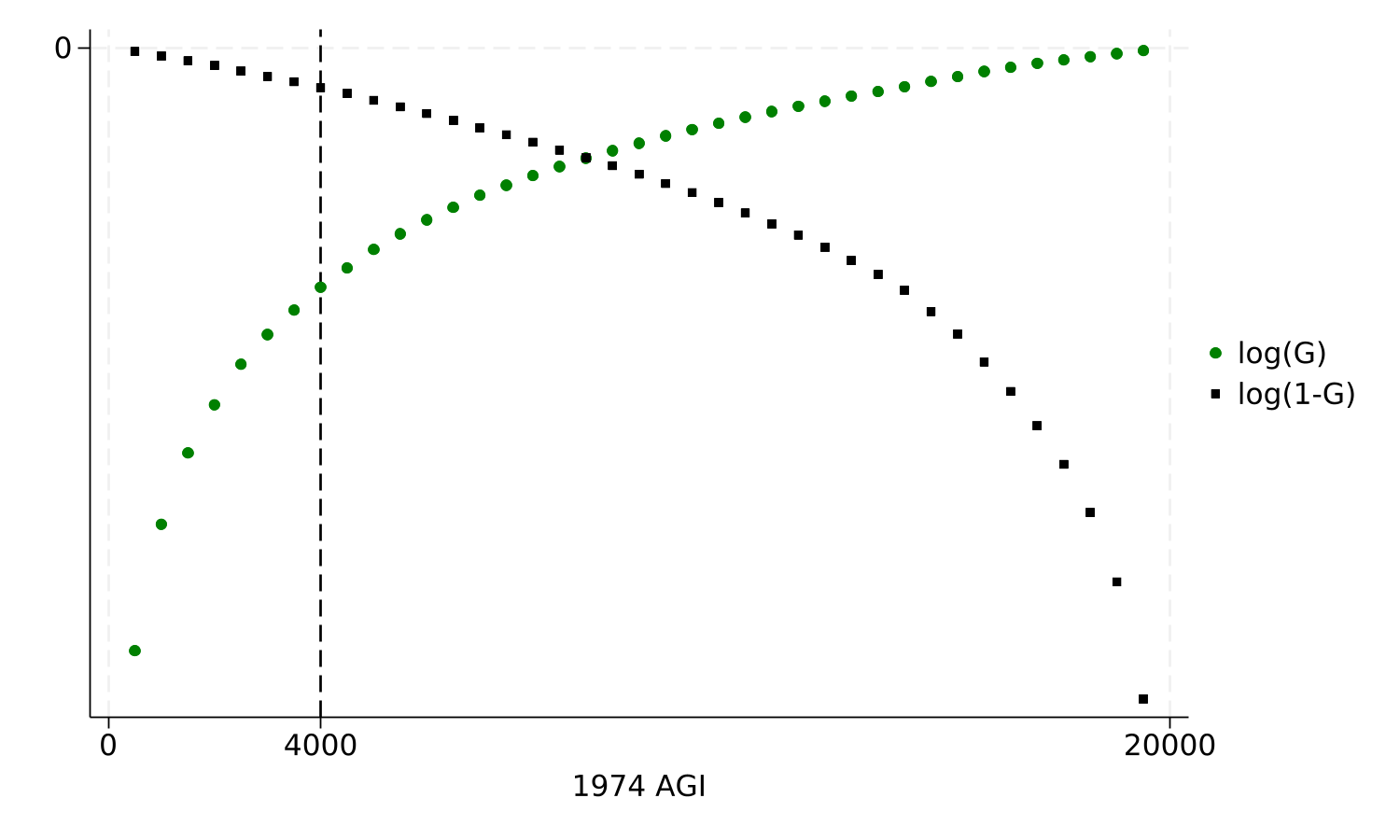}
	\end{subfigure}
	\caption{QQ plot and BLC visualization for the EITC application. Left panel: Q-Q plot of the 1974 vs 1975 distributions of adjusted gross income (with 45 degree line for reference). Right panel: $\ln G$ and $\ln(1-G)$ with $G$ the distribution of AGI for incomes below $\$20,000$ in 1974, before the kink was introduced.} \label{eitcfigures}
\end{figure}

Figure \ref{eitcfigures} provides an example of how Proposition \ref{prop:sandwich} can be used. I use data from US tax filers near the kink in the Earned Income Tax Credit (EITC) studied by \citet{saez2010}. Details on the data are provided below in Section \ref{sec:eitc}. Though not studied by \citet{saez2010}, these data contain information just before the EITC's introduction in 1974. I let $G$ in Proposition \ref{prop:sandwich} be the distribution of adjusted gross income (AGI) in 1974, a year before the EITC was introduced, and $F$ the distribution of AGI $Y$ in 1975-1978.\footnote{I pool 1975-1978 rather than look at 1975 alone because no bunching at $k=\$4,000$ is evident yet in 1975. I do not use population weights in this part of the analysis because they are not yet available in 1974-1975. For this reason I also condition $G$ on the range $[\$0,\$20,000]$, roughly the 75th percentile of family income in 1974.} The EITC introduced a convex kink at \$4,000 in 1975 that remained there through 1978 (and was the same for filers with one child or more than one child).

The left panel shows a Q-Q plot between $F$ and $G$ on the range 0-\$8,000, revealing that $T(z)$ is indeed quite linear near the kink, with any perceptible nonlinearities falling outside of a range of roughly \$2,500-\$6,000. Below $k$ (blue circles), the 1975-1978 distribution reveals $F_0$, while above it (orange squares) $F$ reveals $F_1$. By Proposition \ref{prop:sandwich}, this implies that $F_1$ and $F_0$ are each locally BLC provided that (i) $G$ is locally BLC on each side of the kink and (ii) linearity of the Q-Q relationship between $G$ and each of these random variables persists sufficiently far across the kink for both counterfactuals. The right panel verifies (i) directly: the distribution $G$ appears to be BLC globally. The statistical test for BLC of $G$ reported in the next section confirms this observation.

In Section \ref{sec:litreview} I apply Proposition \ref{prop:sandwich} with a second counterfactual from the EITC application. In \cite{saez2010}, there is no visible bunching among wage earners without self-employment income. I test across the entire distribution for wage earners (with one child) and fail to reject BLC. The Q-Q relationship with actual income is again close to linear, though not quite as closely as in Figure \ref{eitcfigures} for the 1974 counterfactual before the introduction of the kink. Both are included in Table \ref{tab:blctest} there.

\begin{remark}
	\citet{bertanha_better_2018} propose assuming that $\ln y_0$ is Lipschitz continuous with a known constant. An analagous result to Proposition \ref{prop:sandwich} can be derived under this property, but the effect of $T$'s shape is not isolated to its linearity. In particular $f$ has a Lipschitz constant of $L^*$ if $- \frac{L^* \cdot T'(z)^2}{g(x)} \le \frac{d}{dz} \ln T'(z) - \frac{d}{dz}\ln g(z) \le \frac{L^* \cdot T'(z)^2}{g(x)}$. Therefore if the reference density $g$ is Lipschitz with constant $L$ and $T'(z) \ge b > 0$ for all $z$, then $f$ is Lipschitz with constant $L^*=(L+||g||_\infty \cdot ||\frac{d}{dz} \ln T'(z)||_\infty)/b^2$. Linearity of $T$ yields no prediction without knowing $b$.
\end{remark}

\subsubsection{A proposed workflow for researchers to validate BLC empirically}
The results of the preceding two sections suggest a workflow for applied researchers, based on testing BLC in $F$ and/or in a reference distribution $G$. In the EITC application above, Figure \ref{eitcfigures} provided an ``eyeball test'' of BLC for the reference distribution $G$ provided by 1974 AGI. In the next section, I propose a statistical test as well, where BLC is the null hypothesis.

Given such a test, the proposed workflow is:
\begin{enumerate}[noitemsep]
	\item Find a suitable counterfactual density $G$ for $F_0$ or $F_1$. If none is available, skip to step 4. 
	\item Inspect the QQ-plot between $G$ and $F$ for approximate linearity on the relevant range (e.g. below the kink/notch for $F_0$ and above for $F_1$)
	\item Test for BLC and/or monotonicity of $G$, on a range that plausibly includes the bunching region e.g. from $k-E$ to substantially above $k-E$ in the case of $F_0$, where $E$ reflects an upper bound on optimization errors (see Section \ref{sec:optimizationerror})
	\item If no suitable counterfactual is found, test on the actual distribution, away from the kink/notch
	\item If the test fails, one can consider smaller values of $s$ and $t$ in Assumption BC(s,t). Alternatively, one can consider the weaker restriction of \textit{conditional} BLC on a truncated distribution over a narrower range (see Proposition \ref{prop:blctruncation}). Both come at the expense of wider bounds.
\end{enumerate}
If the researcher is primarily interested in an elasticity, they should test on distributions $F$ and/or $G$ for the log of $Y$, rather than $Y$. However the data may be consistent with BLC of $Y$ and not of $\ln Y$, in which case BLC bounds on treatment effects can still be estimated.

\subsubsection{Testing of BLC across the published literature} \label{sec:litreview}
To illustrate the two approaches to validating BLC described above, I implement a survey of empirical bunching papers published since 2010 in the \textit{American Economic Review}, the \textit{Journal of Political Economy}, the \textit{Review of Economic Studies}, \textit{Econometrica}, the \textit{Quarterly Journal of Economics}, the \textit{Review of Economics and Statistics}, and the \textit{American Economic Journal}. The main set of included papers was built from a comprehensive survey of papers containing the terms ``bunching'' ``kink'', or ``notch'' in their title or abstract. The papers were then limited to those that implement the bunching design to identify an elasticity or similar causal parameter, resulting in 27 studies. 

I digitized the distribution of $Y$ as reported in figures from each paper in its published form. In most cases, underlying vector information from the PDF images and was retrieved using the Python package \texttt{PyMuPDF}.\footnote{For one paper (\citealt{lebarbanchon2025}), the image quality was not sufficient for \texttt{PyMuPDF} and I instead used the microdata from the authors' published replication package. For another paper (\citealt{ring2025}), I use a working paper version as the published version is still pending at the time of writing.} Though some papers examine multiple kinks or notches and on multiple sub-samples of the data, I extract one distribution of $Y$ per paper. When available, I also extract one reference distribution that the authors present as a reasonable counterfactual to the distribution featuring the kink/notch. When multiple such comparison distributions are available, I choose one based on (i) approximate sample size; then (ii) spatial or temporal proximity to the actual distribution; and finally (iii) arbitrarily in order of presentation. Details of the inclusion criteria for papers, and the selection of distributions from within the paper, are provided in Appendix \ref{app:litdetails}.

Figures \ref{fig:blcplots} and \ref{fig:blcplots2} show empirical plots of $\ln F$ and $\ln(1-F)$ for each of the tested distributions (the underlying histogram counts extracted are provided in the Online Supplement for reference). When a reference distribution is available in a given paper, I test on it. When one is not, I test on a portion of the actual distribution. Appendix Figure \ref{fig:blcplots} focuses on the papers in which the actual distribution of $Y$ is used for the BLC test. In gray I depict the portion of the distribution around the bunching point which I exclude from the BLC test. When possible, I follow the authors' own excluded region to define this range.\footnote{In the case of \citet{seibold2021}, there is no evident bunching so the entire range of $Y$ is used.} Within the ranges to the left or to the right of the excluded region, $\ln F$ and $\ln(1-F)$ pass a visual ``test'' of (near) concavity in all cases.

\begin{figure}[H]
	\begin{center}
		\includegraphics[width=6.5in]{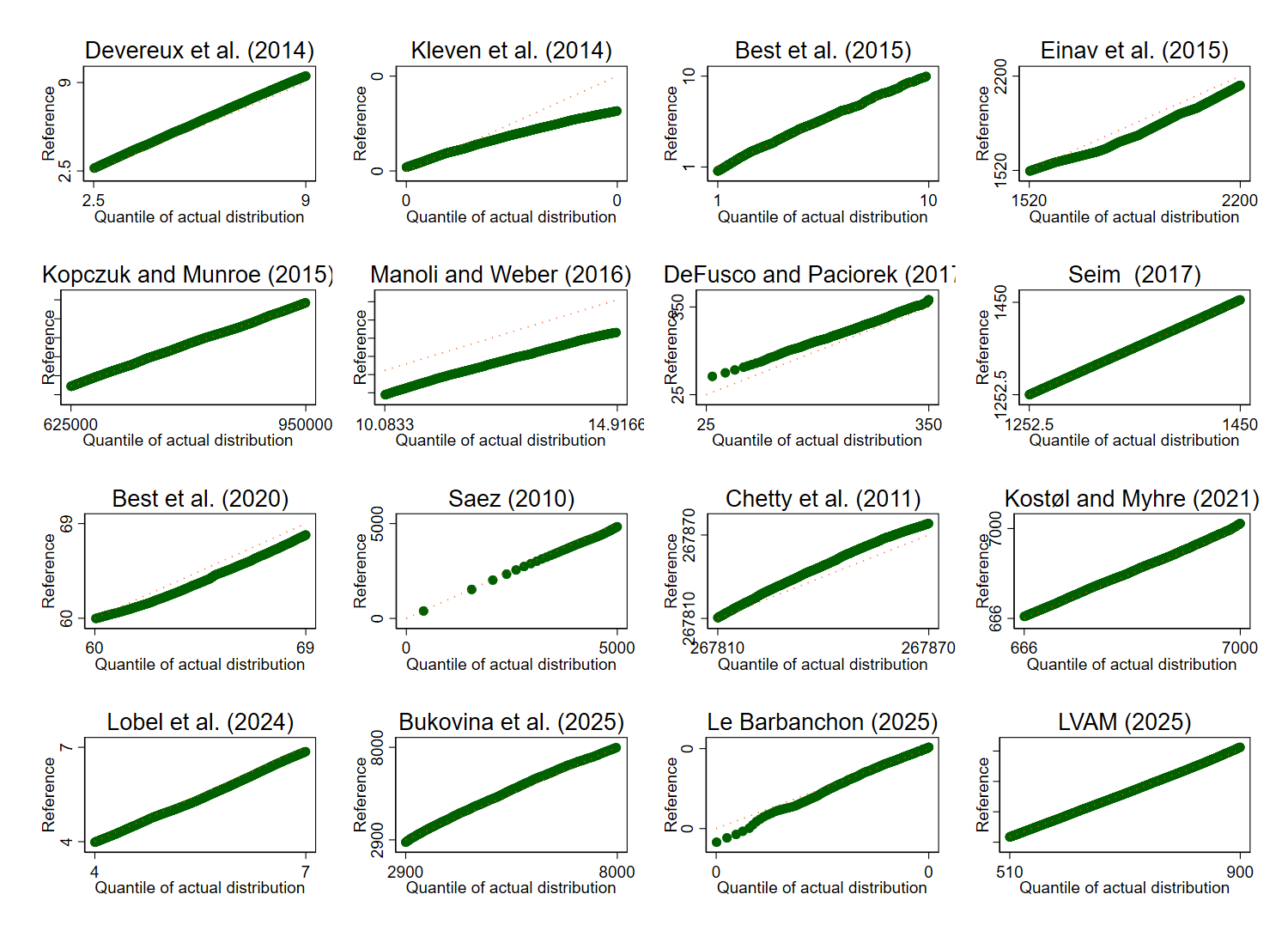}	
	\end{center}
	\caption{Q-Q plots between reference and actual distribution over the relevant range (e.g. on one side of $k$). I show the Q-Q plots of $\ln Y$, but with labels giving the associated level of $Y$ for legibility. Q-Q plots for $Y$ are similarly linear and are shown in  the Online Supplement.}
	\label{fig:qqplots}
\end{figure}

Appendix Figure \ref{fig:blcplots2} turns to the subset of 18 papers in which a suitable reference distribution was found to pursue the strategy of Proposition \ref{prop:sandwich}. In many cases, I am able to test BLC over the whole support of $Y$, but in other cases I again exclude a range of values because the kink or notch has only moved, rather than being entirely absent in the reference distribution. However even these cases allow a direct test of BLC across a range of $G$ that plausibly includes the bunching regions of $F_0$ and $F_1$. Excluded regions are again depicted in grey, and their definitions are given in Appendix \ref{app:litdetails}. To the eye, $\ln F$ and $\ln(1-F)$ again appear generally concave. Figure \ref{fig:qqplots} reports QQ plots between this reference distribution and the actual distribution. These relationships appear to be linear in nearly all cases, with non-linearity most detectable to the eye in \citet{lebarbanchon2025, einav2015}. In these cases, $G$ does not provide as compelling a reference for BLC of $F$ via Proposition \ref{prop:sandwich}, though this does not constitute evidence that $F$ is not BLC.

Table \ref{tab:blctest} reports p-values for a statistical test of BLC for both $\ln Y$ and $Y$ on the tested region. The test converts concavity of $\ln F$ and $\ln 1-F$ into a set of moment inequalities: $\Delta^2 \ln \hat{F}_j \le 0$ and $\Delta^2 \ln(1-\hat{F}_j) \le 0$, where $\Delta^2$ denotes the second difference operator and $F_j$ is the empirical CDF evaluated at the histogram bin edges. I then use generalized moment selection (\citealt{AS}, henceforth AS) to test these moment inequalities, using the ``modified method of moments'' test statistic $S_1$ of AS that sums across studentized moments. An alternative approach would be to use a max-type statistic that counts only the largest studentized violation (e.g. the $S_3$ function of AS). This choice is not very consequential: in Appendix \ref{sec:testing} I describe an alternative max statistic approach that gives similar results. Either test is a demanding one for BLC: since the moment inequalities are defined from the bins at which each distribution is reported in the source paper (typically a quite fine resolution), the test is sensitive to very localized violations of BLC.\footnote{On the other hand, $F(Y)$ is only observed at bin edges from the original paper, so I have no power to reject BLC at a scale smaller than the published bin width. Such failures of BLC are likely inconsequential for $\Delta_k^*$ (see below). The QLR test statistic ($S_2$ of AS) cannot be implemented directly, as the moments are typically not linearly independent.} 

As a hypothesis test for BLC, we can also obtain rejection for a small magnitude deviations from BLC, if the sample size is large. To complement the p-values for BLC, I present a lower 95\% confidence bound for $\kappa := -s^*$, where $s^*$ is defined as the largest non-positive value $s$ for which the tested distribution is BsC. This is obtained by testing BsC for values of $s \in [0,-10]$ and reporting the value at which the test first fails to reject. I thus use the weaker assumption BsC to assess the \textit{size} of deviations from BLC. The relative width of Theorem \ref{thmblc} bounds under $\kappa$ to those under BLC depends on $F(k^-),F(y^{I}),f(k^-),f(y^{I+})$. As a benchmark, the bounds are roughly twice as wide when $\kappa=1$ using the values from the EITC application in Section \ref{sec:eitc}. The test for $\kappa$ is again demanding in the sense that localized violations of BsC can lead to rejection, even in cases where the BsC bounds on the buncher ATE would remain valid because these violations are compensated for by non-binding moment inequalities at other values of $y$. Further details of the test are given in Appendix \ref{sec:testing}.

\begin{table}[!htbp]
\centering
% ---------------------------------------------------------------
% AUTO-GENERATED by make_blctest_table.py -- do not edit by hand.
% Requires: booktabs, array, xcolor. Cites come from the CSV itself.
% ---------------------------------------------------------------
% Highlight for significant p-values -- redefine before the \input to
% restyle (e.g. \newcommand{\blcsig}[1]{\textbf{#1}}).
\providecommand{\blcsig}[1]{\textcolor{purple}{#1}}
% Clickable year for hand-written citations. natbib's \citeyear is not
% hyperlinked, so the link is made against hyperref's bibliography anchor.
% Redefine before the \input to restyle, or to \citeyear alone if not using
% hyperref. Note a manual \hyperlink picks up linkcolor, not citecolor.
\providecommand{\blcyear}[1]{\hyperlink{cite.#1}{\citeyear{#1}}}
\centering\footnotesize
\caption{BLC test results across studies}
\label{tab:blctest}
% The group truncates author lists to one name ("X et al.") for this
% table only, and tightens the intercolumn space. \ifdefined guards the
% biblatex counters so the file still compiles without biblatex.
{\ifdefined\defcounter \defcounter{maxnames}{1}\defcounter{minnames}{1}\fi
\setlength{\tabcolsep}{4pt}%
\begin{tabular}{@{}
  >{\raggedright\arraybackslash\hangindent=1em\hangafter=1}p{1.75in}
  >{\raggedright\arraybackslash}p{1.05in}
  l c c c c c r@{}}
\toprule
 &  &  & \multicolumn{3}{c}{$\log Y$} & \multicolumn{2}{c}{$Y$} &  \\
\cmidrule(lr){4-6} \cmidrule(lr){7-8}
 &  &  & \multicolumn{2}{c}{p-val} & \multicolumn{1}{c}{$\kappa$} & \multicolumn{1}{c}{p-val} & \multicolumn{1}{c}{$\kappa$} &  \\
\cmidrule(lr){4-5} \cmidrule(lr){6-6} \cmidrule(lr){7-7} \cmidrule(lr){8-8}
 & Choice variable & Design & (near $k$) & (global) & (global) & (global) & (global) & $N$ \\
\midrule
\multicolumn{9}{@{}l}{\emph{Using actual choice distribution:}} \\
\addlinespace[0.2em]
\citet{kleven2013} & Taxable income & Notch & \blcsig{$<$0.001} & \blcsig{$<$0.001} & 7.3 & \blcsig{$<$0.001} & 7.4 & 60k \\
\citet{best2018} & House price & Notch & \blcsig{$<$0.001} & \blcsig{$<$0.001} & $>$10 & \blcsig{$<$0.001} & $>$10 & 1.9mil \\
\citet{ruh2019} & Earnings & Notch & 0.204 & \blcsig{$<$0.001} & $>$10 & \blcsig{$<$0.001} & $>$10 & 270k \\
\citet{ring2025} & Charitable giving & Kink & \blcsig{$<$0.001} & \blcsig{$<$0.001} & $>$10 & \blcsig{$<$0.001} & $>$10 & 170k \\
\citet{kleven2011} & Income & Kink & 0.924 & 0.868 & 0 & 0.856 & 0 & 4.8k \\
\citet{almunia2018} & Revenue & Notch & 0.979 & 1.000 & 0 & 1.000 & 0 & 290k \\
\citet{ito2018} & Vehicle weight & Notch & 0.350 & 0.272 & 0 & 0.268 & 0 & 4.6k \\
\citet{bachas2021} & Revenue & Notch & 0.998 & 0.998 & 0 & 0.939 & 0 & 140k \\
\citet{seibold2021} & Job exit age & Kink & 0.996 & 0.999 & 0 & 0.999 & 0 & 58k \\
\addlinespace[0.6em]
\multicolumn{9}{@{}l}{\emph{Using a reference distribution:}} \\
\addlinespace[0.2em]
\citet{devereux2014} & Taxable profit & Kink & \blcsig{$<$0.001} & \blcsig{$<$0.001} & 4.9 & \blcsig{$<$0.001} & 3.5 & 390k \\
\citet{best2015} & Profit/turnover & Kink & \blcsig{$<$0.001} & \blcsig{0.035} & 0.029 & 0.531 & 0 & 3.9k \\
\citet{kopczuk2015} & House price & Notch & \blcsig{$<$0.001} & \blcsig{$<$0.001} & $>$10 & \blcsig{$<$0.001} & $>$10 & 16k \\
\citet{manoli2016} & Years of tenure & Notch & 0.090 & \blcsig{0.010} & 0.42 & \blcsig{0.007} & 0.54 & 13k \\
\citet{defusco2017} & Loan amount & Notch & \blcsig{$<$0.001} & \blcsig{$<$0.001} & 2.8 & \blcsig{$<$0.001} & 4.2 & 240k \\
\citet{best2020} & LTV ratio & Notch & \blcsig{0.006} & \blcsig{0.005} & 0.21 & \blcsig{0.004} & 0.23 & 180k \\
\citet{mortenson2020} & Wage income & Kink & \blcsig{$<$0.001} & \blcsig{$<$0.001} & 2.4 & \blcsig{$<$0.001} & 2.8 & 3.7mil \\
\citet{lebarbanchon2025} & Earnings/benefit & Kink & \blcsig{$<$0.001} & \blcsig{$<$0.001} & $>$10 & \blcsig{$<$0.001} & $>$10 & 110k \\
\citet{saez2010} (wage earners) & Income & Kink & 1.000 & 1.000 & 0 & 1.000 & 0 & 39k \\
\citet{saez2010} (1974 AGI) & Income & Kink & 0.494 & 0.605 & 0 & 0.955 & 0 & 34k \\
\citet{chetty2011} & Income & Kink & 1.000 & 1.000 & 0 & 1.000 & 0 & 900k \\
\citet{kleven2014} & Earnings & Notch & 1.000 & 1.000 & 0 & 1.000 & 0 & 1.2k \\
\citet{einav2015} & Drug spending & Kink & 0.943 & 1.000 & 0 & 1.000 & 0 & 440k \\
\citet{seim2017} & Wealth & Kink & 1.000 & 1.000 & 0 & 1.000 & 0 & 770k \\
\citet{shaffer2020} & Energy use & Kink & 1.000 & 1.000 & 0 & 1.000 & 0 & 290k \\
\citet{kostol2021} & Earnings & Notch & 0.330 & 0.820 & 0 & \blcsig{0.008} & 0.31 & 20k \\
\citet{lobel2024} & Gross revenue & Notch & 0.612 & 0.779 & 0 & 0.623 & 0 & 10k \\
\citet{bukovina2025} & Taxable income & Kink & 0.578 & 0.881 & 0 & 0.319 & 0 & 37k \\
\citet{londonovelez2025} & Net worth & Notch & 0.631 & 0.377 & 0 & 0.111 & 0 & 100k \\
\bottomrule
\end{tabular}}

\vskip 5pt % 
\footnotesize p-values written in purple indicate values below $5\%$. Rounded sample sizes given in final column. See text for details.
\end{table}

The rows in Table \ref{tab:blctest} are organized with the studies in which we reject BLC of $\ln Y$ across the entire tested region (marked as ``(global)'') coming first. I also report p-values where the tested region is additionally restricted to the third of the support of $Y$ that straddles the location of the kink/notch $k$, labeled ``(near k)''. This leads to two less rejections, and suggests that there are applications in which BLC holds near $k$ while BLC fails in the tails of the distribution.\footnote{I note that the p-values in the ``(near k)'' column are not systematically higher than those in the ``(global)'', so the fewer rejections does not appear to be explained by a loss of power.} At the same time, the fact that there are \textit{only} two additional rejections suggests that failures of BLC in empirically relevant settings may be quite dispersed across the support of $Y$, lending some additional credibility to extrapolating the shape constraint from one side of the $k$ to the other.

Table \ref{tab:blctest} suggests broad empirical support for BLC, especially in applications where the choice variable is income or wealth. We can reject BLC at all conventional significance levels in several applications, though these settings have some distinctive features. Among 11 applications in which the choice variable is individual income, earnings or wealth, BLC for $\ln Y$ is rejected in only three in the global column: \cite{ruh2019}, \cite{mortenson2020}, and \cite{kleven2013}. In \citet{ruh2019}, rejection seems to be driven by the tails far from $k$, given that we do not also reject in the ``(near k)'' column.\footnote{\cite{ruh2019} is also one of three papers (along with \citealt{ring2025} and \citealt{seim2017}) in which reconstructing $\ln Y$ from the published figures is imperfect because the authors follow a normalize-and-pool strategy where the choice variable is measured relative to different groups' thresholds (normalizing $k$ to 0).} In the case of \cite{mortenson2020}, I obtain a confidence bound of $2.4$ for $\kappa$, indicating a mild departure from BLC for $\ln Y$. In the case of \cite{kleven2013}, the histogram follows a jagged pattern (see Online Supplement) that could be explained by rounding effects (\citet{lebarbanchon2025} has a similar feature). Large departures from BLC, evidenced by large values $\kappa > 10$, occur for charitable giving, house prices, years of tenure and other non-standard running variables. Overall, I find that BLC is generally well-supported in the tested distribution, especially for $\ln Y$ and in standard tax applications. Though it would be straightforward to implement with a similar test, I do not test for monotonicity of the density.\footnote{As a benchmark to the BLC results, I found that a test for linearity of the density of $Y$ (a Wald test that third differences of the CDF are equal to zero) over the same bins as my ``global'' BLC test rejected in all cases except \citet{einav2015}. This does not directly test the \citet{saez2010} approach, which hinges on the isoelastic model in addition to local linearity over the bunching interval. Rather it suggests that my failures to reject BLC do not reflect a lack of power. An adaptation of the same test for a degree-7 polynomial fit also rejects in all cases where I reject BLC of $\ln Y$.} Many of the distributions that fail BLC appear visually to have declining densities, suggesting that Corollary \ref{corr:blc} could still be used to construct lower bounds on the buncher ATE. 

The general support for BLC found across applications above raises the question of whether there are general economic principles that might explain its prevalence. In the Online Supplement, I discuss theoretical reasons to expect BLC to hold, particularly for distributions of log income or wealth. In Appendix Proposition \ref{prop:blcconvolution} I discuss a closure result for BLC that provides reason to expect BLC to be roughly persistent over time (e.g. in the case of panel data or repeated cross-sections), and to translate from one potential outcome (e.g. $\ln Y_0$) to the other (e.g. $\ln Y_1$).

\section{Application to the EITC} \label{sec:eitc}
This section considers the kink application from \citet{saez2010}. In the Online Supplement, I provide an empirical application of my results to a notch \citep{londonovelez2025}.

Following \citet{saez2010}, I use the IRS Statistics of Income (SOI) dataset, pooling over the years 1995-2004. As $Y$ I use adjusted gross income (AGI). Inflating AGI to 2008 values maintains the first kink in the EITC schedule at its 2008 value of \$8,580 for those with one child and \$12,060 for those with two or more children. At the kink the marginal tax rate increases from -34\% to 0\% in the former case, and from -40\% to 0\% in the latter. To reduce crowding in Table \ref{table:eitcestimates}, I focus on estimates for the population with $2+$ children, which is larger than those with one child.

\begin{figure}[h!]
	\begin{center}
		\includegraphics[width=4.5in]{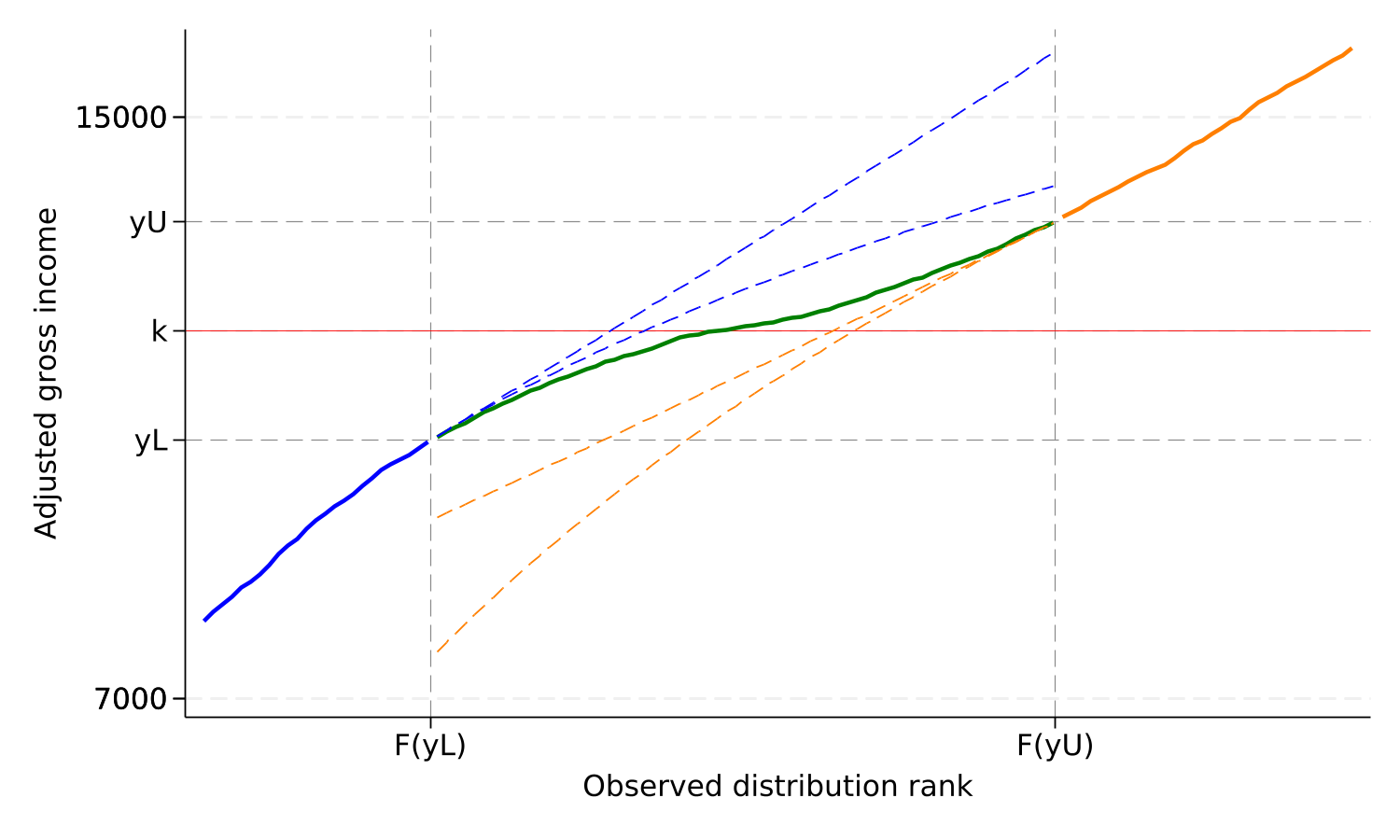}
		\caption{Empirical quantile plot (cf. Figure \ref{figquantiles}) in the EITC application, for family with 2+ children reporting self employment income, and setting $y_L=k-E$, $y_U=k+E$ with $E=1,500$. Graph is zoomed into the range $\$8,000-\$16,000$ for readability. Thick solid lines depict the observed quantile function for $y<y_L$ (blue) and $y>y_U$ (orange). Dashed lines depict estimated envelope functions from assuming BLC. The thinner solid curve (green) depicts the observed distribution in the bunching region.} \label{eitcquantiles}
	\end{center}
\end{figure}

Given that the bunching in this application is diffuse around $k=\$12,060$, I employ Theorem \ref{thm:blcwitherrors} from Appendix \ref{sec:optimizationerror} that replaces $k$ in estimation by values $y_L \le k - E$ and $y_U \ge k+E$ where $E$ is an upper bound on the magnitude of optimization errors. I use $E=\$1,500$, the bandwidth employed by \citet{saez2010} in his baseline estimates. To estimate $f(y_L^-)$ and $f(y_U^+)$, I use the local polynomial density estimator of \citet*{cattaneo_simple_2017} (CJM), which is well-suited to boundary points. I use a triangular kernel, and use CJM's mean-squared error minimizing bandwidth selector on a window within \$10,000 of the kink. I construct confidence intervals for the underlying parameter $\Delta_k^*$ \citep{imbens_confidence_2004, stoye_more_2009} using 
nonparametric bootstrap.

\begin{center}
	\small
	{
\def\sym#1{\ifmmode^{#1}\else\(^{#1}\)\fi}
\begin{tabular}{l*{4}{c}}
\hline\hline
 & \multicolumn{2}{c}{All} & \multicolumn{2}{c}{With self-employment income} \\
 & Estimate & 95\% CI & Estimate & 95\% CI \\
\hline \\
\multicolumn{5}{c}{\underline{Bunching estimates}} \\
Bunching probability (Saez method) & 0.005 & [0.003, 0.007] & 0.031 & [0.026, 0.035] \\
Bunching probability (BLC bounds) & [0.000, 0.037] & [0.000, 0.038] & [0.000, 0.067] & [0.000, 0.070] \\
Mass in buncher ATE & 0.037 & [0.036, 0.039] & 0.067 & [0.063, 0.070] \\
                            &        &        &        &        \\
\multicolumn{5}{c}{\underline{Elasticity estimates}} \\
Saez method & 0.12 & [0.07, 0.17] & 0.53 & [0.39, 0.67] \\
BLC buncher avg. elasticity & [0.1, 0.24] & [0.02, 0.34] & [0.27, 0.58] & [0.09, 0.82] \\
BLC with M=1000/k & [0.1, 0.33] & [0.02, 0.42] & [0.27, 0.67] & [0.10, 0.90] \\
BLC with M=4000/k & [0.1, 0.58] & [0.02, 0.67] & [0.27, 0.91] & [0.10, 1.15] \\
BLC Proposition 5 & [0.1, 29.01] & [0.02, 32.40] & [0.27, 21.54] & [0.10, 26.62] \\
                            &        &        &        &        \\
\multicolumn{5}{c}{\underline{Treatment effect estimates}} \\
BLC buncher ATE & [ 373,  937] & [  57, 1310] & [1041, 2283] & [ 365, 3189] \\
BLC with M=1000 & [ 373, 1937] & [  57, 2310] & [1041, 3283] & [ 365, 4189] \\
BLC with M=4000 & [ 373, 4937] & [  57, 5310] & [1041, 6283] & [ 365, 7189] \\
BLC Proposition 5 & [ 373, 104220] & [  57, 116220] & [1041, 77453] & [ 365, 95396] \\
                            &        &        &        &        \\
Num observations & 328918 &  & 125544 &  \\
\hline\hline
\end{tabular}
}

	\captionof{table}{Elasticity and treatment effect estimates from the EITC application. \label{table:eitcestimates}}	
\end{center}

Table \ref{table:eitcestimates} reports bunching, elasticity, and treatment effect estimates separately for all filers and for filers that report self-employment income, under a series of alternate assumptions. Each estimate is followed by 95\% confidence intervals in gray on the following line. The ``Saez method'' rows report the bunching probability estimated by a simple excess mass calculation introduced in \citet{saez2010} (Fig 2.), as well as a point estimate of the elasticity $e$ from the isoelastic model using this excess mass and following a linear approximation to $f_0$ in the bunching region.\footnote{Note that the Saez method estimates do not replicate the values reported in \cite{saez2010} exactly. To facilitate comparability across assumptions I use my density estimates throughout, rather than replicating his density method.} The third row of estimates reports the mass $\mathcal{B} = F(y_U)-F(y_L)$ captured by the modified buncher ATE parameter with optimization error: $\mathbbm{E}[Y_0 - Y_1|Y_0 \ge y_L, Y_1 \le y_U]$ (Appendix \ref{sec:optimizationerror}). As Proposition \ref{prop:Bpartialid} shows, the ``intended'' bunching probability is only partially identified with optimization errors. These bounds are estimated in the second row and the buncher ATE mass $\mathcal{B} = F(y_U)-F(y_L)$ is equal to the upper bound (see Proposition \ref{prop:posTEs}). Figure \ref{eitcquantiles} displays an empirical version of Figure \ref{figquantiles}, showing this extended bunching interval.

The remaining rows report bounds on the buncher ATE (in levels) under an assumption that $Y_0$ and $Y_1$ are locally BLC, and bounds on the buncher average elasticity are reported under an assumption that $\ln Y_0$ and $\ln Y_1$ are locally BLC (following Eq. \eqref{eq:blcelasticity} and Remark \ref{remark:logs}). The baseline estimates assume RANK to obtain the upper bound, but the following row reports relaxed estimates based upon Propositions \ref{prop:bATEUB1} and \ref{prop:bATEUB2}. I find that the upper bound delivered by Proposition \ref{prop:bATEUB2} is essentially uninformative in this setting. To construct an upper bound based on Proposition \ref{prop:bATEUB1}, I allow for treatment effects to vary by as much as $\$1,000$ among taxpayers with a fixed value of either $Y_0$ or $Y_1$, and then show sensitivity to a much more conservative bound of $M=\$4,000$. For elasticity estimates the same values are divided by $k$ to make them interpretable as proportional responses. 

\singlespacing

\printbibliography

@article{GelmanImbens,
author = {Andrew Gelman and Guido Imbens},
title = {Why High-Order Polynomials Should Not Be Used in Regression Discontinuity Designs},
journal = {Journal of Business \& Economic Statistics},
volume = {37},
number = {3},
pages = {447--456},
year = {2019},
publisher = {Taylor \& Francis},
doi = {10.1080/07350015.2017.1366909}
}

@article{AS,
	author = {Andrews, Donald W. K. and Soares, Gustavo},
	title = {Inference for Parameters Defined by Moment Inequalities Using Generalized Moment Selection},
	journal = {Econometrica},
	volume = {78},
	number = {1},
	pages = {119-157},
	doi = {https://doi.org/10.3982/ECTA7502},
	url = {https://onlinelibrary.wiley.com/doi/abs/10.3982/ECTA7502},
	eprint = {https://onlinelibrary.wiley.com/doi/pdf/10.3982/ECTA7502},
	year = {2010}
}

@article{KKKL18,
author = {Kim, Wooyoung and Kwon, Koohyun and Kwon, Soonwoo and Lee, Sokbae},
title = {The identification power of smoothness assumptions in models with counterfactual outcomes},
journal = {Quantitative Economics},
volume = {9},
number = {2},
pages = {617-642},
doi = {https://doi.org/10.3982/QE545},
url = {https://onlinelibrary.wiley.com/doi/abs/10.3982/QE545},
eprint = {https://onlinelibrary.wiley.com/doi/pdf/10.3982/QE545},
year = {2018}
}

@article{MilgromWeber,
author = {Paul R. Milgrom AND Robert J. Weber},
title = {A Theory of Auctions and Competitive Bidding},
journal = {Econometrica},
volume = {50},
number = {5},
pages = {1089-1122},
url = {https://onlinelibrary.wiley.com/doi/abs/0012-9682(198209)50:5&lt;1089:ATOAAC&gt;2.0.CO;2-E},
year = {1982}
}

@ARTICLE{durot2008,
  author={Durot, CÉcile},
  journal={IEEE Transactions on Reliability}, 
  title={Testing Convexity or Concavity of a Cumulated Hazard Rate}, 
  year={2008},
  volume={57},
  number={3},
  pages={465-473},
  doi={10.1109/TR.2008.928181}}

@article{boltzmanngibbs,
    author = {Ludwig, Danial and Yakovenko, Victor M.},
    title = {Physics-inspired analysis of the two-class income distribution in the USA in 1983–2018},
    journal = {Philosophical Transactions of the Royal Society A: Mathematical, Physical and Engineering Sciences},
    volume = {380},
    number = {2224},
    pages = {20210162},
    year = {2022},
    month = {04}
}

@misc{song2025identificationinferencegeneralbunching,
      title={Identification and Inference in General Bunching Designs}, 
      author={Myunghyun Song},
      year={2025},
      eprint={2411.03625},
      archivePrefix={arXiv},
      primaryClass={econ.EM},
      url={https://arxiv.org/abs/2411.03625}, 
}

@article{lobel2024,
  author = {Lobel, Felipe and Scot, Thiago and Z{\'u}niga, Pedro},
  title = {Corporate Taxation and Evasion Responses: Evidence from a Minimum Tax in Honduras},
  journal = {American Economic Journal: Economic Policy},
  year = {2024},
  volume = {16},
  number = {1},
  pages = {482--517},
  doi = {10.1257/pol.20210587}
}

@article{bukovina2025,
  author = {Bukovina, Jaroslav and Lichard, Tom{\'a}{\v{s}} and Palguta, J{\'a}n and {\v{Z}}{\'u}del, Branislav},
  title = {Corporate Minimum Tax and the Elasticity of Taxable Income: Evidence from Administrative Tax Records},
  journal = {American Economic Journal: Economic Policy},
  year = {2025},
  volume = {17},
  number = {2},
  pages = {358--387},
  doi = {10.1257/pol.20230108}
}

@article{londonovelez2025,
  author = {Londo{\~n}o-V{\'e}lez, Juliana and {\'A}vila-Mahecha, Javier},
  title = {Behavioural Responses to Wealth Taxation: Evidence from Colombia},
  journal = {Review of Economic Studies},
  year = {2025},
  volume = {92},
  number = {4},
  pages = {2624--2655},
  doi = {10.1093/restud/rdae076}
}

@article{kostol2021,
  author = {Kost{\o}l, Andreas R. and Myhre, Andreas S.},
  title = {Labor Supply Responses to Learning the Tax and Benefit Schedule},
  journal = {American Economic Review},
  year = {2021},
  volume = {111},
  number = {11},
  pages = {3733--3766},
  doi = {10.1257/aer.20201877}
}

@article{seibold2021,
  author = {Seibold, Arthur},
  title = {Reference Points for Retirement Behavior: Evidence from German Pension Discontinuities},
  journal = {American Economic Review},
  year = {2021},
  volume = {111},
  number = {4},
  pages = {1126--1165},
  doi = {10.1257/aer.20191136}
}

@article{lebarbanchon2025,
  author = {Le Barbanchon, Thomas},
  title = {Taxes Today, Benefits Tomorrow},
  journal = {American Economic Journal: Applied Economics},
  year = {2025},
  volume = {17},
  number = {4},
  pages = {155--183},
  doi = {10.1257/app.20210550}
}

@article{shaffer2020,
  author = {Shaffer, Blake},
  title = {Misunderstanding Nonlinear Prices: Evidence from a Natural Experiment on Residential Electricity Demand},
  journal = {American Economic Journal: Economic Policy},
  year = {2020},
  volume = {12},
  number = {3},
  pages = {433--461},
  doi = {10.1257/pol.20180061}
}

@article{mortenson2020,
  author = {Mortenson, Jacob A. and Whitten, Andrew},
  title = {Bunching to Maximize Tax Credits: Evidence from Kinks in the US Tax Schedule},
  journal = {American Economic Journal: Economic Policy},
  year = {2020},
  volume = {12},
  number = {3},
  pages = {402--432},
  doi = {10.1257/pol.20180054}
}

@article{ring2025,
  author = {Ring, Marius A. K. and Thoresen, Thor O.},
  title = {Wealth Taxation and Charitable Giving},
  journal = {Review of Economics and Statistics},
  year = {2025},
  pages = {1--45},
  doi = {10.1162/rest_a_01562},
  note = {Advance article (online); volume, issue, and final pages not yet assigned}
}

@article{bachas2021,
  author = {Bachas, Pierre and Soto, Mauricio},
  title = {Corporate Taxation under Weak Enforcement},
  journal = {American Economic Journal: Economic Policy},
  year = {2021},
  volume = {13},
  number = {4},
  pages = {36--71},
  doi = {10.1257/pol.20180564}
}

@article{best2020,
  author = {Best, Michael Carlos and Cloyne, James S and Ilzetzki, Ethan and Kleven, Henrik J},
  title = {Estimating the Elasticity of Intertemporal Substitution Using Mortgage Notches},
  journal = {The Review of Economic Studies},
  year = {2020},
  volume = {87},
  pages = {656--690},
  doi = {10.1093/restud/rdz025}
}

@article{ruh2019,
  author = {Ruh, Philippe and Staubli, Stefan},
  title = {Financial Incentives and Earnings of Disability Insurance Recipients: Evidence from a Notch Design},
  journal = {American Economic Journal: Economic Policy},
  year = {2019},
  volume = {11},
  number = {2},
  pages = {269--300},
  doi = {10.1257/pol.20160076}
}

@article{almunia2018,
  author = {Almunia, Miguel and Lopez-Rodriguez, David},
  title = {Under the Radar: The Effects of Monitoring Firms on Tax Compliance},
  journal = {American Economic Journal: Economic Policy},
  year = {2018},
  volume = {10},
  number = {1},
  pages = {1--38},
  doi = {10.1257/pol.20160229}
}

@article{ito2018,
  author = {Ito, Koichiro and Sallee, James M.},
  title = {The Economics of Attribute-Based Regulation: Theory and Evidence from Fuel Economy Standards},
  journal = {The Review of Economics and Statistics},
  year = {2018},
  volume = {100},
  number = {2},
  pages = {319--336},
  doi = {10.1162/rest_a_00704}
}

@article{best2018,
  author = {Best, Michael Carlos and Kleven, Henrik Jacobsen},
  title = {Housing Market Responses to Transaction Taxes: Evidence From Notches and Stimulus in the U.K.},
  journal = {The Review of Economic Studies},
  year = {2018},
  volume = {85},
  number = {1},
  pages = {157--193},
  doi = {10.1093/restud/rdx032}
}

@article{seim2017,
  author = {Seim, David},
  title = {Behavioral Responses to Wealth Taxes: Evidence from Sweden},
  journal = {American Economic Journal: Economic Policy},
  year = {2017},
  volume = {9},
  number = {4},
  pages = {395--421},
  doi = {10.1257/pol.20150290}
}

@article{defusco2017,
  author = {DeFusco, Anthony A. and Paciorek, Andrew},
  title = {The Interest Rate Elasticity of Mortgage Demand: Evidence from Bunching at the Conforming Loan Limit},
  journal = {American Economic Journal: Economic Policy},
  year = {2017},
  volume = {9},
  number = {1},
  pages = {210--240},
  doi = {10.1257/pol.20140108}
}

@article{manoli2016,
  author = {Manoli, Day and Weber, Andrea},
  title = {Nonparametric Evidence on the Effects of Financial Incentives on Retirement Decisions},
  journal = {American Economic Journal: Economic Policy},
  year = {2016},
  volume = {8},
  number = {4},
  pages = {160--182},
  doi = {10.1257/pol.20140209}
}

@article{kopczuk2015,
  author = {Kopczuk, Wojciech and Munroe, David},
  title = {Mansion Tax: The Effect of Transfer Taxes on the Residential Real Estate Market},
  journal = {American Economic Journal: Economic Policy},
  year = {2015},
  volume = {7},
  number = {2},
  pages = {214--257},
  doi = {10.1257/pol.20130361}
}

@article{einav2015,
  author = {Einav, Liran and Finkelstein, Amy and Schrimpf, Paul},
  title = {The Response of Drug Expenditure to Nonlinear Contract Design: Evidence from Medicare Part D},
  journal = {The Quarterly Journal of Economics},
  year = {2015},
  volume = {130},
  number = {2},
  pages = {841--899},
  doi = {10.1093/qje/qjv005}
}

@article{best2015,
  author = {Best, Michael Carlos and Brockmeyer, Anne and Kleven, Henrik Jacobsen and Spinnewijn, Johannes and Waseem, Mazhar},
  title = {Production versus Revenue Efficiency with Limited Tax Capacity: Theory and Evidence from Pakistan},
  journal = {Journal of Political Economy},
  year = {2015},
  volume = {123},
  number = {6},
  pages = {1311--1355},
  doi = {10.1086/683849}
}

@article{kleven2014,
  author = {Kleven, Henrik Jacobsen and Landais, Camille and Saez, Emmanuel and Schultz, Esben},
  title = {Migration and Wage Effects of Taxing Top Earners: Evidence from the Foreigners' Tax Scheme in Denmark},
  journal = {The Quarterly Journal of Economics},
  year = {2014},
  volume = {129},
  number = {1},
  pages = {333--378},
  doi = {10.1093/qje/qjt033}
}

@article{devereux2014,
  author = {Devereux, Michael P. and Liu, Li and Loretz, Simon},
  title = {The Elasticity of Corporate Taxable Income: New Evidence from UK Tax Records},
  journal = {American Economic Journal: Economic Policy},
  year = {2014},
  volume = {6},
  number = {2},
  pages = {19--53},
  doi = {10.1257/pol.6.2.19}
}

@article{kleven2013,
  author = {Kleven, Henrik J. and Waseem, Mazhar},
  title = {Using Notches to Uncover Optimization Frictions and Structural Elasticities: Theory and Evidence from Pakistan},
  journal = {The Quarterly Journal of Economics},
  year = {2013},
  volume = {128},
  number = {2},
  pages = {669--723},
  doi = {10.1093/qje/qjt004}
}

@article{kleven2011,
  author = {Kleven, Henrik Jacobsen and Knudsen, Martin B. and Kreiner, Claus Thustrup and Pedersen, S{\o}ren and Saez, Emmanuel},
  title = {Unwilling or Unable to Cheat? Evidence From a Tax Audit Experiment in Denmark},
  journal = {Econometrica},
  year = {2011},
  volume = {79},
  number = {3},
  pages = {651--692},
  doi = {10.3982/ecta9113}
}

@article{chetty2011,
  author = {Chetty, R. and Friedman, J. N. and Olsen, T. and Pistaferri, L.},
  title = {Adjustment Costs, Firm Responses, and Micro vs. Macro Labor Supply Elasticities: Evidence from Danish Tax Records},
  journal = {The Quarterly Journal of Economics},
  year = {2011},
  volume = {126},
  number = {2},
  pages = {749--804},
  doi = {10.1093/qje/qjr013}
}

@article{saez2010,
  author = {Saez, Emmanuel},
  title = {Do Taxpayers Bunch at Kink Points?},
  journal = {American Economic Journal: Economic Policy},
  year = {2010},
  volume = {2},
  number = {3},
  pages = {180--212},
  doi = {10.1257/pol.2.3.180}
}

@misc{goff2024treatmenteffectsbunchingdesigns,
      title={Treatment Effects in Bunching Designs: The Impact of Mandatory Overtime Pay on Hours}, 
      author={Leonard Goff},
      year={2024},
      eprint={2205.10310},
      archivePrefix={arXiv},
      primaryClass={econ.EM},
      url={https://arxiv.org/abs/2205.10310}, 
}

@book{csorgo2014strong,
  title={Strong Approximations in Probability and Statistics},
  author={Cs{\"o}rgo, M. and R{\'e}v{\'e}sz},
  isbn={9781483268040},
  series={Probability and mathematical statistics},
  url={https://books.google.ca/books?id=TL3iBQAAQBAJ},
  year={1981},
  publisher={Academic Press}
}

@article{GMarXiv,
      title={Testing the Solvability of Systems of Linear Inequalities}, 
      author={Leonard Goff and Eric Mbakop},
      journal={arXiv preprint},
      volume={2506.06776},
      year={2025},
      eprint={2506.06776},
      archivePrefix={arXiv},
      primaryClass={econ.EM},
      url={https://arxiv.org/abs/2506.06776}, 
}

@article{lahamiaowellner,
title = {Bi-s-concave distributions},
journal = {Journal of Statistical Planning and Inference},
volume = {215},
pages = {127-157},
year = {2021},
issn = {0378-3758},
doi = {https://doi.org/10.1016/j.jspi.2021.03.001},
url = {https://www.sciencedirect.com/science/article/pii/S0378375821000331},
author = {Nilanjana Laha and Zhen Miao and Jon A. Wellner},
}

@misc{goff2024testingidentifyingassumptionsparametric,
      title={Testing Identifying Assumptions in Parametric Separable Models: A Conditional Moment Inequality Approach}, 
      author={Leonard Goff and Désiré Kédagni and Huan Wu},
      year={2024},
      eprint={2410.12098},
      archivePrefix={arXiv},
      primaryClass={econ.EM},
      url={https://arxiv.org/abs/2410.12098}, 
}

@article{bertanha_better_2018,
title = {Better bunching, nicer notching},
journal = {Journal of Econometrics},
volume = {237},
number = {2, Part A},
pages = {105512},
year = {2023},
issn = {0304-4076},
doi = {https://doi.org/10.1016/j.jeconom.2023.105512},
url = {https://www.sciencedirect.com/science/article/pii/S0304407623002282},
author = {Marinho Bertanha and Andrew H. McCallum and Nathan Seegert}
}

@article{cattaneo_simple_2017,
	title = {Simple Local Polynomial Density Estimators},
	volume = {115},
	pages = {1449--1455},
	number = {531},
	journaltitle = {Journal of the American Statistical Association},
	author = {Cattaneo, Matias D and Jansson, Michael and Ma, Xinwei},
	date = {2020}
}

@article{kleven_bunching_2016,
	title = {Bunching},
	volume = {8},
	pages = {435--464},
	journaltitle = {Annual Review of Economics},
	author = {Kleven, Henrik Jacobsen},
	date = {2016}
}

@techreport{bklnnber,
 title = "On Bunching and Identification of the Taxable Income Elasticity",
 author = "Blomquist, Sören and Newey, Whitney and Kumar, Anil and Liang, Che-Yuan",
 institution = "National Bureau of Economic Research",
 type = "Working Paper",
 series = "Working Paper Series",
 number = "24136",
 year = "2017",
 month = "December",
 doi = {10.3386/w24136},
 URL = "http://www.nber.org/papers/w24136",
}

@article{saumard_bi-log-concavity_2019-1,
	title = {Bi-log-concavity: some properties and some remarks towards a multi-dimensional extension},
	shorttitle = {Bi-log-concavity},
	journaltitle = {{arXiv}:1903.07347},
	author = {Saumard, Adrien},
	urldate = {2020-08-16},
	date = {2019-03-18},
	langid = {english}
}

@article{dumbgen_bi-log-concave_2017,
	title = {Bi-log-concave distribution functions},
	volume = {184},
	issn = {03783758},
	url = {https://linkinghub.elsevier.com/retrieve/pii/S0378375816301379},
	doi = {10.1016/j.jspi.2016.10.005},
	pages = {1--17},
	journaltitle = {Journal of Statistical Planning and Inference},
	shortjournal = {Journal of Statistical Planning and Inference},
	author = {Dümbgen, Lutz and Kolesnyk, Petro and Wilke, Ralf A.},
	urldate = {2020-08-16},
	date = {2017-05},
	langid = {english}
}

@article{blomquist_bunching_2019,
	title = {On Bunching and Identification of the Taxable Income Elasticity},
	volume = {129},
	number = {8},
	journaltitle = {Journal of Political Economy},
	author = {Blomquist, Soren and Kumar, Anil and Liang, Che-Yuan and Newey, Whitney},
	date = {2021}
}

@article{blomquist_individual_2015,
	title = {Individual heterogeneity, nonlinear budget sets and taxable income},
	volume = {{CWP}21/15},
	journaltitle = {The Institute for Fiscal Studies Working Paper},
	author = {Blomquist, Soren and Kumar, Anil and Liang, Che-Yuan and Newey, Whitney},
	date = {2015}
}

@article{imbens_confidence_2004,
	title = {Confidence Intervals for Partially Identified Parameters},
	volume = {72},
	pages = {14},
	journaltitle = {Econometrica},
	author = {Imbens, Guido W and Manski, Charles F},
	date = {2004},
	langid = {english}
}

@article{stoye_more_2009,
	title = {More on Confidence Intervals for Partially Identified Parameters},
	volume = {77},
	issn = {0012-9682},
	url = {http://doi.wiley.com/10.3982/ECTA7347},
	doi = {10.3982/ECTA7347},
	pages = {1299--1315},
	number = {4},
	journaltitle = {Econometrica},
	shortjournal = {Econometrica},
	author = {Stoye, Jörg},
	urldate = {2020-08-28},
	date = {2009},
	langid = {english}
}

@article{chernozhukov_iv_2005,
	title = {An {IV} Model of Quantile Treatment Effects},
	volume = {73},
	pages = {245--261},
	number = {1},
	journaltitle = {Econometrica},
	author = {Chernozhukov, Victor and Hansen, Christian},
	date = {2005},
	langid = {english}
}

@article{pollinger_kinks_2020,
  author={Stefan Pollinger},
  title={{Kinks Know More: Policy Evaluation Beyond Bunching with an Application to Solar Subsidies}},
  year=2023,
  month=Nov,
  institution={HAL},
  type={Working Papers},
  url={https://ideas.repec.org/p/hal/wpaper/hal-04182085.html},
  number={hal-04182085},
  doi={},
}

@article{klinetartari_bounding_2016,
Author = {Kline, Patrick and Tartari, Melissa},
Title = {Bounding the Labor Supply Responses to a Randomized Welfare Experiment: A Revealed Preference Approach},
Journal = {American Economic Review},
Volume = {106},
Number = {4},
Year = {2016},
Month = {April},
Pages = {972–1014},
DOI = {10.1257/aer.20130824},
URL = {https://www.aeaweb.org/articles?id=10.1257/aer.20130824}}

\onehalfspacing

\appendix
\renewcommand\thefigure{\thesection.\arabic{figure}}
\renewcommand\thetable{\thesection.\arabic{table}}
\renewcommand{\theequation}{\thesection.\arabic{equation}}
\setcounter{figure}{0}
\setcounter{table}{0}
\setcounter{equation}{0}

\section{Accommodating optimization error and frictions} \label{sec:optimizationerror}

When decision-makers can perfectly manipulate $Y_i$, the results of Section \ref{sec:idsetup} show that their choice subject to a kink or notch can be characterized by their counterfactual choices $Y_{0i}$ and $Y_{1i}$. A generalization of this relationship between choices and outcomes allows for an ``optimization error'' $\epsilon_i$ in the choice of each individual $i$.

Let $Y_i = Y_i^* + \epsilon_i$, where $Y_i^*$ reflects the ``intended'' choice of individual $i$. ``Intended'' choices can refer literally to choices the agent intended to make but were subject to optimization error, or choices that they would have made in the absence of frictions that prevent adjustment to the cost schedule (see Remark below), or simply their actual choices absent measurement error in what is observed by the econometrician. Throughout this section, I make assumption CONVEX (and NOTCH in the case of a notch) regarding $Y_i^*$, rather than for actual outcomes $Y_i$. Intended choices $Y_i^*$ are then related to counterfactual intended choices $Y^*_{0i}$ and $Y^*_{1i}$ by Eq. \eqref{hcasesranknotch}.

Let $Y_{0i} = Y_{0i}^* + \epsilon_i$ and $Y_{1i} = Y_{1i}^* + \epsilon_i$. $Y_{0i}$ and $Y_{1i}$ constitute counterfactual outcomes if the cost function is changed, but $\epsilon_i$ is held fixed.\footnote{\label{fn:teerror} If optimization error were itself affected by a switch from $T_{0i}$ to $T_{1i}$, then $Y_{0i}-Y_{1i}=Y_{0i}^*-Y^*_{1i}$ captures the ``direct effect'' of the cost function change on intended choices, and not any indirect effects via optimization error.} Then by Eq. \eqref{hcasesranknotch}, we have:
\begin{equation} \label{hcasesgeneralwitherror}
	Y_i=
	\begin{cases}
		Y_{0i} & \hspace{.2cm} \textrm{if } \hspace{.2cm} Y_{0i}^*<k\\
		k+\epsilon_i & \hspace{.2cm} \textrm{if } \hspace{.2cm} Y_{0i}^* \ge k \textrm{ and } Y_{1i}^* \le y^I\\
		Y_{1i} & \hspace{.2cm} \textrm{if } \hspace{.2cm} Y_{1i}^*> y^I
	\end{cases} = \begin{cases}
		Y_{0i} & \hspace{.2cm} \textrm{if } \hspace{.2cm} Y_{0i}<k+\epsilon_i\\
		k+\epsilon_i & \hspace{.2cm} \textrm{if } \hspace{.2cm} Y_{0i} \ge k + \epsilon_i \textrm{ and } Y_{1i} \le y^I + \epsilon_i\\
		Y_{1i} & \hspace{.2cm} \textrm{if } \hspace{.2cm} Y_{1i}> y^I+\epsilon_i
	\end{cases}
\end{equation} 
By Eq. \eqref{hcasesgeneralwitherror}, the econometrician observes from $Y_i$ either $Y_{0i}$ or $Y_{1i}$, or might reflect $k+\epsilon_i$ in cases where the decision-maker attempted to choose an $\mathbf{x}$ that targeted $k$ (i.e. $Y_i^*=k$). Note that \eqref{hcasesgeneralwitherror} nests Eq. \eqref{hcasesranknotch} when $\epsilon_i=0$ and Eq. \ref{hcases} for kinks when additionally $N=0$ so that $y^I = k$.

\begin{remark*}
	While \citet{saez2010} discusses the possibility that individuals may not be able to perfectly adjust their income, the literature since \citet{kleven2013} has also emphasized the possibility of ``frictions''. Frictions occur when some individuals are non-optimizers whose choices are unaffected by the presence of the notch, perhaps due to inattention or an inability to change their earnings. What I call ``optimization error'' nests this when $P(\epsilon_i = Y_i(0)-Y_i)>0$. For any individual with $\epsilon_i = Y_i(0)-Y_i$, we have that $Y_i = Y_i(0)$ regardless of their value of $Y_i$.
\end{remark*}

\noindent Eq. \eqref{hcasesgeneralwitherror} reveals the additional challenge for identification in the presence of optimization errors. Let $\mathcal{B}^* := P(Y_i^* = k) = P(Y^*_{0i} \ge k, Y_{1i}^* \le y^I) = F_1^*(y^I)-F_0^*(k)$. Even if $\mathcal{B}^*>0$, the researcher may observe no point mass at $Y_i=k$ in the data, if $\epsilon_i$ is continuously distributed.

To make progress, I assume that $\epsilon_i$ has limited support:
\begin{assumption*}[ERRORBOUND (bounded optimization error)]
	$P(-\epsilon_l < \epsilon_i < \epsilon_u)=1$ for some $\epsilon_l, \epsilon_u \ge 0$ with $E:=\max\{\epsilon_l, \epsilon_u\} < \infty$.
\end{assumption*}
\noindent Adding Assumption ERRORBOUND, we have that 
\begin{equation} \label{hcasesgeneralwithboundederror}
	Y_i=
	\begin{cases}
		Y_{0i} & \hspace{.2cm} \textrm{if } \hspace{.2cm} Y_{0i}<k-\epsilon_l\\
		Y_{0i} & \hspace{.2cm} \textrm{if } \hspace{.2cm} Y_{0i} \in [k-\epsilon_l, k + \epsilon_i)\\
		k+\epsilon_i & \hspace{.2cm} \textrm{if } \hspace{.2cm} Y_{0i} \ge k+\epsilon_i \textrm{ and } Y_{1i} \le y^I+\epsilon_i\\
		Y_{1i} & \hspace{.2cm} \textrm{if } \hspace{.2cm} Y_{1i} \in (y^I + \epsilon_i,y^I+\epsilon_u]\\
		Y_{1i} & \hspace{.2cm} \textrm{if } \hspace{.2cm} Y_{1i}> y^I+\epsilon_u
	\end{cases}
\end{equation}

\begin{remark*}
	Assumption ERRORBOUND is implicit in the polynomial approach of \cite{chetty2011} that excludes a region around the kink/notch. However, it differs from existing approaches that assume that $\epsilon_i$ only affects choices within a window around the kink/notch \citep{blomquist_bunching_2019,song2025identificationinferencegeneralbunching}. In contrast to that approach, I do not assume that $Y_{1}=Y_{1}^*$ when $Y_1^* \ge k + \epsilon_u$ and that $Y_{0}=Y_{0}^*$ when $Y_1^* \le k - \epsilon_l$. I instead use that far enough from $k$, it is by \eqref{hcasesgeneralwithboundederror} knowable that observations of $Y$ reflect $Y_1$ or $Y_0$ (and not $k+\epsilon_i$), in this sense uncontaminated by the kink/notch.
\end{remark*}

As Eq. \ref{hcasesgeneralwithboundederror} makes clear, Assumption ERRORBOUND allows the distributions of $Y_1$ and $Y_0$ to be identified outside of the region $[k-\epsilon_l,y^I+\epsilon_u]$. Inside this region however, are a mix of intended bunchers (having $Y_i = k+ \epsilon_i$) and non-bunchers (having $Y_i=Y_{0i}$ or $Y_i = Y_{1i}$). The two cases cannot be distinguished by the econometrician, given that $\epsilon_i$ is not observed. The Online Supplement depicts a simulated example for a notch. The ``intended'' bunching probability $\mathcal{B}^* = P(Y_i^*=k) = F_1^*(y^I)-F^*_0(k)$ becomes only partially identified:
\begin{proposition} \label{prop:Bpartialid}
	If ERRORBOUND, CONVEX and NOTCH hold: 
	$$\mathcal{B}^* \in \left[\left[F_1(y^I-\epsilon_u)-F_0(k+\epsilon_l)\right]_{+},F_1(y^I+\epsilon_l)-F_0(k-\epsilon_u)\right]$$
	where $[x]_+$ denotes the positive part of $x$.
\end{proposition}

\begin{remark*}
Note that the quantities entering the lower bound in Proposition \ref{prop:Bpartialid} are themselves generally only partially identified. Under BLC of $Y_0$ and $Y_1$, we obtain the bound 
$$\left[F_1(y^I-\epsilon_u)-F_0(k+\epsilon_l)\right]_{+} \ge \left[1-(1-F_{1}(y^I+\epsilon_u))e^{\frac{f_{1}(y^I+\epsilon_u)}{1-F_{1}(y^I+\epsilon_u)}2\epsilon_u}-F_{0}(k-\epsilon_l) e^{\frac{f_{0}(k-\epsilon_l)}{F_{0}(k-\epsilon_l)}2\epsilon_l}\right]_+$$
For the upper bound, if $\epsilon_u > \epsilon_l$, then $F_1(y^I+\epsilon_l)$ is not point identified and if $\epsilon_u < \epsilon_l$ then $F_0(k-\epsilon_u)$ is not point identified. Rather than again extrapolating by BLC, this can be accommodated in a simple way (at the expense of slightly wider bounds) by observing that if ERRORBOUND is satisfied with $\epsilon_l, \epsilon_u$, then it is also satisfied with $\epsilon_l=\epsilon_u=E$.
\end{remark*}

Crucially under Eq. \eqref{hcasesgeneralwithboundederror}, $F_0(y)$ is still identified for $y<k-\epsilon_l$ and $F_1(y)$ is identified for $y> y^I+\epsilon_u$, because there is no censoring effect from the kink or notch there. The values of $k-\epsilon_l$ and $y^I+\epsilon_u$ can in principle be identified by observing where $F(y)$ ``breaks trend'' for $y<k$, often where excess mass begins to pile up above the distribution $F(y)$ for very small $y$, and similarly on the right side of the kink/notch.

I therefore assume that the researcher obtains by inspection values $y_L \le k-\epsilon_l$ and $y_U \ge y^I+\epsilon_u$. Any slack in these inequalities will widen bounds, but they remain valid. Note that $F_0(y)$ is identified for all $y \le y_L$ and $F_1(y)$ for all $y \ge y_U$. In the case of a kink, an alternative to determining $y_L$ and $y_U$ by visual inspection is to compute estimates for $y_L=k-E$ and $y_U = k+E$ to explore sensitivity to various values of $E>0$, since there is no additional unknown parameter $y^I$. I do not recommend obtaining $y_U$ by the ``convergence method'' of \citet{kleven2013} as this makes the assumption that $F(y)=F_0(y)$ for $y$ sufficiently far above the kink/notch. This is not generally a model-consistent assumption (see Online Supplement).

Given that we can  identify portions of the distributions of $Y_1$ and $Y_0$, but cannot for $Y_1^*$ and $Y_0^*$ (absent knowledge of the distribution of $\epsilon$), I define the parameter of interest in the presence of optimization error to be $\Delta_k^* = \mathbbm{E}[Y_{0i}-Y_{1i}|Y_{0i} \ge y_L, Y_{1i} \le y_U]$ and accordingly define the associated bunching probability as $\mathcal{B} = P(Y_{0i} \ge y_L, Y_{1i} \le y_U)$. 

Note that the event $Y_{0i} \ge y_L, Y_{1i} \le y_U$ no longer corresponds to the event $Y_i = k$ given the presence of $\epsilon_i$. This group $Y_{0i} \ge y_L, Y_{1i} \le y_U$ nevertheless contains all of the ``intended bunchers'' $Y_i^*=k$ given ERRORBOUND (these individuals have $Y_{0i}^* \ge k, Y^*_{1i} \le y^I$ and hence $Y_{0i} \ge k-E \ge y_L, Y_{1i} \le y^I+E \le y_U$). The quantity $Y_{0i}-Y_{1i}$ that $\Delta_k^*$ averages over captures direct effects of a switch between cost functions $T_0$ and $T_1$ on intended choices (see footnote \ref{fn:teerror}). As $\mathbbm{E}[Y_{0i}-Y_{1i}|Y_{0i} \ge y_L, Y_{1i} \le y_U]$ nests our previous definition of the buncher ATE when $\epsilon_i = 0$ with probability one and $y_L=k,y_U=y^I$ (provided CONVEX and NOTCH hold), I use the same notation $\Delta_k^*$ and $\mathcal{B}$ in this section.

By contrast with the intended bunching probability $\mathcal{B}^*$, $\mathcal{B}$ is point identified:
\begin{proposition} \label{prop:posTEs}
	If ERRORBOUND, CONVEX, and NOTCH hold, then $\mathcal{B} = F_1(y_U)-F_0(y_L) = F(y_U)-F(y_L)$ for any $y_L \le k-E$ and $y_U\ge y^I+E$. 
\end{proposition}
\noindent Note that when $y_L=k-E$ and $y_U=y^I+E$ exactly, $\mathcal{B}$ is simply the upper bound on $\mathcal{B}^*$ from Proposition \ref{prop:Bpartialid}.

I now turn to partially identifying $\Delta_k^* = \mathbbm{E}[Y_{0i}-Y_{1i}|Y_{0i} \ge y_L, Y_{1i} \le y_U]$. Define:
\begin{align}
	Q^*_k&:= \frac{1}{\mathcal{B}}\int_{F(y_L)}^{F(y_U)} \{Q_0(u)-Q_1(u)\}\cdot du \nonumber \\
	&=\left\{\frac{1}{\mathcal{B}}\int_{F_0(y_L)}^{F_0(y_L)+\mathcal{B}} Q_0(u) \cdot du-y_L\right\} + \left\{y_U -\frac{1}{\mathcal{B}}\int^{F_1(y_U)}_{F_1(y_U)-\mathcal{B}} Q_1(u) \cdot du\right\} - (y_U-y_L)  \label{eq:qkgen}
\end{align}
where the second equality follows using Proposition \ref{prop:posTEs}. Define $\Delta_0^*:= Q_0(F_1(y_U))-y_L = Q_0(F_0(y_L)+\mathcal{B})-y_L$ and $\Delta_1^*:= y_U-Q_1(F_0(y_L))=y_U-Q_1(F_1(y_U)-\mathcal{B})$, which nests Eq. \eqref{eq:deltasdef} when $y_L=k, y_U=y^I$ (in which case $\mathcal{B}$ recovers the bunching probability $P(Y=k)$).
\begin{theorem}[(bounds on the buncher QTE with bounded error)] \label{thm:blcwitherrors}
	Assume CONVEX and NOTCHBOUND hold for $(Y^*_0,Y_1^*)$, and $BC(s,t)$ holds with the ranges $[k,k+\Delta_0^*]$ and $[y^I-\Delta_1^*, y^I]$ replaced by $[y_L,y_L+\Delta_0^*]$ and $[y_U-\Delta_0^*,y_U]$, respectively, then: $Q_k^* \in \left[\Delta^{L}_k,\Delta^{U}_k\right]$, where
	$$ \Delta^{L}_k:= g_0(F(y_L^-), f(y_L^{-}), F(y_U)-F(y_L)) + g_0\left(1-F(y_U),f(y_U^{+}),F(y_U)-F(y_L)\right)-(y_U-y_L)$$
	$$ \Delta^{U}_k:= -g_0(1-F(y_L^{-}),f(y_L^{-}), F(y_L)-F(y_U) ) -g_0\left(F(y_U),f(y_U^{+}), F(y_L)-F(y_U)\right)-(y_U-y_L)$$
\end{theorem}
\noindent The conclusion of Theorem \ref{thm:blcwitherrors} is in exact analogy to the case without optimization error (Theorem \ref{thmblc}), but with $y_L$ replacing $k$ and $y_U$ replacing $y^I$. Note that if $\mathcal{B}$ is small, then $\Delta^{L}_k \approx \Delta^{U}_k \approx \frac{\mathcal{B}}{2 f(y_L^-)}+\frac{\mathcal{B}}{2 f(y_L^+)} - (y_U-y_L)$, and the bounds exclude zero provided that the mass $\mathcal{B} = F(y_U)-F(y_L)$ in the region $[y_L,y_U]$ exceeds what the harmonic mean $(0.5/f(y_L^-)+0.5/f(y_U^+))^{-1}$ of the densities that flank that region would predict, i.e. if there is ``excess mass''. Given Theorem \ref{thm:blcwitherrors}, Remark \ref{remark:logs} for the buncher ATE in logs generalizes with $y^U$ replacing $y^I$ and $y_L$ replacing $k$.

The price of Theorem \ref{thm:blcwitherrors} in terms of assumptions is that the BC(s,t) now requires the concavity assumptions on $F_0$ and $1-F_0$ to hold on the extended interval $[y_L,y_L+\Delta_0^*]$, and $F_1$ and $1-F_1$ on the extended interval $[y_U-\Delta_1^*,y_U]$, which are wider than in the case without optimization error, because $y_L \le k$, $y_U \ge y^I$ and $\mathcal{B} \ge \mathcal{B}^*$.

The role of Assumptions CONVEX and NOTCHBOUND in Theorem \ref{thm:blcwitherrors} are only to yield the formulas given for $\Delta^{L}_k$ and $\Delta^{U}_k$ in terms of observables, rather than in terms of $F_0$ and $F_1$ (note that NOTCHBOUND and CONVEX are sufficient for $F(y)=F_1(y)$ for all $y > y^I$).

The proof of Proposition \ref{prop:buncherATEnotch} shows explicitly that the conclusion that $\Delta_k^* \ge Q_k^*$ with equality under RANK holds unchanged in the case of optimization error, with $y_U$ replacing $y^I$ and $y_L$ replacing $k$, and with Assumption RANK updated to reflect the extrapolation points $y_L$ and $y_L$ replacing $k$ and $y^I$: with probability one, $Y_{0i} \in [y_L, y_L+\Delta_0^*] \textrm{ iff } Y_{1i} \in [y_U-\Delta_1^*, y_U]$. A sufficient condition for RANK is rank invariance over the now larger range of ranks $[F_0(y_L),F_1(y_U)]$. Under this extension of Assumption RANK, we thus have that $\Delta_k^* \in \left[\Delta^{L}_k,\Delta^{U}_k\right]$ (given NOTCHBOUND, CONVEX, and for suitable values of $y_L \le k-E ,y_U \ge y^I+E$ given ERRORBOUND).

However, a caveat is in order regarding RANK in the presence of optimization error. Even if we begin with perfect rank invariance between $Y_{0}^*$ and $Y_{1}^*$ (as in the isoelastic model), continuously distributed error $\epsilon$ will typically break RANK if treatment effects are heterogeneous. The importance of Propositions \ref{prop:bATEUB1} and \ref{prop:bATEUB2}, which give upper bounds on $\Delta_k^*$ without RANK, is thus heightened with optimization error. In the proof of Proposition \ref{prop:bATEUB1}, I treat the case with optimization error and ERRORBOUND explicitly, and show that if $Y_0^*,Y_1^*$ satisfy SI, then $\mathbbm{E}[Y_0-Y_1|Y_0 \ge y_L, Y_1 \le y_U] \le Q_k^* + 2M_\Delta$ where $M_\Delta$ is the overall width of the support of treatment effects in the population.  This expansion arises because the distribution of treatment effects conditional on $Y_d=y$ mixes together the distribution of treatment effects for individuals with $Y_d^* \in [y-E,y+E]$. The bound $\Delta_k^* \le Q_k^* + 2M_\Delta$ obtained in the proof of Proposition \ref{prop:bATEUB1} is conservative in that it uses the global bound $M_\Delta$ regardless of whether $E$ is large or small. The researcher can explore sensitivity of their results to hypothesized values of $M_\Delta$, where $M_\Delta = 0$ in the case of homogenous treatment effects (e.g. the isoelastic model with the outcome taken to be $\ln Y$). 

The above result does not require Assumption SI to hold for $Y_1,Y_0$, but does for $Y_0^*,Y_1^*$. To take a concrete example without RANK, SI of $Y_0^*,Y_1^*$ holds e.g. in the isoelastic model even with heterogeneous elasticities, if $\eta_i \indep e_i$ and both $e_i$ and $\ln \eta_i$ have log-concave densities. In this case note that $\ln Y^*_{0i}$ and $\ln Y^*_{1i}$ are also each globally BLC. Suppose optimization error is additive with respect to $\ln Y$ and also has a log-concave density. Since a convolution between a BLC random variable and a log concave density of $\epsilon$ is again BLC, we furthermore have that $\ln Y_{0i}$ and $\ln Y_{1i}$ are also BLC. This yields a simple concrete example of a sufficient set of conditions that allows for both heterogeneity in responses and optimization error.

\clearpage

\setcounter{page}{1}
\vspace*{.3cm}
\begin{center}
	{\LARGE \textbf{SUPPLEMENTARY APPENDIX}}
\end{center}

\renewcommand{\thesection}{S.\arabic{section}}
\renewcommand{\thesubsection}{S.\arabic{section}.\arabic{subsection}}
\renewcommand{\theequation}{S.\arabic{equation}}
\setcounter{equation}{0}
\renewcommand{\theHequation}{S\arabic{equation}}
\setcounter{section}{0} % ensure starts at S.1
\renewcommand{\theHsection}{S\arabic{section}}

\phantomsection

\section{Details on bi-log-concavity and bi-s-concavity} \label{app:blc}

The notion of bi-log-concavity (BLC) was introduced by \citep{dumbgen_bi-log-concave_2017}:
\begin{definition*}[(BLC)] A distribution function $F$ is bi-log-concave (BLC) if both $\ln F$ and $\ln(1-F)$ are concave functions.
\end{definition*}
\noindent For a CDF function $F(y)$, the definition of BLC treats $F(y)$ and $1-F(y)$ as functions on the whole real line, and concavity is interpreted as concavity along the whole real line. We'll say that a random variable $Y$ is BLC if it's CDF is BLC.

\citet{lahamiaowellner} generalize BLC to define the notion of bi-s-concavity (BsC):
\begin{definition*}[(BsC)] For a given $s \le 1$, a distribution function $F$ is bi-s-concave (BsC) if both $F^s$ and $(1-F)^s$ are concave functions (in the case of positive $s$), or are convex functions (in the case of negative $s$). We call $F$ bi-s-concave with $s=0$ if $F$ is BLC.
\end{definition*}
\noindent If $F$ is BsC for a given value $s$, then it is also BsC for any smaller value of $s$ \citep{lahamiaowellner}. Thus $s=1$ represents the most restrictive case, and $s \downarrow -\infty$ the weakest. 

Theorem 3 of \citep{lahamiaowellner} gives several equivalent characterizations of BsC, generalizing analogous results proved for BLC in \citet{dumbgen_bi-log-concave_2017}. A particularly useful one is the following. For any CDF $F$, let $J(F) = \{y: 0 \le F(y) < 1\}$ be the interior of the support of $F$.
\begin{proposition} \label{prop:blcequivLipshitz}
	For any $s \le 1$, $Y$ is BsC if and only if $f=F'$ exists and is bounded and strictly positive on $J(F)$, and furthermore $f'=F''$ exists almost everywhere on $J(F)$ satisfying
	\begin{equation} \label{eq:bisLipschitz}
		-(1-s)\frac{f(y)^2}{1-F(y)} \le f'(y) \le (1-s) \frac{f(y)^2}{F(y)}
	\end{equation}
\end{proposition}
\noindent Proposition \ref{prop:blcequivLipshitz} shows that BsC is equivalent to $F$ admitting a density with a density whose derivative is bounded from above and below by functions that depend on $f(y)$ and $F(y)$. The ``almost everywhere'' statement on $J(F)$ in Proposition \ref{prop:blcequivLipshitz} is with respect to the Lebesgue measure.

It follows that $F$ is BsC for some finite $s \le 1$ if and only if $\bar{\gamma}(F):=\operatorname{ess sup}_{y \in J(F)} \max\{F(y),1-F(y)\} \cdot \frac{|f'(y)|}{f^2(y)}$ is finite, where the essential supremum is also with respect to the Lebesgue measure. In particular, it is clear from Eq. \ref{eq:bisLipschitz} that $F$ is BsC for $s=1-\bar{\gamma}(F)$ and any lower value of $s$. Corollary 4 of \citet{lahamiaowellner} shows in the other direction if F is BsC then $\bar{\gamma}(F) \le 1-s$. Finiteness of $\bar{\gamma}(F)$ is a mild regularity condition that also arises as a sufficient condition for strong approximation of quantile processes \citep{csorgo2014strong}.
I extend the notion of BsC introduced by \citet{lahamiaowellner} by defining a weaker notion of BsC, imposing concavity only on an open interval $I=(a,b) \subseteq \mathbbm{R}$ rather than all of $J(F)$: 
\begin{definition*}[(local BsC)] A distribution function $F$ is \textit{locally BsC} on an interval $I$, if the required concavity/convexity property from BsC holds on the interval $I$. 
\end{definition*}
\noindent For example, $F$ is BsC with $s=-1$ on the $I$ if $1/F(y)$ and $1/(1-F(y))$ are both convex functions on the interval $I$. $F$ is BLC on $I$ if $\ln F(y)$ and $\ln(1-F(y))$ are concave on $I$.

It is straightforward to verify that $F$ is BsC on $I$ if and only if Eq. \ref{eq:bisLipschitz} holds almost everywhere in $I$, which amounts to the quantity  $\bar{\gamma}_I(F):=\operatorname{ess sup}_{y \in I} \max\{F(y),1-F(y)\} \cdot \frac{|f'(y)|}{f^2(y)}$ (with the supremum over $J(F)$ in $\bar{\gamma}(F)$ replaced by the supremum over $I$) being finite. An implication is:
\begin{proposition} \label{prop:blcequivLipshitzInterval}
	A random variable $Y$ with differentiable density $f(y)$ on the interval $I=(a,b) \subseteq J(F)$ is BsC on $I$ if and only if Eq. \eqref{eq:bisLipschitz} holds almost everywhere in $I$.
\end{proposition} 
\noindent Proofs for this Appendix are given in the Online Supplement. From this we can also obtain the following simple sufficient condition for BsC for some $s$:
\begin{proposition} \label{prop:somecontdiff}
	Suppose that $Y$ admits a density that is non-zero on interval $I$ and either (i) continuously differentiable on $I$ or more generally (ii) differentiable with a bounded derivative on $I$. Then $Y$ is BsC on $I$ for some finite $s$.
\end{proposition}
\noindent I note that truncation preserves BsC (and in general increases the highest $s$ for which it holds). Let $F_J$ denote the CDF of a random variable truncated to interval $J$, i.e. the CDF of $Y|Y \in J$. For example if $J=(a,b)$, then $F_J(y) = \frac{F(y)-F(a)}{F(b)-F(a)}$ for any $y \in (a,b)$, $F(y) = 0$ for any $a \le y$, and $F(y) = 1$ for any $y>b$. BsC of $F_J$ is a testable implication of BsC:
\begin{proposition} \label{prop:blctruncation}
	If $F$ is BsC on $I$ and $P(Y \in J)>0$ for $J=(a,b)$, then $F_J$ is BsC on $I \cap J$. The reverse is not true.
\end{proposition}
\noindent BsC has useful closure properties under convolution. The following result is due to \cite{saumard_bi-log-concavity_2019-1}:
\begin{proposition} \label{prop:blcconvolution}
	If $X$ is BLC and $Y$ has a log-concave density with $X \indep Y$, then $X+Y$ is BLC.
\end{proposition}
\noindent Proposition \ref{prop:blcconvolution} has some useful implications, for example that BLC of $Y_0$ implies BLC of $Y_1$ if treatment effects are independent of $Y_0$, and treatment effects are log concave.  This suggests that if treatment effects are nearly homogeneous (implying that the log concavity and independence conditions hold approximately), we can infer BLC of $Y_1$ from BLC of $Y_0$, or vice versa. Similarly if changes over time are log concave and independent of $Y_{d,t-1}$, then BLC of $Y_{d,t-1}$ implies BLC of $Y_{d,t}$. This provides an alternative to Proposition \ref{prop:sandwich} to use changes over time to infer BLC. See the Online Supplement for the general $s$ case and further discussion of using Proposition \ref{prop:blcconvolution}.

\section{Details on the literature review and empirical survey of BLC} \label{app:litdetails}

The 27 studies listed in Table \ref{tab:blctest} are selected from an initial census of all papers published in the journals listed in Section \ref{sec:validate} (the traditional top-5 economics journals along with the \textit{Review of Economics and Statistics} and the journals of \textit{American Economic Journal}) that contain any of the terms ``bunching'' ``kink'', or ``notch'' in their title or abstract.\footnote{I did not include the \textit{Journal of Public Economics} in the search. This would have increased the number of eligible studies, but 27 comprises a manageable number of papers allowing me to consider some of the nuances of each one.} This list was cross-referenced against an AI-assisted literature review using Claude to follow the citation trails for papers citing \citet{saez2010}, \citet{chetty2011}, \citet{kleven2013},\citet{kleven_bunching_2016} or \citet{blomquist_bunching_2019}. From the union of these two sets, I manually selected all papers that follow the approaches of \citet{saez2010,chetty2011,kleven2013} or an adaptation thereof to estimate an elasticity-type parameter, using bunching at a kink or notch having known statutory parameters. I exclude papers that e.g. simply document or quantify manipulation from bunching.

Within each paper, I extracted one distribution of choices subject to the kink/notch. If the study analyzed multiple kinks/notches or subsamples, I attempted to choose the distribution leading to their ``primary'' estimate, if there was one. Otherwise I chose based on sample size, or if similar, then on arbitrary order of presentation in the paper. Similarly, I chose one reference distribution from each paper, if one was available. Typically this reference distribution was an additional cross-section before or after the kink/notch was introduced, removed, or its threshold moved. In \cite{seim2017} and \cite{mortenson2020}, administratively reported distributions were available that do not feature visible bunching, despite bunching in the self-reported distribution. In these cases I used the administratively-reported distribution as the reference distribution. When multiple reference distributions were available, I first chose one related to the actual distribution by a threshold change if one was available. Next, if more than one candidate reference distribution seemed to represent similarly credible ``counterfactuals'', I chose on the basis of spatial or temporal proximity. I otherwise chose on sample size, and by arbitrary order of presentation to break any remaining ties.

Table \ref{tab:studydetails} describes the distributions extracted. A brief description of the ``actual'' distribution and the reference distribution (if one was available) samples is provided along with the figure and series from which it was drawn. To generate usable data from the figures, Claude was used to generate Python scripts to extract numeric data to CSV files using the \texttt{pyMuPDF} package. In most cases, the PDF contained vector image data allowing for nearly exact extraction of numeric data. To verify the extracted data, each series was plotted on top of the original image for manual verification by eye. The statistical test requires counts within each bin of $Y$. In some cases, the relevant figure from a given paper did not report counts directly, reporting instead the proportion of the sample in each bin, or density estimates. In these cases I use the text of the paper to determine the relevant sample size and convert these into counts.

Details of the statistical test for BLC and BsC are provided below, in Appendix \ref{sec:testing}. In settings where only the actual distribution is available, I test BLC/BsC on the region excluding any kinks or notches. Since most authors already report an excluded region when implementing a polynomial fit to the distribution, I use their excluded region where possible. When none is explicitly reported, I choose the excluded region based on visual inspection of where the distribution breaks trend, separately on either side of the kink/notch. The definition of the excluded region for each paper is given in the rightmost column of Table \ref{tab:studydetails}.

In cases where a reference distribution was available, Table \ref{fig:qqplots} reports Q-Q plots between the actual and reference distributions. When the reference distribution itself represents choices subject to a kink or notch, the Q-Q relationship is shown only for the range of values on one side of $k$ in the actual distribution. For example in the case of \cite{kostol2021}, a disability insurance notch moves from approximately \$12,000 in 2014 to \$8,000 in 2015. The Q-Q plot is shown for values under \$10,000, which is above the notch in the ``actual'' distribution (2015) but excludes values that are close enough in the 2014 distribution to the notch for the distribution to break trend. For testing of BLC/BsC on the reference distribution, I exclude only regions near budget set discontinuities.

\begin{table}
	\centering\footnotesize
	\caption{Details on studies used for BLC tests} \vspace{-.1cm}
	% ---------------------------------------------------------------
% AUTO-GENERATED by make_studies_table.py -- do not edit by hand.
% Requires: booktabs, array. Cite commands come from the CSV itself.
% Wrap in your own table/sidewaystable environment and add the caption.
% ---------------------------------------------------------------
{\setlength{\tabcolsep}{4pt}%  
\begin{tabular}{@{}
  >{\raggedright\arraybackslash}p{1.5in}
  >{\raggedright\arraybackslash}p{2.15in}
  >{\raggedright\arraybackslash}p{1.60in}
  >{\raggedright\arraybackslash}p{0.80in}@{}}
\toprule
Study & Distribution used for BLC test &
Source of distribution(s) & Excluded region \\
\midrule
\multicolumn{4}{@{}l}{\emph{Actual distribution}} \\
\addlinespace[0.2em]
\citet{almunia2018} & Main sample & Figure 2 & 5.7 to 6.53 \\
\citet{bachas2021} & Around first threshold & Figure 5, left panel & 45 to 58.8 \\
\citet{best2018} & Near 250,000 GBP notch & Figure 3, panel A & 245,000 to 275,000 \\
\citet{ito2018} & Near notch at 1520 kg & Figure 4, panel B & 1500 to 1550 \\
\citet{kleven2011} & Self employed: after audit & Figure 4, panel A & 309,700 to 327,700 \\
\citet{kleven2013} & Notch around 300k & Figure VI, panel C & 295 to 336 \\
\citet{ring2025} & Main sample & Figure 6, panel B & -100 to 100 \\
\citet{ruh2019} & Main sample & Figure 3 & -128 to 464 \\
\citet{seibold2021} & Pure financial incentive kink has no bunching & Figure 3 panel B2 & 63 \\
\addlinespace[0.6em]
\multicolumn{4}{@{}l}{\emph{Reference distribution}} \\
\addlinespace[0.2em]
\citet{best2015} & 2008: no kink in 2008 & Figure 2, panel C & $\leq$ 0 \\
\citet{best2020} & Passive LTV & Figure 4, panel B & --- \\
\citet{bukovina2025} & 2018, after abolishing minimum tax & Reference.: Figure 2, 2018 panel; Actual.: Figure 2, 2017 panel & --- \\
\citet{chetty2011} & 1997: kink moves up between 1997 and 2001 & Reference.: Figure IV, panel (h); Actual.: Figure IV, panel (d) & 0 to 270 \\
\citet{defusco2017} & 2004: increase in threshold from 2000 to 2004 & Figure 5 & $\geq$ 350 \\
\citet{devereux2014} & 2008/2009: kink was abolished in 2006 & Figure 2, panel B & --- \\
\citet{einav2015} & 2009: kink moves up between 2007 and 2009 & Figure III & 2400 to 3000 \\
\citet{kleven2014} & 1980-1990 sample: before implementation & Figure 4 & --- \\
\citet{kopczuk2015} & New Jersey, before notch & Figure 3 & --- \\
\citet{kostol2021} & 2014, when threshold was higher & Figure 6, panel A & $\geq$ 10,000 \\
\citet{lebarbanchon2025} & Missouri has a kink much higher & Reference.: Figure 2, Missouri panel (bottom right); Actual.: Figure 2, Louisiana panel (top right) & --- \\
\citet{lobel2024} & 2011-2013, before introduction & Figure 3 & --- \\
\citet{londonovelez2025} & 2009, before introduction & Figure 2 & --- \\
\citet{manoli2016} & Construction workers, not subject to the notch & Reference.: Figure 4; Actual.: Figure 2 & --- \\
\citet{mortenson2020} & Employer reported income & Figure 7, panel A & --- \\
\citet{saez2010} & Wage earners with one child, have no bunching & Figure 4, panel A & --- \\
\citet{seim2017} & Third party reported wealth & Figure 2 & --- \\
\citet{shaffer2020} & 2007, prior to kink & Figure 6, panel A & --- \\
\bottomrule
\end{tabular}}
 \label{tab:studydetails}
\end{table}

\begin{figure}[!h]
	\begin{center}
		\includegraphics[width=6in]{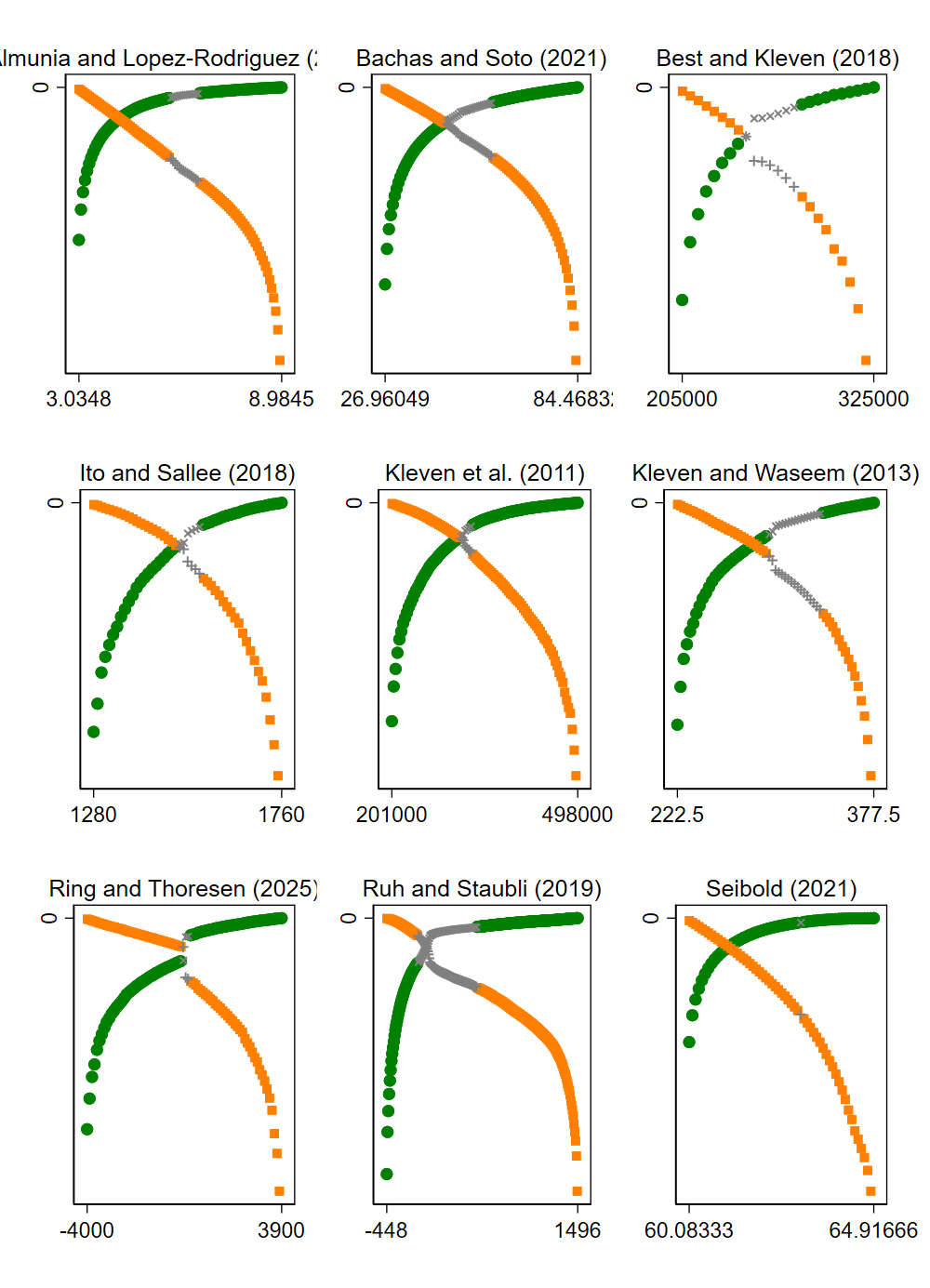}	
	\end{center}
	\caption{Plots of $\ln F$ and $\ln(1-F)$ for the distributions tested in Table \ref{tab:blctest}: papers in which the actual distribution is tested. Excluded regions for BLC test are depicted with gray $\times$ and $+$ symbols.}
	\label{fig:blcplots}
\end{figure}

\begin{figure}[!h]
	\begin{center}
		\includegraphics[width=6in]{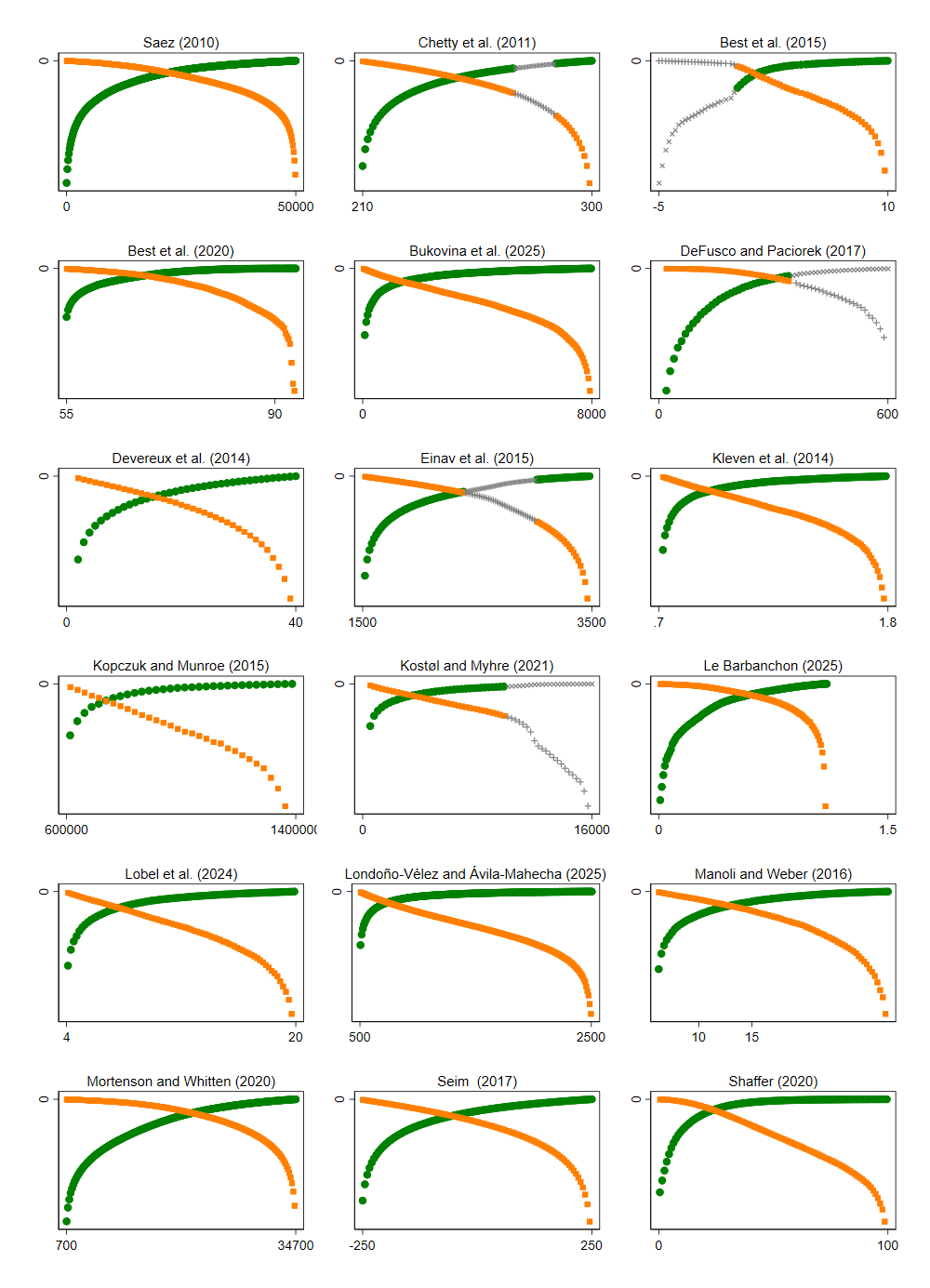}	
	\end{center}
	\caption{Plots of $\ln F$ and $\ln(1-F)$ for the distributions tested in Table \ref{tab:blctest}: papers in which a reference distribution is tested. Excluded regions for BLC test are depicted with gray $\times$ and $+$ symbols.}
	\label{fig:blcplots2}
\end{figure}

\subsection{Statistical testing of bi-log-concavity and BsC on an interval} \label{sec:testing}

To review the plausibility of BLC across applications in the literature, I propose a test for BLC and BsC on an \textit{observed} interval that uses information that can be gleaned from histograms depicting the distribution of $Y$. That it also holds within the unobserved bunching interval remains an assumption, which can be motivated by the methods discussed in Section \ref{sec:validate}.

Specifically, given a set of $J$ non-overlapping bins that partition the support of $Y$, I take as data the counts $c_j$ of observations in each bin. Let  $y_j$ denote the top of bin $j$. The data provides a consistent estimate $\hat{F}_j = (\sum_{k \le j} c_j)/n$ of $F(y_j)$, where $n=\sum_j c_j$. Let $J$ be the total number of bins. Define sequences $q_j = \ln F_{j+1} - \ln F_j$ and $r_j = \ln (1-F_{j+1}) - \ln (1-F_j)$. The hypothesis of BLC implies the null that $q_j$ and $r_j$ are each weakly decreasing in $j$. Accordingly, let $b$ be a $2(J-2)$ vector with components $b_j = q_{j+1}-q_j$ for $j=1 \dots J-2$ and $b_j = r_{j+1}-r_j$ for $j=J-2+1 \dots 2(J-2)$. If the $y_j$ are not equally spaced, then BLC instead yields decreasingness of the sequences $q_j = \frac{\ln F_{j+1} - \ln F_j}{y_{j+1}-y_j}$ and $r_j = \frac{\ln(1-F_{j+1}) - \ln(1-F_j)}{y_{j+1}-y_j}$, and I construct $b$ using these moments instead. Tests for BsC for an $s \ne 0$ use moments $q_j = s\cdot \frac{e^{s \cdot \ln F_{j+1}} - e^{s \cdot  \ln F_j}}{y_{j+1}-y_j}$ and $r_j = s\cdot  \frac{e^{s \cdot \ln(1-F_{j+1})} - e^{s\cdot \ln(1-F_j)}}{y_{j+1}-y_j}$. I drop from the test any moments that include the 1\% of mass on each end, which avoids floating point issues near the tails of the distribution.

The null to be tested is that $b_j \le 0$ for all $j$. I test this hypothesis using the inference method of \citet{AS}, with their $S_1$ test statistic.\footnote{To make use of \citet{AS} (AS), note that $\hat{b}$ is asymptotically linear with $\sqrt{n}(\hat{b}-b) \stackrel{d}{\rightarrow} N(0,\Omega)$ and $\Omega = J \Sigma J'$ where $\Sigma$ is the asymptotic variance of the stacked empirical CDFs $\hat{F}$ and $J$ a Jacobian available in closed form. Their Section 10.2 nests my case: take $\hat{\tau}:=\hat{F}$, $\tau_0:=F$, $m_j(Y,\theta,\tau):= -b_j(\theta, \tau)$, $\bar{m}_j(\theta):= m_j(Y,\theta,\hat{\tau}) = -\hat{b}_j(\theta)$ and $\theta_0=s^*$, where $b(s,F)$ are the s-dependent stacked double differences $q$ and $r$. AS's Theorem 1 applies under their high-level conditions A.2-A.3, which hold given the fixed trimming of $F$ from 0 and 1.} I use their recommended tuning parameter value of $\sqrt{\ln n}$ to control moment selection and bootstrap critical values. For robustness, I report in the Online Supplement results instead using the inference procedure of \citet{GMarXiv}, which exploits an equivalent formulation of the null as $v=0$ against the alternative that $v \ge 0$, where $v:=\max \{b' \lambda: \lambda \in \mathbbm{R}^{2{J-2}}\}$. The results are similar to those obtained by \citet{AS}, except that \citet{GMarXiv} rejects BLC of $\ln Y$ in the cases of \cite{ito2018} and \cite{londonovelez2025}, but does not in the case of \cite{best2015}. The test-statistic of \citet{GMarXiv} is the maximum value of $\frac{(\hat{b}_m)_+}{\hat{\sigma}_m}$ across moments $m$, where $\hat{\sigma}_m^2$ is an estimate of the asymptotic variance of $\hat{b}_m$. That is, the test looks for the maximum positive studentized moment, so the single ``largest'' (relative to sampling error) spike or hole in the empirical distribution will drive whether the test rejects. This is similar to the $S_3$ function in the \citet{AS} method, but I use the \citet{GMarXiv} implementation of a max-type statistic to also show robustness with respect to two ways in which critical values might be computed. 

When BLC is rejected at the 95\% level for a paper in Table \ref{tab:blctest}, I construct a 95\% confidence bound for $\kappa=-s^*$, where $s^*$ is the smallest value of $s$ for which the distribution does not violate the above implications of BsC (i.e. at the discrete bin steps available). I implement this by testing BsC at the 95\% level for values of $-s$ between 0 and 15, iterating by the bisection method. This procedure yields an upper confidence bound for $s^*$, and hence a lower confidence bound for $\kappa$. Thus $\kappa$ should be interpreted as informative about the size of the deviation of BLC ``mandated'' by the data, rather than as a conservative estimate of how far from BLC the distribution is. Nevertheless, the difference between these two will be minimal for studies with a sufficiently large sample.

I use the native bin width at which histograms are reported in each paper. Larger bandwidths could be preferable in some cases, as they would increase the number of observations involved in each moment being tested. However, they also make the test more conservative in the sense of testing fewer inequalities. Improvements to BLC testing represent an interesting avenue for future work. For example, given the full microdata with continuous $Y$ an alternative to using moment inequalities to test BLC would be to extend tests that compare the empirical CDF to its least concave majorant \citep{durot2008}.

\section{Sufficient conditions for Assumptions NOTCH and RANK} \label{sec:ranknotchsuff}
In this section I give a set of primitive sufficient conditions for both RANK and NOTCH to hold, which represent a nonparametric relaxation of the isoelastic model.
\begin{assumption*}[QL (preferences quasilinear in the running variable)]
	For each $i$, the function $u_i(z,\mathbf{x})$ from Assumption CONVEX can be taken to be additively separable and linear in $y_i(\mathbf{x})-z$, i.e. $u_i(z,\mathbf{x}) = y_i(\mathbf{x})-z+\pi_i(\mathbf{x})$ for some function $\pi_i(\cdot)$.
\end{assumption*}
\noindent Note that Assumptions QL and CONVEX together imply that $\pi_i$ is strictly concave in $\mathbf{x}$.\footnote{To see this, note that strict quasi-concavity of $u_i(z,\mathbf{x}) = z+\pi_i(\mathbf{x})$ says that for any $\lambda \in (0,1)$,
	$$\lambda z_1 + (1-\lambda)z_2+\pi_i(\lambda \mathbf{x}_1 + (1-\lambda)\mathbf{x}_2) > z_1+\pi_i(\mathbf{x}_1) = (1-\lambda)(z_1 + \pi_i(\mathbf{x}_1)) +\lambda (z_1+\pi_i(\mathbf{x}_1)) $$
	where without loss we take $u_i(z_1,\mathbf{x}_1) \ge u_i(z_2,\mathbf{x}_2)$. Using this latter condition, we have that the first term on the RHS above is greater than or equal to $(1-\lambda)(z_2+\pi_i(\mathbf{x}_2))$. Combining the two inequalities, we then have that 
	$\pi_i(\lambda \mathbf{x}_1 + (1-\lambda)\mathbf{x}_2) > \lambda \pi_i(\mathbf{x}_1) + (1-\lambda)\pi_i(\mathbf{x}_2)$, i.e. $\pi_i$ is strictly concave.}

\begin{assumption*}[UNI (simple univariate heterogeneity)]
	The following hold:
	\begin{enumerate}[noitemsep]
		\item CONVEX and QL hold.
		\item There is a single margin of choice $y$, i.e. $\pi_i(\mathbf{x}) = \pi_i(y_i(\mathbf{x}))$, so we can simply write $\pi_i(y)$. $\pi_i(\cdot)$ is differentiable for each $i$.
		\item For some scalar $\eta_i$, $\pi_i(y)=\pi(y;\eta_i)$ where $\frac{d}{dy} \pi(y;\eta_i)$ is strictly increasing in $\eta_i$ for all $y$.
		\item The cost functions $T_{0i}$ and $\tilde{T}_{1i}$ and $N_i$ are common to all individuals $i$, depend on $\mathbf{x}$ only through $y_i(\mathbf{x})$, and $T_0(y)$ and $\tilde{T}_1(y)$ are differentiable in $y$.
	\end{enumerate}
\end{assumption*}
\begin{proposition} \label{prop:unisuff}
	Under UNI, Assumptions RANK and NOTCH hold.
\end{proposition}
\noindent Compared with the general utility maximization model under CONVEX, UNI is restrictive in a few ways. Quasilinearity in $y$ might be seen as mild, but by 2., individual margins of choice $\mathbf{x}$ cannot affect utility (or cost functions) apart from their effects through $y_i(\mathbf{x})$. Item 3. is strong, requiring rank invariance between $Y_0$ and $\tilde{Y}_1$: the proof of Proposition \ref{prop:unisuff} shows 3. implies that $Y_{0i}=g_0(\eta_i)$ for some strictly increasing function $g_0$, while $\tilde{Y}_{1i}=Y_{1i}=g_1(\eta_i)$ for a strictly increasing $g_1$.

Nevertheless, UNI is still substantially more general than existing literature considering bunching at notches. $\pi_i(y)$ need not be monotonically decreasing in $y$. In an income taxation setting such as \citet{kleven2013}, individuals may receive positive or negative utility from work, as long as they have declining marginal utility in a consumption metric. Their utility from work may also have different local elasticities at different values of $y$: a dimension of flexibility that could explain estimating different elasticity values for notches located at different income values, in settings with point identification. Treatment effects $Y_{0i}-Y_{1i}$ need not be homogeneous, even after applying a transformation (such as the logarithm in the isoelastic model) to $Y$. Finally, the cost functions $T_0$ and $\tilde{T}_1$ need not be linear in $y$, extending beyond simple linear tax schemes.

\begin{example}[(isoelastic model at a tax notch or kink)]
	The isoelastic model applied to a standard tax setting with linear $T_0$ and $\tilde{T}_1$ satisfies Assumption UNI. In this case, for example $\frac{d}{dy} \pi(y;\eta_i) = -(y/\eta_i)^{1/e}$, which is non-positive and is strictly increasing in $\eta$ given $e > 0$, satisfying UNI.
\end{example}

\begin{example}[(notch in a wealth tax)]
	Consider the setting of \citet{londonovelez2025}, in which individuals face a tax on reported wealth $y$. \citet{londonovelez2025} do not include an explicit choice model to underpin their empirical strategy, but here I give one that satisfies Assumption UNI. Suppose that utility is $a \cdot \eta_i-T(y)-\phi(\eta_i-y)$, where $a \cdot \eta_i$ represents an consumption annuity from $i$'s actual wealth, denoted $\eta_i$, $T(y)$ indicates tax on reported wealth $y$ (net of any evasion), and $\phi(x)$ represents an evasion cost that is increasing and strictly convex in the amount evaded $x=\eta_i-y$. Write utility in the form of QL as $u_i(z,y)=z- \tilde{\phi}(\eta_i-y)$ where $\tilde{\phi}(x) = \frac{\phi(x)}{a}-x$ is strictly convex in $x$ (thus satisfying item 2. of UNI), and we replace the tax schedule $T(y)$ with $\tilde{T}(y)=T(y)/a$. In this model $i$'s wealth $\eta_i$ is taken as given, and the single margin of choice is how much to report $y$ (with the remainder being e.g. sheltered from the tax authorities). Item 3. of UNI holds because $\frac{d}{dy} \pi_i(y; \eta_i) = \frac{d}{dy} \left\{-\frac{\phi(\eta_i-y)}{a}+\eta_i-y\right\} = \frac{\phi'(\eta_i-y)}{a} -1$, which is increasing in $\eta_i$ given that $\phi''(x) > 0$.
\end{example}

\section{Proofs}

\subsection{Proof of Lemma \ref{propobserve}}

\noindent The proof proceeds in the following two steps:
\begin{enumerate}[label=\roman*)]
	\item First, I show that $Y_{0i}\le k$ implies that $Y_i = Y_{0i}$, and similarly $Y_{1i}\ge k$ implies that $Y_i = Y_{1i}$. This holds under CONVEX but also under the weaker assumption of WARP.
	\item Second, I show that under CONVEX $Y_{i} < k \implies Y_{i} = Y_{0i}$ and $Y_{i} > k \implies Y_{i} = Y_{1i}$.
\end{enumerate}
Item i) above establishes the first and third cases of Lemma \ref{propobserve}. The only remaining possible case is that $Y_{1i} \le k \le Y_{0i}$ (note $Y_{0i} \le k \le Y_{1i}$ with at least one inequality strict yields a contradiction). However, to finish establishing Lemma \ref{propobserve}, we also need the reverse implication: that $Y_{1i} \le k \le Y_{0i}$ implies $Y_i=k$. This comes from taking the contrapositive of each of the two claims in item ii).

For any cost function $T$, let $(z_{Ti},\mathbf{x}_{Ti})$ denote the choices decision-maker $i$ would make under $T$. In this notation Assumption WARP says that if $z_{T'i} \ge T(\mathbf{x}_{T'i})$, then $(z_{Ti},\mathbf{x}_{Ti})=(z_{T'i},\mathbf{x}_{T'i})$.\\

\noindent \textbf{Proof of i): } Let $\mathcal{X}_{0i} = \{\mathbf{x}: y_i(\mathbf{x}) \le k\}$ and $\mathcal{X}_{1i} = \{\mathbf{x}: y_i(\mathbf{x}) \ge k\}$. If $Y_{0i}\le k$, then $\mathbf{x}_{T_{0i}}$ is in $\mathcal{X}_0$. Since $T_i(\mathbf{x})=T_{0i}(\mathbf{x})$ for all $\mathbf{x} \in \mathcal{X}_0$, it follows that $z_{T_{0i}i} \ge T_i(\mathbf{x}_{T_{0i}i})$, i.e. the decision-maker's choice under $T_0$ is feasible under the kinked budget constraint $T$. Note that $T_{i}(\mathbf{x}) \ge T_{0i}(\mathbf{x})$ for all $\mathbf{x}$. By WARP then $(z_{T_ii},\mathbf{x}_{T_ii})=(z_{T_{0i}i},\mathbf{x}_{T_{0i}i})$. Thus $Y_i = y_i(\mathbf{x}_{T_ii}) = y_i(\mathbf{x}_{T_{0i}i}) = Y_{0i}$. So $Y_{0i} \le k \implies Y_i=Y_{0i}$. By the same logic we can show that $Y_{1i} \ge k \implies Y_i=Y_{1i}$.\\

\noindent \textbf{Proof of ii): } For any convex cost function $T(\mathbf{x})$, $(z_{Ti},\mathbf{x}_{Ti})=\textrm{argmax}_{z,\mathbf{x}} \left\{ u_i(z,\mathbf{x}) \textrm{ s.t. } z \ge T(\mathbf{x})\right\}$. If $u_i(z,\mathbf{x})$ is strictly quasi-concave, then the RHS exists and is unique since it maximizes $u_i$ over the convex domain $\{(z,\mathbf{x}): z \ge T(\mathbf{x})\}$. Furthermore, by monotonicity of $u(z,\mathbf{x})$ in $z$ we may substitute in the constraint $z=T(\mathbf{x})$ and write $\mathbf{x}_{Ti} = \textrm{argmax}_\mathbf{x} u_i(T(\mathbf{x}),\mathbf{x})$. Suppose that $y_i(\mathbf{x}_{Ti}) \ne k$, and consider any $\mathbf{x} \ne \mathbf{x}_{Ti}$ such that $y_i(\mathbf{x}) \ne k$. Let $\tilde{\mathbf{x}} = \theta \mathbf{x} + (1-\theta)\mathbf{x}^*$ where $\mathbf{x}^* = \mathbf{x}_{Ti}$ and $\theta \in (0,1)$. Since $T(\mathbf{x})$ is convex in $\mathbf{x}$ and $u_i(z,\mathbf{x})$ is weakly decreasing in $z$:
\begin{equation} \label{eqsinglpeak} u_i(B(\tilde{\mathbf{x}}),\tilde{\mathbf{x}}) \ge u_i(\theta T(\mathbf{x}) + (1-\theta) T(\mathbf{x}^*),\tilde{\mathbf{x}}) > \min\{u_i(T(\mathbf{x}), \mathbf{x}), u_i(T(\mathbf{x}^*), \mathbf{x}^*)\} = u_i(T(\mathbf{x}), \mathbf{x}) \end{equation}
where I have used CONVEX in the second step, and that $\mathbf{x}^*$ is a maximizer in the third. This result implies that for any such $\mathbf{x} \ne \mathbf{x}^*$, if one draws a line between $\mathbf{x}$ and $\mathbf{x}^*$, the function $u_i(T(\mathbf{x}),\mathbf{x})$ is strictly increasing as one moves towards $\mathbf{x}^*$. When $\mathbf{x}$ is a scalar, this argument is used by \citet{blomquist_individual_2015} (see Lemma A1 therein) to show that $u_i(T(\mathbf{x}),\mathbf{x})$ is strictly increasing to the left of $\mathbf{x}^*$, and strictly decreasing to the right of $\mathbf{x}^*$. Note that for any (binding) linear budget constraint $T(\mathbf{x})$, the result still holds without monotonicity of $u_i(z,\mathbf{x})$ in $z$.

For any function $T$, let $u_{Ti}(\mathbf{x}) = u_i(T(\mathbf{x}),\mathbf{x})$, and note that
$$u_{T_ii}(\mathbf{x})=
\begin{cases}
	u_{T_{0i}i}(\mathbf{x}) & \hspace{.2cm} \textrm{if } \hspace{.2cm} \mathbf{x} \in \mathcal{X}_{0i}\\
	u_{T_{1i}i}(\mathbf{x}) & \hspace{.2cm} \textrm{if } \hspace{.2cm} \mathbf{x} \in \mathcal{X}_{1i}
\end{cases}$$
where $T_i$ is the actual, kinked budget constraint faced by $i$. Let $\mathbf{x}_{ki}$ be the unique maximizer of $u_{T_ii}(\mathbf{x})$, where $Y_{i} = y_i(\mathbf{x}_{ki})$. Suppose that $Y_{i} < k$. Suppose furthermore that $Y_{0i} \ne Y_{i}$, with $Y_{0i}=y_i(\mathbf{x}_{0i})$ and $\mathbf{x}_{0i}$ the maximizer of $u_{T_{0i}i}(\mathbf{x})$. Note that we must have that $\mathbf{x}_{0i} \notin \mathcal{X}_{0i}$, because $T_{0i}=T_i$ in $\mathcal{X}_{0i}$ so we can't have $u_{T_{0i}i}(\mathbf{x}_{0i}) > u_{T_{0i}i}(\mathbf{x}_{ki})$ (since $\mathbf{x}_{ki}$ maximizes $u_{T_i}(\mathbf{x})$). Thus $Y_{0i} > k$.

\begin{figure}[H]
	\begin{center}
		\includegraphics[height=2.5in]{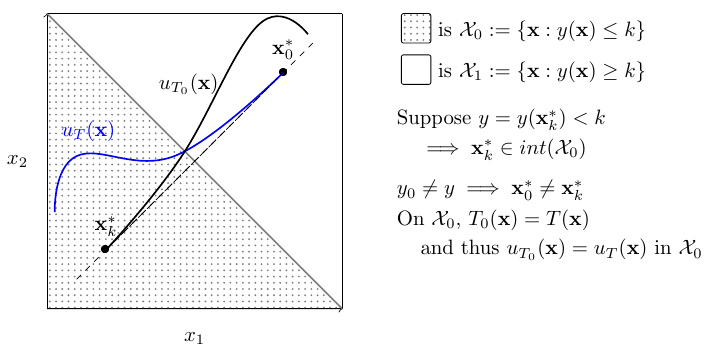}
		\caption{Depiction of the step establishing $(Y < k) \implies (Y = Y_{0})$ in the proof of Lemma \ref{propobserve}. Since the result considers a single decision-maker $i$, I suppress this index in the Figure. In this example $z=(x_1,x_2)$ and $y(\mathbf{x}) = x_1+x_2$. If $Y<k$ and $Y_0 \ne Y$ then $u_{T_0}(\mathbf{x})$ is strictly increasing as one moves along the dotted line from $\mathbf{x}_k^*$ towards $\mathbf{x}^*_0$. But $u_{T}(\mathbf{x})$ is strictly increasing as one moves in the opposite direction along the same line, from $\mathbf{x}_0^*$ towards $\mathbf{x}^*_k$. Since $Y<k$, there is some interval of the dotted line that is within $\mathcal{X}_0$. On this interval, the functions $T_0$ and $T$ are equal, and thus so are the functions $u_{T_0}$ and $u_{T}$.}\label{figcurves}
	\end{center}
\end{figure}

By $Y_i<k$ and continuity of $y_i(\mathbf{x})$, $\mathcal{X}_{0i}$ is a closed set and $\mathbf{x}_{ki}$ belongs to the interior of $\mathcal{X}_{0i}$. Thus, while $\mathbf{x}_{0i}$ is not in $\mathcal{X}_{0i}$, there exists a point $\tilde{\mathbf{x}} \in \mathcal{X}_{0i}$ along the line between $\mathbf{x}_{0i}$ to $\mathbf{x}_{ki}$. Since $Y_i\ne k$ and $Y_{0i}\ne k$, Eq. (\ref{eqsinglpeak}) then implies that $u_{T_i i}(\tilde{\mathbf{x}}) > u_{T_i i}(\mathbf{x}_{0i})$. Since $u_{T_{0i}i}(\mathbf{x})=u_{T_i i}(\mathbf{x})$ for all $\mathbf{x}$ in $\mathcal{X}_{0i}$, it follows that $u_{T_{0i}i}(\tilde{\mathbf{x}}) > u_{T_{0i}i}(\mathbf{x}_{0i})$. However, this contradicts the premise that $\mathbf{x}_{0i}$ maximizes $u_{T_{0i}i}(\mathbf{x})$. Thus, $Y_{i} < k$ implies $Y_{i} = Y_{0i}$. Figure \ref{figcurves} depicts the logic visually. The proof that $Y_{i} > k$ implies $Y_{i} = Y_{1i}$ is analogous.

\subsection{Proof of Proposition \ref{thmstraddle}}
Item i) in the proof of Lemma \ref{propobserve} establishes that under WARP $Y_i=k$ implies $Y_{1i} \le k \le Y_{0i}$, since i) implies that $Y_{0i}<k \implies Y_{i} < k$ and $Y_{1i}>k \implies Y_i > k$ (taking contrapositives, we have that $\{Y_i\ge k$ and $Y_i \le k\}$ implies $Y_{1i} \le k \le Y_{0i}$). Item ii) in hat proof shows that under CONVEX $Y_{1i} \le k \le Y_{0i}$ also implies $Y_i=k$, thus $Y_{1i} \le k \le Y_{0i}$ and $Y_i = k$ are equivalent. Note that by adding $\Delta_i = Y_{0i}-Y_{1i}$ to both sides of the inequality $Y_{1i} \le k$, we have that $Y_{0i} \le k+\Delta_i$. Combining with the other inequality that $Y_{0i} \ge k$, we can thus rewrite the event $Y_{1i} \le k \le Y_{0i}$ as $Y_{0i} \in [k, k+\Delta_i]$ (or equivalently to $Y_{1i} \in [k-\Delta_i, k]$). We thus have that $\mathcal{B} \le P(Y_{0i} \in [k,k+\Delta_i])$ under WARP, and that $\mathcal{B} = P(Y_i=k)= P(Y_{1i} \le k \le Y_{0i})$ under CONVEX.

\subsection{Proof of Proposition \ref{corr:dens}}
Consider first a pure kink under WARP. By item i) in the proof of Proposition \ref{thmstraddle}, we have that $Y_{0i} \le k \implies Y_i=Y_{0i}$. Thus, for any $\delta > 0$ and $y < k$: $Y_{0i} \in [y - \delta, y]$ implies that $Y_{i} \in [y - \delta, y]$ and hence $P(Y_{0i} \in [y - \delta, y]) \le P(Y_{i} \in [y - \delta, y])$. This implies that $f_0(y) - f(y) = \lim_{\delta \downarrow 0} \frac{P(Y_{0i} \in [y - \delta, y]) - P(Y_{i} \in [y - \delta, y])}{\delta} \le 0$, i.e. that $f(y) \ge f_0(y)$ when both exist. An analogous argument holds for $Y_1$, where we consider the event $Y_{1i} \in [y, y+\delta]$ any $y>k$. By CONT, the limits $f(k^-)$ and $f(k^+)$ exist, and thus upper bound $f_0(k)$ and $f_1(k)$, respectively. Under CONVEX instead of WARP, and NOTCHBOUND, the inequalities become equalities.

\subsection{Proof of Lemma \ref{propobservenotch}}
The implications of CONVEX in Lemma \ref{propobserve} do not require that $T_{0i}(\mathbf{x})=\tilde{T}_{1i}(\mathbf{x})$ when $y_i(\mathbf{x})$. That is, while $T_0$ and $\tilde{T}_1$ must be continuous, the function $T_{0i}(\mathbf{x}) \cdot \mathbbm{1}(y_i(\mathbf{x}) < k) + T_{1i}(\mathbf{x}) \cdot \mathbbm{1}(y_i(\mathbf{x}) \ge k)$ need not be. Therefore replacing $T_1$ from Lemma \ref{propobserve} with $\tilde{T}_1$, we have the following implications of CONVEX: $Y_{i} < k \implies Y_{i} = Y_{0i}$ and $Y_{i} > k \implies Y_{i} = Y_{1i}$. In particular, this implies that $Y_{1i} \le k \le Y_{0i} \implies Y_i = k$.

Meanwhile, by WARP, we have that $Y_{0i}\le k \implies Y_i = Y_{0i}$ as in Lemma \ref{propobserve}. Thus we also have that $Y_i=k$ implies $Y_{0i} \ge k$. However, we do not have that $Y_{1i}\ge k \implies Y_i = Y_{1i}$, because we do not necessarily have that $T_{1i}(\mathbf{x}) < T_{0i}(\mathbf{x})$ when $y_i(\mathbf{x})<k$ (because of the addition of $N>0$ in $T_1$--see Figure \ref{fig:kinknotch}). This represents the main departure from Lemma \ref{propobserve} for notches.

Let $A_i$ be the event that $Y_i=k$ but $Y_{1i} > k$. We can summarize our results thus far as:
\begin{equation} \label{hcasesnotchproof}
Y_i=
\begin{cases}
	Y_{0i} & \hspace{.2cm} \textrm{if } \hspace{.2cm} Y_{0i}<k\\
	k & \hspace{.2cm} \textrm{if } \hspace{.2cm} Y_{0i} \ge k \textrm{ and } \{Y_{1i} \le k \textrm{ or } A_i\}  \hspace{.2cm}\\ 
	Y_{1i} & \hspace{.2cm} \textrm{if } \hspace{.2cm} Y_{1i}>k \textrm{ and not } A_i
\end{cases}
\end{equation}
Writing $A_i$ in terms of preferences (rather than including the event $Y_i=k$ in the definition of $A_i$), we have:
$$\{Y_{1i} > k \textrm{ and } \max_{\mathbf{x}: y_i(\mathbf{x})=k} u_i(\tilde{T}_{1i}(\mathbf{x}),\mathbf{x}) \ge u_i(\tilde{T}_{1i}(\mathbf{x}_{1i})+N_i,\mathbf{x}_{1i})\}$$
where $\mathbf{x}_{1i}$ is $i$'s chosen value of $\mathbf{x}$ under $T_{1i}$, and note that $\tilde{T}_{1i}(\mathbf{x}_{1i})+N_i$, and $\tilde{T}_{1i}(\mathbf{x})=T_{0i}(\mathbf{x})$ for any $\mathbf{x}$ such that $y_{i}(\mathbf{x})=k$. 

\subsection{Proof of Proposition \ref{prop:buncherATEnotch}}

I use the generalized definitions of $\Delta_k^* = \mathbbm{E}[Y_0-Y_1|Y_0 \ge y_L,Y_1 \le y_U]$, $\mathcal{B} = P(Y_0 \ge y_L, Y_1 \le y_U) = F_1(y_U)-F_0(y_L)$ and $Q_k^*$ as defined in Eq. \eqref{eq:qkgen} from Section \ref{sec:optimizationerror}, with $\Delta_0^*:= Q_0(F_1(y_U))-y_L$ and $\Delta_1^*:= y_U-Q_1(F_0(y_L))$. This allows us to accommodate optimization error under ERRORBOUND with $y_L \le k-E$ and $y_U \ge y^I-E$. We can also handle the relaxation of NOTCH to NOTCHBOUND by choosing $y_L=k$ and $y_U=y^I$ as defined in NOTCHBOUND. 

In this general setting let $\mathcal{B} = F(y_U)-F(y_L)$ and define $E_0 := \mathbbm{E}[Y_0|Y_0 \in [y_L,y_L+\Delta_0^*]]$ and $E_1 := \mathbbm{E}[Y_1|Y_1 \in [y_U-\Delta_1^*, y_U]]$, where recall that $\Delta^*_0 := Q_0(F_0(y_L)+\mathcal{B}) - y_L$ and $\Delta_1^* := y_U-Q_1(F_0(y_U))$. Let $\mathcal{B}>0$ so that $\Delta_k^*$ and $Q_k^*$ are each unambiguously defined.

The main results below does not make use of Assumptions CONVEX or NOTCH. They are only necessary in order to equate $\mathbbm{E}[Y_0-Y_1|Y_0 \ge y_L,Y_1 \le y_U]$ with $\mathbbm{E}[Y_0-Y_1|Y=k]$, the definition of $\Delta_k^*$ used in the main text, and to make $\mathcal{B} = P(Y=k)$. At the end of the proof, I separately treat the case of a pure kink under WARP.\\

\noindent \textit{Preliminaries. }I begin by establishing some basic implications of CONT: namely that $E_0$ coincides with $\frac{1}{\mathcal{B}}\int_{F_0(y_L)}^{F_0(y_L)+\mathcal{B}} Q_0(u)\cdot du$, while $E_1$ coincides with $\frac{1}{\mathcal{B}}\int_{F_1(y_U)-\mathcal{B}}^{F_1(y_U)} Q_1(u)\cdot du$. This establishes $Q_k^* = E_0 - E_1$, given $\mathcal{B}>0$.

As a quantile function, $Q_0(u)$ has the property that $y \ge Q_0(u) \iff F_0(y) \ge u$, and thus $y \in [y_L, y_L+\Delta_0^*) \iff F_0(y) \in [F_0(y_L), F_0(y_L+\Delta_0^*))$. Since $F_0$ has no point mass at $y_L+\Delta_0^*$ (by part iii of CONT), $Y_0 \in [y_L, y_L+\Delta_0^*] \iff F_0(Y_0) \in [F_0(y_L), F_0(y_L+\Delta_0^*)]$, with probability one. Note that $F_0(y_L+\Delta_0^*)=F_0(y_L)+\mathcal{B}$. Another property of the quantile function is that $Q_0(U) = Y_0$ almost surely, where $U:=F_0(Y_0)$ is a uniform random variable on $[0,1]$ (see e.g. Lemmas 3-4 of \citet{goff2024testingidentifyingassumptionsparametric}). Therefore given $\mathcal{B} > 0$, $E_0$ can be written:
\begin{align*}
\mathbbm{E}[Q_0(U) | U \in [F_0(y_L), F_0(y_L+\Delta_0^*)] = \frac{\int_{F_0(y_L)}^{F_0(y_L+\Delta_0^*)} Q_0(u) \cdot du}{F_0(y_L+\Delta_0^*)-F_0(y_L)} = \frac{1}{\mathcal{B}}\int_{F_0(y_L)}^{F_0(y_L)+\mathcal{B}} Q_0(u)\cdot du
\end{align*}
where in the last step we have used $F_0(k+\Delta_0^*) = F_0(Q_0(F_0(k)+\mathcal{B})) = F_0(k)+\mathcal{B}$, by continuity of $F_0$ (which implies that $F_0(Q_0(u))=u$) at $y_L+\Delta_0^*$. Note that this also establishes that $P(Y_0 \in [y_L, y_L+\Delta_0^*])= P(F_0(Y_0) \in [F_0(y_L), F_0(y_L+\Delta_0^*)]) = F_0(y_L)+\mathcal{B} - F_0(y_L) = \mathcal{B}$.

An analogous argument applies for $Y_1$ by writing $\Delta_1^* = y_U-Q_1(F_1(y_U)-\mathcal{B})=Q_1(F_0(y_L))$, establishing that $E_1=\frac{1}{\mathcal{B}}\int_{F_1(y_U)-\mathcal{B}}^{F_1(y_U)} Q_1(u)\cdot du$ and $P(Y_1 \in [y_U-\Delta_1^*, y_U])=\mathcal{B}$.\\

\noindent \textit{Main proof. } By the law of iterated expectations:
\begin{align}
E_0 &= p_0 \cdot \mathbbm{E}[Y_0|Y_0 \in [y_L,y_L+\Delta_0^*], Y_1 \le y_U] + (1-p_0)\cdot \mathbbm{E}[Y_0|Y_0 \in [y_L,y_L+\Delta_0^*], Y > y_U] \nonumber \\
&\le  p_0 \cdot \mathbbm{E}[Y_0|Y_0 \in [y_L,y_L+\Delta_0^*], Y_1\le y_U]+(1-p_0)\cdot(y_L+\Delta_0^*) \label{eq:E0lt}
\end{align}
where $p_0:=P(Y_1 \le y_U|Y_0 \in [y_L,y_L+\Delta_0^*])$, and using that $\mathbbm{E}[Y_0|Y_0 \in [y_L,y_L+\Delta_0^*], Y > y_L] \le k+\Delta_0^*$. Meanwhile
\begin{align}
\mathbbm{E}[Y_0|Y_0 \ge y_L,Y_1 \le y_U]	&=q_0\cdot \mathbbm{E}[Y_0|Y_0 \in [y_L, y_L+\Delta_0^*],Y_1 \le y_U] \nonumber \\
& \hspace{2in}+ (1-q_0)\cdot \mathbbm{E}[Y_0|Y_0 > y_L+\Delta_0^*,Y_1 \le y_U] \nonumber \\
&\ge  q_0 \cdot \mathbbm{E}[Y_0|Y_0 \in [y_L,y_L+\Delta_0^*], Y \le y_U]+(1-q_0)\cdot(k+\Delta_0^*) \label{eq:E0gt} 
\end{align}
where $q_0:= P(Y_0 \le y_L + \Delta_0^*|Y_0 \ge y_L,  Y_1 \le y_U)$, and $\mathbbm{E}[Y_0|Y_0 > y_L+\Delta_0^*,Y_1 \le y_U] > y_L + \Delta_0^*$ when $q_0 < 1$ in the third. 

I now show that $q_0=p_0$. Recall that $P(Y_0 \in [y_L,y_L+\Delta_0^*])=\mathcal{B}$ by CONT. Thus
$$p_0:=P(Y \le y_U|Y_0 \in [y_L,y_L+\Delta_0^*]) = \frac{P(Y_1 \le y_U,Y_0 \in [y_L,y_L+\Delta_0^*])}{P(Y_0 \in [y_L,y_L+\Delta_0^*])}$$
Meanwhile
$$q_0:= P(Y_0 \le y_L + \Delta_0^*|Y_0 \ge y_L, Y_1 \le y_U) = \frac{P(Y_0 \in [y_L, y_L + \Delta_0^*], Y_1 \le y_U)}{\mathcal{B}}  = p_0$$
since $\mathcal{B} = P(Y_0 \in [y_L,y_L+\Delta_0^*])$. Combining with \eqref{eq:E0lt} and \eqref{eq:E0gt}, we have that $\mathbbm{E}[Y_0|Y_0 \ge y_L, Y_1 \le y_U] \ge E_0$. Equality holds when $q_0$ and $p_0$ are equal to one.

Similarly for $Y_1$, we have that
\begin{align}
E_1 &= p_1 \cdot \mathbbm{E}[Y_1|Y_1 \in [y_U-\Delta_1^*, y_U], Y_0 \ge y_L] + (1-p_1)\cdot \mathbbm{E}[Y_1|Y_1 \in [y_U-\Delta_1^*, y_U], Y_0 < y_L] \nonumber \\
&\ge  p_1 \cdot \mathbbm{E}[Y_1|Y_1 \in [y_U-\Delta_1^*, y_U], Y_0 \ge y_L]+(1-p_1)\cdot(k-\Delta_1^*) \label{eq:E1lt}
\end{align}
where $p_1:=P(Y_0 \ge y_L|Y_1 \in [y_U-\Delta_1^*, y_U]) = \frac{P(Y_0 \ge y_L,Y_1 \in [y_U-\Delta_1^*, y_U])}{P(Y_1 \in [y_U-\Delta_1^*, y_U])}$. Meanwhile:
\begin{align}
\mathbbm{E}[Y_1|Y_0 \ge y_L,Y_1 \le y_U]
&=q_1\cdot \mathbbm{E}[Y_1|Y_1 \in [y_U-\Delta_1^*, y_U],Y_0 \ge y_L] \nonumber \\
& \hspace{2in}+ (1-q_1)\cdot \mathbbm{E}[Y_1|Y_1 < k-\Delta_1^*,Y_0 \ge y_L] \nonumber \\
&\le  q_1 \cdot \mathbbm{E}[Y_1|Y_1 \in [y_U-\Delta_1^*, y_U],Y_0 \ge y_L]+(1-q_1)\cdot(k-\Delta_1^*) \nonumber \\
&=  p_1 \cdot \mathbbm{E}[Y_1|Y_1 \in [y_U-\Delta_1^*, y_U],Y_0 \ge y_L]+(1-p_1)\cdot(k-\Delta_1^*) \label{eq:E1gt}
\end{align}
where $q_1:= P(Y_1 \ge y_U - \Delta_1^*|Y_0 \ge y_L,Y_1 \le y_U) = \frac{P(Y_0 \ge y_L,Y_1 \in [y_U - \Delta_1^*,y_U])}{\mathcal{B}}=p_1$ using that $P(Y_1 \in [y_U-\Delta_1^*, y_U]) = \mathcal{B}$ and $\mathbbm{E}[Y_1|Y_1 < k-\Delta_1^*,Y_0 \ge y_L] < k- \Delta_1^*$ when $q_1 < 1$. Combining \eqref{eq:E0lt} and \eqref{eq:E1gt}, we have that $\mathbbm{E}[Y_1|Y_0 \ge y_L, Y_1 \le y_U] \le E_1$ (with equality when $q_1=p_1=1$). 

All together: $\Delta_k^*=\mathbbm{E}[Y_0-Y_1|Y_0 \ge y_L, Y_1 \le y_U] \ge E_0 - E_1$. Assumption RANK from the main text is a special case of the assumption that $Y_{0i} \in [y_L, y_L+\Delta_0^*] \textrm{ iff } Y_{1i} \in [y_U-\Delta_1^*, y_U]$ with probability one. Then $p_1=p_0=1$, and the inequality becomes and equality: $\Delta_k^* = E_0 - E_1$.\\

Now I turn to establishing $\Delta_k^* \ge Q_k^*$ in the case of a pure kink under WARP, without optimization error (with $y_L=y_U=k$). Now we have $\mathbbm{E}[Y_0|Y=k] = \mathbbm{E}[Y_0|Y \ge y_L, Y \ge y_U]$ and $\mathbbm{E}[Y_1|Y=k] = \mathbbm{E}[Y_1|Y \ge y_L, Y \ge y_U]$. By the law of iterated expectations: 
\begin{align}
E_0 &= p_0 \cdot \mathbbm{E}[Y_0|Y_0 \in [k,k+\Delta_0^*], Y=k] + (1-p_0)\cdot \mathbbm{E}[Y_0|Y_0 \in [k,k+\Delta_0^*], Y\ne k] \nonumber \\
&\le  p_0 \cdot \mathbbm{E}[Y_0|Y_0 \in [k,k+\Delta_0^*], Y=k]+(1-p_0)\cdot(k+\Delta_0^*) \label{eq:E0ltWARP}
\end{align}
as in Eq \eqref{eq:E0lt}, where now $p_0:=P(Y=k|Y_0 \in [k,k+\Delta_0^*])$. Meanwhile Eq. \ref{eq:E0gt} yields $\mathbbm{E}[Y_0|Y=k] \ge  q_0 \cdot \mathbbm{E}[Y_0|Y_0 \in [k,k+\Delta_0^*], Y=k]+(1-q_0)\cdot(k+\Delta_0^*)$ with $q_0= P(Y_0 \le k + \Delta_0^*|Y=k)$, using that $Y=k \implies Y_0 \ge k$ by WARP (see Lemma \ref{propobservenotch}). Again $q_0=p_0$, since
$$p_0 = \frac{P(Y=k,Y_0 \in [k,k+\Delta_0^*])}{P(Y_0 \in [k,k+\Delta_0^*])}=\frac{P(Y_0 \le k + \Delta_0^*, Y=k)}{\mathcal{B}} = q_0$$
using again that $Y=k \implies Y_0 \ge k$ by WARP. We then have that $\mathbbm{E}[Y_0|Y=k] \ge E_0$.

\noindent Similarly for $Y_1$, we have that
\begin{align}
E_1 &\ge  p_1 \cdot \mathbbm{E}[Y_1|Y_1 \in [k-\Delta_1^*, k], Y=k]+(1-p_1)\cdot(k-\Delta_1^*) \label{eq:E1ltWARP}
\end{align}
as in Eq \eqref{eq:E1lt}, where $p_1:=P(Y = k|Y_1 \in [k-\Delta_1^*, k]) = \frac{P(Y=k,Y_1 \in [k-\Delta_1^*, k])}{P(Y_1 \in [k-\Delta_1^*, k])}$. Using $Y=k \implies Y_1 \le k$ we have that:		
\begin{align}
\mathbbm{E}[Y_1|Y=k] &\le  q_1 \cdot \mathbbm{E}[Y_1|Y_1 \in [k-\Delta_1^*, k],Y=k]+(1-q_1)\cdot(k-\Delta_1^*) \nonumber \\
&=  p_1 \cdot \mathbbm{E}[Y_1|Y_1 \in [k-\Delta_1^*, k],Y=k]+(1-p_1)\cdot(k-\Delta_1^*) \label{eq:E1gt}
\end{align}
where $q_1:= P(Y_1 \ge k - \Delta_1^*|Y=k) = \frac{P(Y=k,Y_1 \ge k - \Delta_1^*)}{\mathcal{B}}=\frac{P(Y=k,Y_1 \in [k-\Delta_1^*, k])}{P(Y_1 \in [k-\Delta_1^*, k])}=p_1$. Thus $\mathbbm{E}[Y_1|Y=k] \le E_1$ and $\Delta_k^*=\mathbbm{E}[Y_0-Y_1|Y=k] = \mathbbm{E}[Y_0|Y=k]-\mathbbm{E}[Y_1|Y=k] \ge E_0 - E_1$.

\subsection{Proof of Proposition \ref{prop:bATEUB1}}
I prove a generalization that accommodates optimization error and an optional strengthening of SI:
\begin{assumption*}[TP2 (total positivity between $Y_0$ and $Y_1$)]
$$f_{Y_0,Y_1}(y_0,y_1) \cdot f_{Y_0,Y_1}(y_0',y_1') \le f_{Y_0,Y_1}(\max\{y_0,y_0'\},\max\{y_1,y_1'\}) \cdot f_{Y_0,Y_1}(\min\{y_0,y_0'\},\min\{y_1,y_1'\})$$
\end{assumption*}
\noindent Assumption TP2 , also known as ``affiliation'' between $Y_0$ and $Y_1$ \citep{MilgromWeber} implies SI and is itself implied by rank invariance. It allows for an improvement in the upper bound from Proposition \ref{prop:bATEUB1}.
\begin{proposition*}
Suppose that $w(\Delta|Y_0^*=k)$ and $w(\Delta|Y_1^*=y^I)$ are both bounded from above by $M$ with probability one, and $w(\Delta)$ is both bounded from above by $M_\Delta$ with probability one, SI holds for $(Y_0^*,Y_1^*)$, and ERRORBOUND holds. Consider values $y_L \le k - E$ and $y_U \ge y^I + E$, and assume $F_0$ is continuous on $[y_L,y_L+\Delta_0^*]$ and $Y_1$ is on  $[y_U-\Delta_0^*,y_U]$. Then, with $Q_k^*$ given by \eqref{eq:qkgen}:
\begin{enumerate}
	\item $\mathbbm{E}[Y_0-Y_1|Y_0 \ge y_L, Y_1 \le y_U] \le Q_k^* + 2M_\Delta$. 
	\item $\mathbbm{E}[Y_0-Y_1|Y^*_0 \ge k, Y^*_1 \le y^I] \le Q_k^* + M \cdot \left( \min \left\{\frac{F(y^I) \cdot (1-F(y^I))}{\mathcal{B}},1\right\} + \min \left\{\frac{F(k^-) \cdot (1-F(k^-))}{\mathcal{B}},1\right\}\right) \\ \le Q_k^* + M \cdot \left\{\frac{1}{2 \mathcal{B}},2\right\}$ under CONVEX and NOTCHBOUND
\end{enumerate}	
Furthermore under CONVEX and NOTCHBOUND, $\mathbbm{E}[Y_0-Y_1|Y^*_0 \ge k, Y^*_1 \le y^I] \le Q_k^* + M \cdot \left(\frac{1-F(y^I)}{1-F(y^I)+\mathcal{B}}+\frac{F(k)}{F(k)+\mathcal{B}}\right)$ if SI is replaced by TP2 of $(Y_0^*,Y_1^*)$, and if we assume directly that $Y_0$ and $Y_1$ satisfy TP2 then $\mathbbm{E}[Y_0-Y_1|Y_0 \ge y_L, Y_1 \le y_U] \le Q_k^* + M_\Delta \cdot \left(\frac{1-F(y_U)}{1-F(y_U)+\mathcal{B}}+\frac{F(y_L)}{F(y_L)+\mathcal{B}}\right)$.
\end{proposition*} 
To begin the proof, iterate expectations and use the definition of $\Delta_0^*$ (cf. \eqref{eq:E0lt} and \eqref{eq:E0gt}):
\begin{align*}
\mathbbm{E}[Y_0|Y_0 \ge y_L, Y_1 \le y_U] &= \mathbbm{E}[Y_0|Y_0 \in [y_L, y_L+\Delta_0^*]]+\frac{A_0}{\mathcal{B}} \cdot \left\{\mathbbm{E}[Y_0|Y_0 > y_L+\Delta_0^*,Y_1 \le y_U]\right.\\
&\left. \hspace{2in} - \mathbbm{E}[Y_0|Y_0 \in [y_L, y_L+\Delta_0^*],Y_1 > y_U]\right\}
\end{align*}
where $A_0:= P(Y_0 \ge y_L+\Delta_0^*, Y_1 \le  y_U)=P(Y_0 \in [y_L,y_L + \Delta_0^*], Y_1 > y_U) = P(Y_0 \le y_L + \Delta_0^*, Y_1 > y_U)$. The second equality holds given that $P(Y_0 \in [y_L,y_L+\Delta_0^*]) = \mathcal{B}$ using continuity, and the third that $P(Y_0 \le k, Y_1 > y^I)=0$ and $y_U \ge y^I \ge k \ge y_L$.

Let $u_0(y) = \operatorname{esssup} \operatorname{supp}\{Y_0|Y_1=y\}$ and $\ell_0(y) = \operatorname{essinf} \operatorname{supp}\{Y_0|Y_1=y\}$. Let $u^*_0(y) = \operatorname{esssup} \operatorname{supp}\{Y_0|Y_1=y\}$ and $\ell^*_0(y) = \operatorname{essinf} \operatorname{supp}\{Y_0|Y_1=y\}$. By SI, $\ell^*_0(y)$ and $u^*_0(y)$ are weakly increasing in $y$. 

Turning to the term in brackets:
\begin{align}
&\mathbbm{E}[Y_0|Y_0 > y_L+\Delta_0^*,Y_1 \le y_U] - \mathbbm{E}[Y_0|Y_0 \in [y_L, y_L+\Delta_0^*],Y_1 > y_U] \nonumber \\
&= \int dF_{Y_1|Y_0 > y_L+\Delta_0^*,Y_1 \le y_U}(y) \cdot \mathbbm{E}[Y_0|Y_0 > y_L+\Delta_0^*,Y_1=y] \nonumber \\
&\hspace{2in}-\int dF_{Y_1|Y_0 \in [y_L, y_L+\Delta_0^*],Y_1 > y_U}(y) \cdot \mathbbm{E}[Y_0|Y_0 \in [y_L, y_L+\Delta_0^*],Y_1=y] \nonumber \\
&\le \int dF_{Y_1|Y_0 > y_L+\Delta_0^*,Y_1 \le y_U}(y) \cdot u_0(y)-\int dF_{Y_1|Y_0 \in [y_L, y_L+\Delta_0^*],Y_1 > y_U}(y) \cdot \ell_0(y) \label{eq:mu0ell0}
\end{align}
by the definitions of $u_0$ and $\ell_0$. Similarly, by iterating expectations we have:
\begin{align*}
\mathbbm{E}[Y_1|Y_0 \ge y_L, Y_1 \le y_U] &= \mathbbm{E}[Y_1|Y_1 \in [y_U-\Delta_1^*,y_U]]+\frac{A_1}{\mathcal{B}} \cdot \left\{\mathbbm{E}[Y_1|Y_1 < y_U-\Delta_1^*,Y_0 \ge y_L]\right.\\
&\left. \hspace{2in} - \mathbbm{E}[Y_1|Y_1 \in [y_U-\Delta_1^*,y_U],Y_0 < y_L]\right\}
\end{align*}
where $A_1 = P(Y_0 < y_L, Y_1 \in [y_U-\Delta_1^*,y_U]) = P(Y_1 < y_U-\Delta_1^*,Y_0 \ge y_L)$ and again
\begin{align}
&\mathbbm{E}[Y_1|Y_1 < y_U-\Delta_1^*,Y_0 \ge y_L]-\mathbbm{E}[Y_1|Y_1 \in [y_U-\Delta_1^*,y_U],Y_0 < y_L]\nonumber \\
&= \int dF_{Y_0|Y_1 < y_U-\Delta_1^*,Y_0 \ge y_L}(y) \cdot \mathbbm{E}[Y_1|Y_1 < y_U-\Delta_1^*,Y_0 = y_L]\nonumber \\
&\hspace{2in}-\int dF_{Y_0|Y_1 \in [y_U-\Delta_1^*,y_U],Y_0 < y_L}(y) \cdot \mathbbm{E}[Y_1|Y_1 \in [y_U-\Delta_1^*,y_U],Y_0 = y] \nonumber \\
&\ge \int dF_{Y_0|Y_1 < y_U-\Delta_1^*,Y_0 \ge y_U}(y) \cdot \ell_1(y)-\int dF_{Y_0|Y_1 \in [y_U-\Delta_1^*,y_U],Y_0 < y_L}(y) \cdot u_1(y) \label{eq:mu1ell1}
\end{align} 
with $u_1(y) = \operatorname{esssup} \operatorname{supp}\{Y_1|Y_0=y\}$ and $\ell_1(y) = \operatorname{essinf} \operatorname{supp}\{Y_1|Y_0=y\}$.

With these preliminaries, I first establish the second claim of the Proposition above. Take $y_L=k$, $y_U=y^I$ and consider the special case in which $E=0$ so that $Y_d=Y_d^*$, and thus that $u_d=u_d^*$ and $\ell_d = \ell_d^*$ are increasing by SI, and $w(\Delta|Y_0=k)=u_0(k) - \ell_0(k) \le M$ and $w(\Delta|Y_1=k)=\ell_1(y^I) - u_1(y^I) \ge -M$ by assumption. Then by the above:
$$\mathbbm{E}[Y_0|Y_0 > k+\Delta_0^*,Y_1 \le y^I] - \mathbbm{E}[Y_0|Y_0 \in [k, k+\Delta_0^*],Y_1 > y^I] \le u_0(y^I) - \ell_0(y^I) \le M$$
$$\mathbbm{E}[Y_1|Y_0 > k+\Delta_0^*,Y_1 \le y^I] - \mathbbm{E}[Y_1|Y_0 \in [k, k+\Delta_0^*],Y_1 > y^I] \ge \ell_1(k) - u_1(k) \ge -M$$
Note that $A_0 \in [0,\mathcal{B}]$. However we also have by SI that:
$$1-F_1(y^I)-A_0=P(Y_1 > y^I, Y_0 > k+\Delta_0^*) \ge P(Y_1 > y^I)\cdot P(Y_0 > k+\Delta_0^*) = (1-F_1(y^I))^2$$
which implies the possibly tighter bound that $A_0 \le F_1(y^I) \cdot (1-F_1(y^I))$. Thus we have $0 \le A_0 \le \min\{\mathcal{B},F(y^I) \cdot (1-F(y^I))\}$, and therefore
\begin{align*}
\mathbbm{E}[Y_0|Y^*_0 \ge k, Y^*_1 \le y^I] &\le
\mathbbm{E}[Y_0|Y_0 \in [k, k+\Delta_0^*]] + \min \left\{\frac{F(y^I) \cdot (1-F(y^I))}{\mathcal{B}},1\right\} \cdot M
\end{align*}
using Eq. \eqref{eq:iddensnotch} as well as CONVEX and NOTCHBOUND in the last step. Again using SI:
$$F_0(k)-A_1=P(Y_1 \le y^I-\Delta_1^*, Y_0 \le k) \ge P(Y_1 \le y^I-\Delta_1^*)\cdot P(Y_0 \le k) = F_0(k)^2$$	
Therefore $A_1 \le \min\{\mathcal{B},F_0(k)(1-F_0(k))\}$ and
\begin{align*}
\mathbbm{E}[Y_1|Y^*_0 \ge k, Y^*_1 \le y^I]&\ge
\mathbbm{E}[Y_1|Y_1 \in [y^I-\Delta_1^*],y^I] - \min \left\{\frac{F(k^-) \cdot (1-F(k^-))}{\mathcal{B}},1\right\} \cdot M
\end{align*}
again using Eq. \eqref{eq:iddensnotch}. Combining, we have: \small
$$\mathbbm{E}[Y_1|Y^*_0 \ge k, Y^*_1 \le y^I] \le Q_k^* + M \cdot \left( \min \left\{\frac{F(y^I) \cdot (1-F(y^I))}{\mathcal{B}},1\right\} + \min \left\{\frac{F(k^-) \cdot (1-F(k^-))}{\mathcal{B}},1\right\}\right)$$ \normalsize
where $Q_k^* =  \mathbbm{E}[Y_0|Y_0 \in [k, k+\Delta_0^*]]-\mathbbm{E}[Y_1|Y_1 \in [y^I-\Delta_1^*],y^I]$ by continuity. Since $p(1-p) \le 1/4$ for any $p \in [0,1]$, the bounds that involve $\mathcal{B}$ always bind when $\mathcal{B} > 1/4$. This leads to the compact form $\Delta_k^* \le Q_k^* + M \cdot \min \left\{\frac{1}{2 \mathcal{B}},2\right\}$ which is always also valid.

To establish the first claim in the Proposition, now consider $Y_0,Y_1$ in the case of ERRORBOUND. Note that conditional on $Y_{1i}=y$, $Y^*_{1i} \in [y-E,y+E]$ with probability one, and thus:
$$ [\ell_0(y),u_0(y)] \subseteq \left[y+\inf_{t \in [y-E,y+E]} \{\ell^*_0(t)-t\}, \quad y+\sup_{t \in [y-E,y+E]} \{u^*_0(t)-t\}\right] := [\bar{\ell}_0(y),\bar{u}_0(y)]$$ 
since any $s \in \operatorname{supp}\{Y_0|Y_1=y\}$ satisfies $s=y+\delta$ for some 
\begin{align*}
\delta &\in \operatorname{supp}\{\Delta|Y_1=y\} = \operatorname{supp}\{Y_0^*-Y_1^*|Y_1=y\}  \subseteq \bigcup_{t \in \operatorname{supp}\{Y_1^*|Y_1=y\}} \operatorname{supp}\{Y_0^*-Y_1^*|Y_1=y,Y_1^*=t\} \\
&\subseteq \bigcup_{t \in \operatorname{supp}\{Y_1^*|Y_1=y\}} \operatorname{supp}\{Y_0^*-t|Y_1^*=t\}  \subseteq \bigcup_{t \in \operatorname{supp}\{Y_1^*|Y_1=y\}} \left[\ell^*_0(t)-t, u^*_0(t)-t\right] 
\end{align*}
We can rewrite
$$\bar{\ell}_0(y) = y+\inf_{t \in [y-E,y+E]} \{u^*_0(t)-t\} = \inf_{e \le |E|} \{u^*_0(y-e)+e\} \textrm{ and } \bar{\ell}_0(y) = \sup_{e \le |E|} \{u^*_0(y-e)+e\}$$
As a sup or inf of increasing functions of $y$, $\bar{\ell}_0(y)$ and $\bar{u}_0(y)$ are also increasing in $y$. This does not require SI to hold for $Y_1^*,Y_0^*$ rather we have shown that this property is inherited by $\bar{\ell}_0(y)$ and $\bar{u}_0(y)$ from SI for $Y_0,Y_1$. Note finally that $\bar{u}_0(y)-\bar{\ell}_0(y) \le M_\Delta$ for any $y \in \operatorname{supp}\{Y_1\}$, since for any $t \in [y-E,y+E]$, both $\ell^*_0(t)-t$ and $u^*_0(t)-t$ are in $\operatorname{supp}\{\Delta\}$. As before, we have $[\ell_1(y),u_1(y)] \subseteq [\bar{\ell}_1(y),\bar{u}_1(y)]$
where again $\bar{\ell}_1(y)$ and $\bar{u}_1(y)$ are increasing in $y$ and $\bar{u}_1(y)-\bar{\ell}_1(y) \le M_\Delta$ for any $y \in \operatorname{supp}\{Y_0\}$.

Combining, the inequalities \eqref{eq:mu0ell0} and \eqref{eq:mu1ell1} both hold with $u_d$ and $\ell_d$ replaced by $\bar{u}_d$ and $\bar{\ell}_d$, and thus $\mathbbm{E}[Y_0|Y_0 \ge k, Y_1 \le y^I] \le \bar{u}_0(y_U)-\bar{\ell}_0(y_U) \le M_\Delta$ while $\mathbbm{E}[Y_1|Y_0 \ge k, Y_1 \le y^I] \ge \bar{\ell}_1(y_L)-\bar{u}_1(y_L) \ge -M_\Delta$. Combining with $A_0,A_1 \in [0, \mathcal{B}]$ we obtain the final result that $\mathbbm{E}[Y_0-Y_1|Y_0 \ge y_L, Y_1 \le y_U] \le Q_k^* + 2M_\Delta$.\\

To establish the final two claims of the Proposition, note that if we strengthen SI to TP2 of $(Y_0,Y_1)$, then we have (see e.g. Thm. 23 of \cite{MilgromWeber}) that
$$P(Y_0 > y_L + \Delta_0^*,Y_1 > y_U|Y_0 \ge y_L) \ge P(Y_0 > y_L + \Delta_0^*|Y_0 \ge y_L) \cdot P(Y_1 > y_U|Y_0 \ge y_L)$$
which rearranges to $(1-F_0(y_L))\cdot (1-F_1(yU)-A_0) \ge (1-F_1(y_U))^2$ yielding the bound that $A_0 \le \mathcal{B} \cdot \frac{1-F(y_U)}{1-F(y_U)+\mathcal{B}}$. By noting that $(-Y_0,-Y_1)$ are also TP2, we obtain similarly that
$$P(Y_1 \le y_U - \Delta_1^*,Y_0 \le y_L|Y_1 \le y_U) \ge P(Y_1 \le y_U - \Delta_1^*|Y_1 \le y_U) \cdot P(Y_0 \le y_L|Y_1 \le y_U)$$
which rearranges to $(F_0(y_L)+\mathcal{B}) \cdot (F_0(y_L)-A_1) \ge F_0(y_L)^2$ yielding the bound that $A_1 \le \mathcal{B} \cdot \frac{F_0(y_L)}{F_0(y_L)+\mathcal{B}}$. These always improve the bounds $A_0,A_1 \le \mathcal{B}$, so we have under TP2 of $(Y_0^*,Y_1^*)$:
$$\mathbbm{E}[Y_0-Y_1|Y^*_0 \ge k, Y^*_1 \le y^I] \le Q_k^* + M \cdot \left(\frac{1-F_1(y_U)}{1-F_1(y_U)+\mathcal{B}}+\frac{F_0(y_L)}{F_0(y_L)+\mathcal{B}}\right)$$
The claim $\mathbbm{E}[Y_0-Y_1|Y^*_0 \ge k, Y^*_1 \le y^I] \le Q_k^* + M \cdot \left(\frac{1-F(y^I)}{1-F(y^I)+\mathcal{B}}+\frac{F(k)}{F(k)+\mathcal{B}}\right)$ 
now follows in the case of no optimization error, with $y_L=k,y_U=y^I$ under CONVEX and NOTCH (allowing $F_0(y_L)$ and $F_1(y_U)$ to be identified) using the improved bounds on $A_0$ and $A_1$ and that TP2 implies SI (Thm 5. of \citealt{MilgromWeber}). If we instead assume TP2 of $(Y_0,Y_1)$ directly, then :
\begin{equation} \label{eq:tp2direct}
\mathbbm{E}[Y_0-Y_1|Y_0 \ge y_L, Y_1 \le y_U] \le Q_k^* + M_\Delta \cdot \left(\frac{1-F(y_U)}{1-F(y_U)+\mathcal{B}}+\frac{F(y_L)}{F(y_L)+\mathcal{B}}\right)
\end{equation}
following the argument above (using again that TP2 implies SI) but with the replacement of $M_\Delta$ instead of $M$, since $w(\Delta|Y_d^*) \le M$ does not directly bound $w(\Delta|Y_d)$, but $w(\Delta) \le M_\Delta$ does. A sufficient condition for $(Y_0,Y_1)$ to inherit TP2 of $(Y_0^*,Y_1^*)$ is given at the end of Appendix \ref{sec:optimizationerror}.

\subsection{Proof of Theorems \ref{thmblc} and \ref{thm:blcwitherrors}}

The proof is structured to accommodate the general structure of Theorem \ref{thm:blcwitherrors} in which there is potential optimization error of bounded magnitude $E$, in which case the results depend on researcher-specified values $y_L \le k - E$ and $y_U \ge y^I + E$. The proof of Theorem \ref{thmblc} is a special case in which $y_L=k$ and $y_U=y^I$ (or $y_L=y_U=k$ in the further special case of a kink with no error).

I first consider the case of local BLC, i.e. BC(0,0). Accordingly, consider a $F_{d}(y)$ that is log-concave on an interval between $y$ and $y+v$. In what follows, we will consider $y=y_L$ and positive $v$ for $d=0$ and negative such $v$ with $y=y_U$ for $d=1$. Concavity implies that a first-order Taylor expansion for $\ln F_{d}(y+v)$ from $y$ overshoots: i.e. $\ln F_{d}(y+v) \le \ln F_{d}(y) + v \cdot \frac{d}{dy} \ln F_{d}(y)$.  Similarly, that $\ln(1-F_{d}(y))$ is concave between $y$ and $y+v$ implies that $\ln(1-F_{d}(y+v)) \le \ln(1-F_{d}(y)) + v \cdot \frac{d}{dy} \ln (1-F_{d}(y))$. These two inequalities can be rearranged to put upper and lower bounds on $F_{d}(y+v)$: 
\begin{equation} \label{eq:cdfsandwich} 1-(1-F_{d}(y))e^{-\frac{f_{d}(y)}{1-F_{d}(y)}v} \le F_{d}(y+v) \le F_{d}(y) e^{\frac{f_{d}(y)}{F_{d}(y)}v}
\end{equation}
An analogous expression to Eq. \eqref{eq:cdfsandwich} is obtained in Theorem 1 of \citet{dumbgen_bi-log-concave_2017}.

Defining $u = F_{0}(y_L+v)$, we can use the substitution $v = Q_{0}(u)-y_L$ to translate \eqref{eq:cdfsandwich} with $y=y_L$ into bounds on the conditional quantile function of $Y_{0i}$, evaluated at $u$:
\begin{equation} \label{eq:quantilesandwich1}
\frac{F_{0}(y_L)}{f_{0}(y_L)}\cdot \ln\left(\frac{u}{F_{0}(y_L)}\right) \le Q_{0}(u) -y_L  \le -\frac{1-F_{0}(y_L)}{f_{0}(y_L)}\cdot \ln\left(\frac{1-u}{1-F_{0}(y_L)}\right)
\end{equation}
And similarly for $Y_1$, letting $y=y_U$ and $u = F_{1}(y_U-v)$:
\begin{equation} \label{eq:quantilesandwich2}
\frac{1-F_{1}(y_U)}{f_{1}(y_U)}\cdot \ln\left(\frac{1-u}{1-F_{1}(y_U)}\right) \le y_U-Q_{1}(u)  \le -\frac{F_{1}(y_U)}{f_{1}(y_U)}\cdot \ln\left(\frac{u}{F_{1}(y_U)}\right)
\end{equation}
\noindent A lower bound for a quantity $Q_k^*$ taking the form of Eq. \ref{eq:qkgen} (i.e. Eq. \ref{eq:twoterms} when $y_L=k,y_U=y^I$):
$$	Q_k^*= \left\{\frac{1}{\mathcal{B}}\int_{F_0(y_L)}^{F_0(y_L)+\mathcal{B}} Q_0(u)\cdot du - y_L\right\} + \left\{y_U- \frac{1}{\mathcal{B}}\int_{F_1(y_U)-\mathcal{B}}^{F_1(y_U)} Q_1(u)\cdot du\right\}-(y_U-y_L)$$
for a quantity $\mathcal{B}>0$ is
\small{\begin{align*}
	&\frac{F_{0}(y_L)}{f_{0}(y_L)\cdot \mathcal{B}}\int_{F_{0}(y_L)}^{F_{0}(y_L)+\mathcal{B}} \ln\left(\frac{u}{F_{0}(y_L)}\right)du + \frac{1-F_{1}(y_U)}{f_{1}(y_U)\cdot \mathcal{B}}\int_{F_{1}(y_U)-\mathcal{B}}^{F_{1}(y_U)}  \ln\left(\frac{1-u}{1-F_{1}(y_U)}\right)du-(y_U-y_L)
	\end{align*}} \normalsize
	We can write this as $g(F_{0}(y_L),f_{0}(y_L),\mathcal{B})+ h(F_{1}(y_U),f_{1}(y_U),\mathcal{B})-(y_U-y_L)$ where
	\begin{align*}
&g(a,b,x) := \frac{a}{bx}\int_a^{a+x} \ln\left(\frac{u}{a}\right)du = \frac{a^2}{bx}\int_1^{1+\frac{x}{a}} \ln\left(u\right)du\\
&= \frac{a^2}{bx}\left.\left\{u \ln(u)-u\right\}\right|_1^{1+\frac{x}{a}}= \frac{a^2}{bx}\left\{\left(1+\frac{x}{a}\right)\ln\left(1+\frac{x}{a}\right)- \frac{x}{a}\right\}= \frac{a}{bx}\left(a+x\right)\ln\left(1+\frac{x}{a}\right)-\frac{a}{b}
\end{align*}
and $h(a,b,x) := \frac{1-a}{bx}\int_{a-x}^{a} \ln\left(\frac{1-v}{1-a}\right)dv = g(1-a,b,x)$. To see that $h(a,b,x)=g(1-a,b,x)$, note that for an arbitrary integrable function $\phi$:
\begin{align} \label{eq:cov1}
\int_{a-x}^{a} \phi\left(\frac{1-v}{1-a}\right)dv=\int_{1-a}^{1-a+x} \phi\left(\frac{u}{1-a}\right)du
\end{align}
by a change of variables $u=1-v$. Similarly, an upper bound for $Q_k^*$ is: 
\begin{align*}
&-\frac{1-F_{0}(y_L)}{f_{0}(y_L)(\mathcal{B})} \int_{F_{0}(y_L)}^{F_{0}(y_L)+\mathcal{B}} \ln\left(\frac{1-u}{1-F_{0}(y_L)}\right) du-\frac{F_{1}(y_U)}{f_{1}(y_U)(\mathcal{B})}\int_{F_{1}(y_U)-\mathcal{B}}^{F_{1}(y_U)}  \ln\left(\frac{u}{F_{1}(k)}\right)du - (y_U-y_L)\\
&=\tilde{g}(F_{0}(y_L),f_{0}(y_L),\mathcal{B})+ \tilde{h}(F_{1}(y_U),f_{1}(y_U),\mathcal{B}) - (y_U-y_L)
\end{align*}
where $\tilde{g}(a,b,x) := -\frac{1-a}{bx}\int_a^{a+x} \ln\left(\frac{1-u}{1-a}\right)du = -g(1-a,b,-x)$ and similarly $\tilde{h}(a,b,x) := -\frac{a}{bx}\int_{a-x}^{a} \ln\left(\frac{v}{a}\right)dv=\tilde{g}(1-a,b,x)=-g(a,b,-x)$, with the middle inequality again following by change of variables as in Eq. \ref{eq:cov1}.

In the case of $BC(s,t)$ with $s,t \ne 0$, we have by analogous logic to Eq. \eqref{eq:cdfsandwich} that
\begin{equation} \label{eq:cdfsandwich_s} 1-(1-F_{d}(y))\left(1-t\cdot \frac{f_{d}(y)}{1-F_{d}(y)}v\right)^{1/t} \le F_{d}(y+v) \le F_{d}(y) \left(1+s \cdot \frac{f_{d}(y)}{F_{d}(y)}v\right)^{1/s}
\end{equation}
for either $d \in \{0,1\}$. Making the same substitution $u = F_{0}(y_L+v)$ as in the BLC case:
\begin{equation} \label{eq:quantilesandwich1_s}
\frac{F_{0}(y_L)}{f_{0}(y_L)}\cdot \frac{\left(\frac{u}{F_{0}(y_L)}\right)^s-1}{s} \le Q_{0}(u) - y_L  \le -\frac{1-F_{0}(y_L)}{f_{0}(y_L)}\cdot \frac{\left(\frac{1-u}{1-F_{0}(y_L)}\right)^t-1}{t}
\end{equation}
and then with $u = F_{1}(y_U-v)$:
\begin{equation} \label{eq:quantilesandwich2_s}
\frac{1-F_{1}(y_U)}{f_{1}(y_U)}\cdot \frac{\left(\frac{1-u}{1-F_{1}(y_U)}\right)^t-1}{t} \le y_U-Q_{1}(v)  \le -\frac{F_{1}(y_U)}{f_{1}(y_U)}\cdot \frac{\left(\frac{v}{F_{1}(y_U)}\right)^s-1}{s}
\end{equation}
The integrals have the same structure as in the case of $s=t=0$, with the function $\phi(x)=\ln(x)$ appearing in Eq. \ref{eq:cov1} replaced with $\phi(x) = (x^s-1)/s$ (which recovers $\ln(x)$ as $s \rightarrow 0$).

A lower bound for $Q_k^*$ is now:
\small{\begin{align*}
	&\frac{F_{0}(y_L)}{f_{0}(y_L)\cdot \mathcal{B}}\int_{F_{0}(y_L)}^{F_{0}(y_L)+\mathcal{B}} \frac{\left(\frac{u}{F_{0}(y_L)}\right)^s-1}{s}du + \frac{1-F_{1}(y_U)}{f_{1}(y_U)\cdot \mathcal{B}}\int_{F_{1}(y_U)-\mathcal{B}}^{F_{1}(y_U)}  \frac{\left(\frac{1-u}{1-F_{1}(y_U)}\right)^t-1}{t}du - (y_U-y_L)\\
	&=g_s(F_{0}(y_L),f_{0}(y_L),\mathcal{B})+ g_t(1-F_{1}(y_U),f_{1}(y_U),\mathcal{B}) - (y_U-y_L)
	\end{align*}} \normalsize
	\noindent where $g_s(a,b,x):= \frac{a^2}{bx} \cdot \left(\frac{(1+\frac{x}{a})^{s+1}-1}{s(s+1)} - \frac{x}{sa} \right)$. Some algebra simplifies this to $g_s(a,b,x)= \frac{a}{b(s+1)}\left[\left(1+ \frac{a}{x}\right) \frac{\left(1+\frac{x}{a}\right)^s-1}{s}-1\right]$ which recovers the BLC result $g=g_0$ in the limit of $s \rightarrow 0$. Similarly, an upper bound is: \small
	\begin{align*}
&-\frac{1-F_{0}(y_L)}{f_{0}(y_L)\cdot \mathcal{B}}\int_{F_{0}(k)}^{F_{0}(k)+\mathcal{B}} \frac{\left(\frac{1-u}{1-F_{0}(y_L)}\right)^t-1}{t}du - \frac{F_{1}(y_U)}{f_{1}(y_U)\cdot \mathcal{B}}\int_{F_{1}(y_U)-\mathcal{B}}^{F_{1}(y_U)}  \frac{\left(\frac{u}{F_{1}(k)}\right)^s-1}{s}du - (y_U-y_L)\\
&=-g_t(1-F_{0}(y_L),f_{0}(y_L),-\mathcal{B})-g_s(F_{1}(y_U),f_{1}(y_U),-\mathcal{B}) - (y_U-y_L).
\end{align*} \normalsize

Turning to the question of sharpness, I focus on the case without errors with $y_L=k$ and $y_U=y^I$, and the identified set that holds under BC(s,t) with CONT: $\Delta^{L}_k:= g_s(F(k^-), f(k^{-}), \mathcal{B}) + g_t\left(1-F(y^I),f(y^{I+}),\mathcal{B}\right) - (y^I-k),  \Delta^{U}_k:= -g_t(1-F(k^{-}),f(k^{-}), -\mathcal{B} ) -g_s\left(F(y^I),f(y^{I+}), -\mathcal{B}\right) - (y^I-k)$. The identified set is sharp if for any $\Delta \in [\Delta_k^{L},\Delta_k^{U}]$, there exists a distribution of potential outcomes consistent with the data and BC(s,t) for which $Q_k^*$ is equal to $\Delta$.

Consider first the lower bound $\Delta_k^{L}$. Let $Q(u)$ denote the quantile function of the data distribution $F$. Define $Q^L_0(u)$ to be a function that is equal to $Q(u)$ for $0 \le u \le F(k^-)$, and equal to $k$ plus the lower bound on the LHS of Eq. \eqref{eq:quantilesandwich1_s} for all $F(k^-) \le u \le 1$. Similarly, let $Q^L_1(u)$ be a function equal $y^I$ minus the lower bound on the LHS of Eq. \eqref{eq:quantilesandwich2_s} for all $0 \le u \le F(y^{I-})$ and equal to $Q(u)$ for $0 \le u \le F(k^-)$ for all $F(y^{I-}) \le u \le 1$. Both of these are valid quantile functions defined on $u \in [0,1]$: they are increasing and left continuous on $[0,1]$. Furthermore, $Q^L_{0}(u)$ and $Q^L_{1}(u)$ are locally BLC inside the bunching region by construction. For concreteness, one can then build a joint distribution $(Y_0,Y_1)$ from these marginal quantile functions above by imposing rank invariance. $Q_k^* = \Delta_k^{*L}$ when $(Y_0,Y_1)$ follows this distribution, by the same steps used above to prove that $Q^*_k \ge \Delta^L_k$ generally, with the weak inequalities replaced as equalities. To build a distribution of $(Y_0,Y_1)$ that meets the upper bound $\Delta_k^U$, we proceed analogously, defining $Q^U_{0}(u)$ and $Q^U_{1}(u)$ from the upper bounds in \eqref{eq:quantilesandwich1_s}-\eqref{eq:quantilesandwich2_s}.

To show that we can satisfy $Q_k^* = \Delta$ also for any intermediate value $\Delta \in (\Delta_k^{L},\Delta_k^{U})$, let $Q^\alpha_{0}(u) = (1-\alpha)\cdot Q^L_{0}(u)+\alpha\cdot Q^U_{0}(u)$ and $Q^\alpha_{1}(u) = (1-\alpha)\cdot Q^L_{1}(u)+\alpha\cdot Q^U_{1}(u)$ for $\alpha \in [0,1]$, again imposing rank invariance to obtain a joint distribution of $Y_0,Y_1$. Under this distribution, BLC remains satisfied via \eqref{eq:quantilesandwich1_s}-\eqref{eq:quantilesandwich2_s}, $Q_k^* = (1-\alpha) \cdot \Delta_k^L + \alpha\cdot \Delta_k^U$. Since we have constructed a distribution delivering $Q_k^*$ at any value in $[\Delta_k^{L},\Delta_k^{U}]$ and satisfying rank invariance, those bounds are also sharp for $\Delta_k^*$ under Assumption RANK, CONVEX and NOTCH (by Proposition \ref{prop:buncherATE}).

\subsection{Proof of Proposition \ref{prop:bATEUB2}}
I prove a generalization of Proposition \ref{prop:bATEUB2} accommodating optimization error, and the t-concavity of $1-F_0$ is assumed to hold everywhere above $k-E$ and s-concavity of $F_1$ is assumed to hold everywhere below $y^I+E$, with NOTCHBOUND and ERRORBOUND. Proposition \ref{prop:bATEUB2} holds as a special case when $E=0$ and NOTCH holds. Consider any values $y_L \le k - E$ and $y_U \ge y^I + E$.

Using SI, we  have that  
\begin{align*}
&\mathbbm{E}[Y_0|Y_0 \ge y_L, Y_1 \le y_U] - \mathbbm{E}[Y_1|Y_0 \ge y_L, Y_1 \le y_U] \le \mathbbm{E}[Y_0|Y_0 \ge y_L]-\mathbbm{E}[Y_1|Y_1 \le y_U]\\
&=\mathbbm{E}[Y_0-y_L|Y_0 \ge y_L]+\mathbbm{E}[y_U-Y_1|Y_1 \le y_U] + (y_U-y_L)\\
&=\frac{1}{1-F_0(y_L)}\int_{F_0(y_L)}^1 \{Q_0(u)-y_L\} \cdot du + \frac{1}{F_1(y_U)}\int_{0}^{F_1(y_U)} \{y_U-Q_1(u)\} \cdot du + (y_U-y_L)
\end{align*}
by CONT. By t-concavity of $1-F_0$ and Eq. \eqref{eq:quantilesandwich1_s}: $Q_{0}(u) - y_L  \le -\frac{1-F_{0}(y_L)}{f_{0}(y_L)}\cdot \frac{\left(\frac{1-u}{1-F_{0}(y_L)}\right)^t-1}{t}$. Similarly by s-concavity of $F_1$ and \eqref{eq:quantilesandwich2_s}: $y_U-Q_{1}(v)  \le -\frac{F_{1}(y_U)}{f_{1}(y_U)}\cdot \frac{\left(\frac{v}{F_{1}(y_U)}\right)^s-1}{s}$.
Evaluating the integrals, we have that $\mathbbm{E}[Y_0|Y_0 \ge y_L, Y_1 \le y_U]$ is upper bounded by
\begin{align*}
&-g_t(1-F_{0}(y_L),f_{0}(y_L),-(1-F_{0}(y_L)))-g_s(F_{1}(y_U),f_{1}(y_U),-F_{1}(y_U))-(y_U-y_L)
\end{align*}
which is equal to $\frac{1-F_0(y_L)}{f_0(y_L)(t+1)}+\frac{F_1(y_U)}{f_1(y_U)(s+1)}-(y_U-y_L)$, using that $g_s(a,b,-a) = -a/b$ for any $s$. Taking $y_L=k$ and $y_U=y^I$ recovers Prop. \ref{prop:bATEUB2} as stated.

\subsection{Proof of Proposition \ref{prop:sandwich}}
Let $I=[a,b]$, and $Z$ be a random variable with distribution $G$. Note that $Y = T(Z)$ and $Z = T^{-1}(Y)$ with probability one, and since $T^{-1}$ is strictly increasing we thus have that $F(y) = P(Y \le y) = P(Z \le T^{-1}(y)) = G(T^{-1}(y))$ and $f(y) = \frac{g(T^{-1}(y))}{T'(T^{-1}(y))}$.

Let $h_F(y) := f(y)/\{1-F(y)\}$ its hazard function and $r_F(y) = f(y)/F(y)$. I begin with log-concavity of $F(y)$ and $1-F(y)$ ($s=0$), and then use this result to extend to the case of general $s$. $\ln F(y)$ is concave on $I$ if and only if $r_F(y)$ is weakly decreasing on $I$. By the above we have that $r_F(y) = r_G(T^{-1}(y))/T'(T^{-1}(y))$, and thus
\begin{align*}
r'_F(y) &= \frac{d}{dy} \frac{r_G(T^{-1}(y))}{T'(T^{-1}(y))} = \frac{r'_G(T^{-1}(y))-T''(T^{-1}(y))/T'(T^{-1}(y))\cdot  r_G(T^{-1}(y))}{T'(T^{-1}(y))^2} \le 0\\
&\iff r'_G(T^{-1}(y)) \le r_G(T^{-1}(y)) \cdot T''(T^{-1}(y))/T'(T^{-1}(y))\\
&\iff r'_G(z)/r_G(z) \le T''(z)/T'(z) \iff \frac{d}{dz} \ln r_G(z) \le \frac{d}{dz} \ln T'(z)
\end{align*}
where we let $z = T^{-1}(y)$. The other inequality is symmetric.

For general $s$, $F$ is s-concave if and only if the function $f(y)/F(y)^{1-s} = r_F(y) \cdot F(y)^s$ is weakly decreasing (see e.g. \citealt{lahamiaowellner}). This condition is $r'_F(y)\cdot F(y)^s+s r_F(y) F(y)^{s-1} f(y) \le 0$ or equivalently $r'_F(y) + s r^2_F(y) \le 0$. Substituting from above, this condition becomes $ \frac{d}{dz} \ln r_G(z) + s\cdot r_G(z)^2 \le \frac{d}{dz} \ln T'(z) $. The other inequality yields $\frac{d}{dz} \ln h_G(z) + -t \cdot h_G(z)^2 \ge \frac{d}{dz} \ln T'(z)$.

\subsection{Proof of Proposition \ref{prop:Bpartialid}}
By ERRORBOUND, $Y_d^* \le y$ implies $Y_d=Y_d^*+\epsilon \le y+\epsilon_u$ for either $d \in \{0,1\}$, and thus $F_{d}(y) \le F^*_{d}(y+\epsilon_u)$. Similarly $Y_d^* > y$ implies $Y_d=Y_d^*+\epsilon > y-\epsilon_l$, and thus $F_{d}(y) \ge F^*_{d}(y-\epsilon_l)$. Evaluating for $d=0$ at $k-\epsilon_u$ and $k+\epsilon_l$ respectively and combining, we have that $F^*_0(k) \in [F_0(k-\epsilon_u),F_0(k+\epsilon_l)]$. Similarly $F^*_1(y^I) \in [F_1(y^I-\epsilon_u),F_1(y^I+\epsilon_l)]$. Thus $\mathcal{B}^* = F_1^*(y^I)-F^*_0(k) \in [F_1(y^I-\epsilon_u)-F_0(k+\epsilon_l, F_1(y^I+\epsilon_l)-F_0(k-\epsilon_u)]$. The final result follows using that as a probability $\mathcal{B}^*$ must be non-negative.

\subsection{Proof of Proposition \ref{prop:posTEs}}
$P(Y_{0i} \ge y_L, Y_{1i} \le y_U) = P(Y_{0i} \ge y_L) + P(Y_{1i} \le y_U)- P(Y_{0i} \ge y_L \textrm{ or } Y_{1i} \le y_U) = 1 - F_0(y_L) + F_1(y_U) - (1-P(Y_{0i} < y_L, Y_{1i}>y_U)) = F_1(y_U)-F_{0}(y_L) + P(Y_{0i} < y_L, Y_{1i}>y_U)$. $P(Y_{0i} < y_L, Y_{1i}>y_U) = 0$, because $Y_{0i} < y_L, Y_{1i}>y_U \implies Y_{0i} < k-E, Y_{1i}>y^I+E$ which in turn implies $Y^*_{0i} < k, Y^*_{1i} > y^I$ by ERRORBOUND. $Y^*_{0i} < k, Y^*_{1i} > y^I$ occurs with probability zero by NOTCH and CONVEX (cf. Eq. \eqref{hcasesnotch}). $F_1(y_U)-F_{0}(y_L)=F(y_U)-F(y_L)$ by Eq. \eqref{hcasesgeneralwithboundederror}.

\subsection{Proof of Proposition \ref{prop:unisuff}}
In what follows take $\tilde{Y}_{1i}$ to represent an interior solution. If for example $y$ were constrained to be non-negative, then an additional assumption that $P(\tilde{Y}_{1i}>0)=1$ would suffice. Given differentiability of $\pi_i$, $Y_{0i}$ and $\tilde{Y}_{1i}$ each satisfy a first order condition, e.g. $Y_{0i}$ satisfies $\frac{d}{dy}(y-T_{0}(y)+\pi(y;\eta_i))=0 \implies -\frac{d}{dy}\pi(y;\eta_i) = 1-T'_0(y)$. By convexity of $T_0$, $1-T_0'(y)$ is weakly decreasing in $y$ while $-\frac{d}{dy}\pi(y;\eta_i)$ is strictly increasing in $y$, leading to a single crossing point $y$ that is strictly increasing in $\eta_i$ (using 3.) Thus $Y_{0i}=g_0(\eta_i)$ for a strictly increasing function $g_1$. The same argument shows that $\tilde{Y}_{1i} = Y_{1i} = g_1(\eta_i)$ for strictly increasing $g_1$, implying RANK.

Now, using items 1.-2. and 4. of UNI, $A_i = \{Y_{1i} > k \textrm{ and } k-\tilde{T}_{1i}(k)+\pi_i(k) \ge Y_{1i}-\tilde{T}_{1i}(Y_{1i})-N+\pi_i(Y_{1i})\}$. Given that $\pi_i$ is differentiable, $Y_{1i}-\tilde{T}_{1}(Y_{1i})+\pi_i(Y_{1i})-\pi_i(k) \le N+k-\tilde{T}_{1}(k)$ iff
$$\int_{k}^{g_1(\eta_i)} \left(1-\tilde{T}'_{1}(y)+\frac{d}{dy} \pi(y; \eta_i)\right)\cdot dy  \le N+k-\tilde{T}_{1}(k)$$
Note that $u_i(y-\tilde{T}_1(y),y)=y-\tilde{T}_1(y)+\pi_i(y)$ is strictly concave in $y$, given CONVEX and weak convexity of $\tilde{T}_1$. Thus for $y$ below the optimum of $Y_{1i}$, $\frac{d}{dy} u_i(y-\tilde{T}_1(y),y) = 1-\tilde{T}'_{1}(y)+\frac{d}{dy} \pi(y; \eta_i) > 0$. Therefore, the integrand above in parenthesis is non-negative. The integral is therefore increasing in $\eta_i$ because, both (i) $g_1(\eta_i))$ is weakly increasing in $\eta_i$; while (ii) the integrand is itself increasing in $\eta_i$ for each $y$ by Item 3. Thus the inequality is satisfied for all $\eta_i$ below some threshold, i.e. iff $Y_{1i} \le y^I$, for some $y^I > k$. This delivers Assumption NOTCH.

\end{document}